%
%
%
%
\input amstex
\magnification=\magstep1
\input epsf.tex
%
\TagsOnRight
\let\footnote=\plainfootnote
\topskip=32pt
\tolerance=1000
\def\normal{\baselineskip=13pt plus 1pt minus 1pt}
\normal
\parskip 10pt
\nopagenumbers
\newdimen\pagewidth  \newdimen\pageheight
\pagewidth=\hsize  \pageheight=\vsize
\parindent 30pt
%
%
\def\monthyear{\ifcase\month\or
 sije\v cnja \or velja\v ce \or o\v zujka \or travnja \or
 svibnja \or lipnja \or srpnja \or kolovoza \or rujna \or
 listopada \or studenog \or prosinca\fi
 \space\number\year .}
\def\danas{\number\day .\space \monthyear}
\def\mydate{\danas}
%
%
\def\mojnaslov{D. Horvat et al.: Nonmesonic Hypernuclear Decays }

\def\firstheadline{\hbox to \pagewidth{%
    {\baselineskip=10pt
     {\devetrm \hbox{\vtop{ \hsize=2.8truecm \parindent=0pt
           \leftline{Preprint from}
           \leftline{University of Zagreb}     }}}
      \hfil
          {\devetrm  \hbox{\vtop{ \hsize=2.8truecm \parindent=0pt
                \rightline{UZG-TH-1246/94}
                \rightline{\mydate}          }}} }%
       }}

\def\otherheadline{\devetit
  \ifodd\pageno  \qquad \hss \mojnaslov \hss P.\, \folio
    \else P.\, \folio\hss \naspog \hss\qquad
    \fi}
\headline={\ifnum\pageno=1
\hfil
\else\otherheadline
    \fi}
%

\def\vsi{\vec{\sigma }}

\def\vp{\vec p}  \def\vr{\vec r}   
\def\vx{\vec x}
  

%
%

\def\hr{\hat{r}}

%

%
\def\x{\bold x}
\def\y{\bold y}
\def\z{\bold z}

\def\q{\bold q}

\def\r{\bold r}
%
%

%
%

%
%
\let\DS=\displaystyle
%
%
%

%

\font\eur=cmr10

%

%
%
\font\petrm=cmr5
\font\sestrm=cmr6

\font\osamrm=cmr8
\font\devetrm=cmr9

%

\font\devetit=cmti9
\font\osamit=cmti8

%

\font\devetmi=cmmi9
\font\osammi=cmmi8

\font\sestmi=cmmi6
\font\petmi=cmmi5
%

\font\devetsy=cmsy9
\font\osamsy=cmsy8

\font\sestsy=cmsy6
\font\petsy=cmsy5
%
\font\osamsl=cmsl8
\font\devetsl=cmsl9

%
\font\osambf=cmbx8
\font\devetbf=cmbx9

\font\sestbf=cmbx6
\font\petbf=cmbx5
%
\font\osamtt=cmtt8
\font\devettt=cmtt9

%
%
\def\uskoosam{%
\textfont0=\osamrm \scriptfont0=\sestrm
\scriptscriptfont0=\petrm \def\rm{\fam0\osamrm}%
\textfont1=\osammi \scriptfont1=\sestmi
\scriptscriptfont1=\petmi \def\oldstyle{\fam1\osammi}%
\textfont2=\osamsy \scriptfont2=\sestsy
\scriptscriptfont2=\petsy
\textfont\itfam=\osamit \def\it{\fam\itfam\osamit}%
\textfont\slfam=\osamsl \def\sl{\fam\slfam\osamsl}%
\textfont\ttfam=\osamtt \def\tt{\fam\osamtt}%
\textfont\bffam=\osambf \scriptfont\bffam=\sestbf
\scriptscriptfont\bffam=\petbf \def\bf{\fam\bffam\osambf}%
\rm}
%
%
\def\uskodevet{%
\textfont0=\devetrm \scriptfont0=\sestrm
\scriptscriptfont0=\petrm \def\rm{\fam0\devetrm}%
\textfont1=\devetmi \scriptfont1=\sestmi
\scriptscriptfont1=\petmi \def\oldstyle{\fam1\devetmi}%
\textfont2=\devetsy \scriptfont2=\sestsy
\scriptscriptfont2=\petsy
\textfont\itfam=\devetit \def\it{\fam\itfam\devetit}%
\textfont\slfam=\devetsl \def\sl{\fam\slfam\devetsl}%
\textfont\ttfam=\devettt \def\tt{\fam\devettt}%
\textfont\bffam=\devetbf \scriptfont\bffam=\sestbf
\scriptscriptfont\bffam=\petbf \def\bf{\fam\bffam\devetbf}%
\rm}

%

\def\beu{\hbox{\eur b}}
\def\deu{\hbox{\eur d}}


%
%

\def\mpik{\vskip 0.5cm\centerline{$\spadesuit \qquad \spadesuit$}\vskip
0.5cm}
\def\mpikk{\vskip 0.5cm\centerline{$\spadesuit \qquad \spadesuit
\qquad \spadesuit\qquad \spadesuit$}\vskip
0.5cm}
%
%
%
\def\gapp{\,{\raise 1.5pt\hbox{$>$}}\!\!\!\!\!{\raise
-3.5pt\hbox{$\sim$}}\,}
\def\lapp{\,{\raise 1.5pt\hbox{$<$}}\!\!\!\!\!{\raise
-3.5pt\hbox{$\sim$}}\,}
%
%

%
%
%
%
%
%
%
%
%
%


\def\beu{\hbox{\eur b}}
\def\deu{\hbox{\eur d}}

%
%

%



\def\beu{\hbox{\eur b}}
\def\deu{\hbox{\eur d}}


\def\beu{\hbox{\eur b}}
\def\deu{\hbox{\eur d}}




%

%

%
\def\bra#1{\langle#1\vert}               
\def\ket#1{\vert#1\rangle}               
\def\brik#1#2#3{\langle#1\vert\, #2\,\vert#3\rangle}

%
%
%
\def\slashchar#1{\setbox0=\hbox{$#1$}  
   \dimen0=\wd0     
   \setbox1=\hbox{/} \dimen1=\wd1  
   \ifdim\dimen0>\dimen1   
      \rlap{\hbox to \dimen0{\hfil/\hfil}} 
      #1     
   \else     
      \rlap{\hbox to \dimen1{\hfil$#1$\hfil}} 
      /      
   \fi}      %
\def\sc#1{\slashchar#1}
%
\def\square#1{\mathop{\mkern0.5\thinmuskip\vbox{\hrule\hbox
        {\vrule\hskip#1\vrule height#1 width 0pt\vrule}\hrule}
        \mkern0.5\thinmuskip}}
\def\Square{\mathchoice
        {\square{6pt}}{\square{5pt}}{\square{4pt}}{\square{3pt}}}

\def\dalamb{\Square\!}
%

%
%

%
%
%
%

%
\magnification=\magstep1
\TagsOnRight
\let\footnote=\plainfootnote
\topskip=32pt
\tolerance=1000
\def\normal{\baselineskip=13pt plus 1pt minus 1pt}
\def\subnormal{\baselineskip=10pt plus 1pt minus 1pt}
\normal
\parskip 10pt
\nopagenumbers
\newdimen\pagewidth  \newdimen\pageheight
\pagewidth=\hsize  \pageheight=\vsize
\parindent 30pt
%
%
\def\monthyear{\ifcase\month\or
 sije\v cnja \or velja\v ce \or o\v zujka \or travnja \or
 svibnja \or lipnja \or srpnja \or kolovoza \or rujna \or
 listopada \or studenog \or prosinca\fi
 \space\number\year .}
\def\danas{\number\day .\space \monthyear}
\def\mydate{\danas}
%
%
\def\mojnaslov{D. Horvat et al.: Hypernuclear Potentials }

\def\firstheadline{\hbox to \pagewidth{%
    {\baselineskip=10pt
     {\devetrm \hbox{\vtop{ \hsize=2.8truecm \parindent=0pt
           \leftline{Preprint from}
           \leftline{University of Zagreb}     }}}
      \hfil
          {\devetrm  \hbox{\vtop{ \hsize=2.8truecm \parindent=0pt
                \rightline{UZG-TH-1246/94}
                \rightline{\mydate}          }}} }%
       }}

\def\otherheadline{\devetit
  \ifodd\pageno  \qquad \hss \mojnaslov \hss P.\, \folio
    \else P.\, \folio\hss \naspog \hss\qquad
    \fi}
\headline={\ifnum\pageno=1
\hfil
\else\otherheadline
    \fi}
%

\def\vsi{\vec{\sigma }}

\def\vp{\vec p}  \def\vr{\vec r}   
\def\vx{\vec x}
  

%
%

\def\hr{\hat{r}}

%

%
\def\x{\bold x}
\def\y{\bold y}
\def\z{\bold z}

\def\q{\bold q}

\def\r{\bold r}
%
%
\let\DS=\displaystyle
%
%
%

%

%

%
%
\font\petrm=cmr5
\font\sestrm=cmr6

\font\osamrm=cmr8
\font\devetrm=cmr9

%

\font\devetit=cmti9
\font\osamit=cmti8

%

\font\devetmi=cmmi9
\font\osammi=cmmi8

\font\sestmi=cmmi6
\font\petmi=cmmi5
%

\font\devetsy=cmsy9
\font\osamsy=cmsy8

\font\sestsy=cmsy6
\font\petsy=cmsy5
%
\font\osamsl=cmsl8
\font\devetsl=cmsl9

%
\font\osambf=cmbx8
\font\devetbf=cmbx9

\font\sestbf=cmbx6
\font\petbf=cmbx5
%
\font\osamtt=cmtt8
\font\devettt=cmtt9

%
%
\def\uskoosam{%
\textfont0=\osamrm \scriptfont0=\sestrm
\scriptscriptfont0=\petrm \def\rm{\fam0\osamrm}%
\textfont1=\osammi \scriptfont1=\sestmi
\scriptscriptfont1=\petmi \def\oldstyle{\fam1\osammi}%
\textfont2=\osamsy \scriptfont2=\sestsy
\scriptscriptfont2=\petsy
\textfont\itfam=\osamit \def\it{\fam\itfam\osamit}%
\textfont\slfam=\osamsl \def\sl{\fam\slfam\osamsl}%
\textfont\ttfam=\osamtt \def\tt{\fam\osamtt}%
\textfont\bffam=\osambf \scriptfont\bffam=\sestbf
\scriptscriptfont\bffam=\petbf \def\bf{\fam\bffam\osambf}%
\rm}
%
%
\def\uskodevet{%
\textfont0=\devetrm \scriptfont0=\sestrm
\scriptscriptfont0=\petrm \def\rm{\fam0\devetrm}%
\textfont1=\devetmi \scriptfont1=\sestmi
\scriptscriptfont1=\petmi \def\oldstyle{\fam1\devetmi}%
\textfont2=\devetsy \scriptfont2=\sestsy
\scriptscriptfont2=\petsy
\textfont\itfam=\devetit \def\it{\fam\itfam\devetit}%
\textfont\slfam=\devetsl \def\sl{\fam\slfam\devetsl}%
\textfont\ttfam=\devettt \def\tt{\fam\devettt}%
\textfont\bffam=\devetbf \scriptfont\bffam=\sestbf
\scriptscriptfont\bffam=\petbf \def\bf{\fam\bffam\devetbf}%
\rm}

%

\def\beu{\hbox{\eur b}}
\def\deu{\hbox{\eur d}}


%
%

\def\mpik{\vskip 0.5cm\centerline{$\spadesuit \qquad \spadesuit$}\vskip
0.5cm}
\def\mpikk{\vskip 0.5cm\centerline{$\spadesuit \qquad \spadesuit
\qquad \spadesuit\qquad \spadesuit$}\vskip
0.5cm}
%
%
%
\def\gapp{\,{\raise 1.5pt\hbox{$>$}}\!\!\!\!\!{\raise
-3.5pt\hbox{$\sim$}}\,}
\def\lapp{\,{\raise 1.5pt\hbox{$<$}}\!\!\!\!\!{\raise
-3.5pt\hbox{$\sim$}}\,}
%
%


%
%
%
%
%
%
%
%
%
%


\def\beu{\hbox{\eur b}}
\def\deu{\hbox{\eur d}}

%
%

%



\def\beu{\hbox{\eur b}}
\def\deu{\hbox{\eur d}}


\def\beu{\hbox{\eur b}}
\def\deu{\hbox{\eur d}}




%

%

%
\def\bra#1{\langle#1\vert}               
\def\ket#1{\vert#1\rangle}               
\def\brik#1#2#3{\langle#1\vert\, #2\,\vert#3\rangle}

%
%
%
\def\slashchar#1{\setbox0=\hbox{$#1$}  
   \dimen0=\wd0     
   \setbox1=\hbox{/} \dimen1=\wd1  
   \ifdim\dimen0>\dimen1   
      \rlap{\hbox to \dimen0{\hfil/\hfil}} 
      #1     
   \else     
      \rlap{\hbox to \dimen1{\hfil$#1$\hfil}} 
      /      
   \fi}      %
\def\sc#1{\slashchar#1}
%
\def\sla{\raise .15ex\hbox{$/$}\kern -.54em\hbox{$a$}}
\def\slp{\raise .15ex\hbox{$/$}\kern -.54em\hbox{$p$}}
\def\slq{\raise .15ex\hbox{$/$}\kern -.52em\hbox{$q$}}
\def\slnab{\raise .15ex\hbox{$/$}\kern -.58em\hbox{$\nabla$}}
\def\sle#1{\raise .15ex\hbox{$/$}\kern -.54em\hbox{$#1$}}
\def\slpart{\raise .15ex\hbox{$/$}\kern -.54em\hbox{$\partial$}}
\def\slB{\raise .15ex\hbox{$/$}\kern -.63em\hbox{$B$}}
\def\slA{\raise .15ex\hbox{$/$}\kern -.68em\hbox{$A$}}
%
%
%
%
\def\ispod#1#2{\raise -2.2ex\hbox{$
        {\vcenter{
        \hrule height 5pt width 0.6pt
        \hrule height .5pt width#2pt
}}$}}
\def\okomic{\raise -2.2ex\hbox{$
        {\vcenter{\hrule height 5pt width 0.6pt}}$}}
\def\ktrk#1#2{
\overline#1\!\!\!\!\!\!\!\!\!\!\!\ispod{4}{21}\okomic\!#2}


\def\kktrk#1#2{
\overline#1\, \!\!\!\!\!\!\!\ispod{4}{12}\okomic\!#2}
\def\kkitrk#1#2{
#1\, \!\!\!\!\!\!\!\ispod{4}{12}\okomic\!\overline#2}
\def\pkditrk#1#2{
#1\!\!\!\!\!\!\!\!\!\!\!\!\!\!\!\!\!\!\!\ispod{4}{34}\okomic\! #2}

%
%
%
\def\kleb#1#2#3#4#5#6{C_{#1\,#2\,#3\,#4}^{#5\,#6}}
%
%
%
\def\square#1{\mathop{\mkern0.5\thinmuskip\vbox{\hrule\hbox
        {\vrule\hskip#1\vrule height#1 width 0pt\vrule}\hrule}
        \mkern0.5\thinmuskip}}
\def\Square{\mathchoice
        {\square{6pt}}{\square{5pt}}{\square{4pt}}{\square{3pt}}}

\def\dalamb{\Square\!}
%

%
%

%
%
%
%

%
%
%
%
\newdimen\tempdim                       
\newdimen\othick   \othick=1.53pt         
\newdimen\ithick   \ithick=.4pt         
\newdimen\dthick   \dthick= 1pt         
\newdimen\jthick   \jthick= 1pt         
\newdimen\spacing  \spacing=9pt         
\newdimen\abovehr  \abovehr=6pt         
\newdimen\belowhr  \belowhr=8pt         
\newdimen\nexttovr \nexttovr=8pt        

\def\r{\hfil&\omit\vrsp\vrule width\othick\cr&}   
\def\rr{\hfil\down{\abovehr}&\omit\vrsp\vrule width\othick\cr
        \noalign{\hrule height\ithick}\up{\belowhr}&} 
\def\jrr{\hfil\down{\abovehr}&\omit\vrsp\vrule width\othick\cr
        \noalign{\hrule height\jthick}\up{\belowhr}&} 
\def\up#1{\tempdim=#1\advance\tempdim by1ex
        \vrule height\tempdim width0pt depth0pt} 
\def\down#1{\vrule height0pt depth#1 width0pt}   
\def\large#1#2{\setbox0=\vtop{\hsize#1 \lineskiplimit=0pt \lineskip=1pt
        \baselineskip\spacing \advance\baselineskip by 3pt \noindent
        #2}\tempdim=\dp0\advance\tempdim by\abovehr\box0\down{\tempdim}}
\def\vrsp{\hskip\nexttovr\relax}
\def\toprule#1{\def\startrule{\hrule height#1\relax}} 
\toprule{\othick}               
\def\nstrut{\vrule height\spacing depth3.5pt width0pt}
\def\preamble#1{\def\startup{#1}}    
\preamble{&##}                  
{\catcode`\!=\active
 \gdef!{\hfil\vrule width0pt\vrsp\vrule width\ithick\relax\vrsp&}}

\def\table #1{\vbox\bgroup \setbox0=\hbox{#1}
    \vbox\bgroup\offinterlineskip \catcode`\!=\active
    \halign\bgroup##\vrule width\othick\vrsp&\span\startup\nstrut\cr
    \noalign{\medskip}
    \noalign{\startrule}\up{\belowhr}&}

\def\caption #1{\down{\abovehr}&\omit\vrsp\vrule width\othick\cr
    \noalign{\hrule height\othick}\egroup\egroup \setbox1=\lastbox
    \tempdim=\wd1 \hbox to\tempdim{\hfil \box0 \hfil} \box1 \smallskip
    \hbox to\tempdim{\advance\tempdim by-20pt\hfil\vbox{\hsize\tempdim
    \noindent #1}\hfil}\egroup}
%

%
%

\rightline{PHY-FER-II/99}  
\rightline{\danas}  
\nopagenumbers
\centerline{\vrule height 1mm width 13cm}
\vskip 0.6truecm
\centerline{\bf HYPERNUCLEAR POTENTIALS AND THE}
\medskip
\centerline{\bf PSEUDOSCALAR MESON EXCHANGE CONTRIBUTION }
\medskip
\vskip 0.8truecm
\centerline{{\bf Cesar Barbero}$^{(1)}$ \footnote{$^{\dag}$}{\uskoosam
e-mail: barbero\@venus.fisica.unlp.edu.ar},
{\bf Dubravko Horvat}$^{(2)}$ \footnote{$^{\flat}$}{\uskoosam
e-mail: dubravko.horvat\@fer.hr},
{\bf Franjo
Krmpoti\' c}$^{(1)}$\footnote{$^{\ddag}$}{\uskoosam
e-mail: krmpotic\@venus.fisica.unlp.edu.ar},}
\smallskip
\centerline{{\bf Zoran Naran\v ci\' c}$^{(2)}$ and
{\bf Dubravko Tadi\' c}$^{(3)}$\footnote{$^{\P}$}{\uskoosam
e-mail: tadic\@phy.hr}}
\bigskip
\vskip 0.5truecm
\centerline{$^{(1)}$Departamento de F\'isica, Facultad de Ciencias,}
\smallskip
\centerline{Universidad Nacional de  La Plata, C. C. 67,}
\smallskip
\centerline{1900 La Plata, Argentina}
\vskip 0.5truecm
\centerline{$^{(2)}$Department of
Physics, Faculty of Electrical Engineering,}
\smallskip
\centerline{University of Zagreb, 10\,000 Zagreb, Croatia}
\vskip 0.5truecm
\centerline{$^{(3)}$Physics Department,}
\smallskip
\centerline{University of Zagreb, 10\,000 Zagreb, Croatia}
\vskip 0.8truecm
\centerline{\vrule height 1mm width 13cm}
\smallskip
\centerline{\bf Abstract}
\smallskip
\midinsert\narrower
\noindent\uskodevet
The pieces of the hypernuclear strangeness violating potential due to the
pseudoscalar meson exchanges are derived using methods which were
successfully applied to hyperon nonleptonic decays. The estimates are
tested by comparison with measured hyperon nonleptonic decay amplitudes.
All isospin changes $\Delta I=1/2$ and $\Delta I=3/2$ are included in
the derived potential. All calculational methods used are reviewed and
described in detail.
\endinsert
\vfill
\eject
\pageno=-1
%
\def\uvo{1}
\def\drug{2}
\def\sep{3}
\def\pol{4}
\def\cur{5}
\def\meso{6}
\def\efe{8}
\def\wbb{7}
\def\efp{9}

\def\apa{\text{A}}
\def\bili{\text{B}}
\def\cpc{\text{C}}
\def\apd{\text{D}}
\def\dode{\text{E}}
\def\apf{\text{F}}
\def\apg{\text{G}}
\def\aph{\text{H}}

%
%
%
%

%
%
%

\def\uvo{1}
%
%
\pageno=1

\noindent{\bf \uvo. Introduction}
\def\naspog{\uvo. Introduction}
\vskip 1cm
Hypernuclei ${}^A_{\Lambda}N$, heavier than ${}^5_{\Lambda}$He, decay,
mainly through nonmesonic channels.
In these channels the $\Lambda$ mass excess of
176\,MeV is converted into kinetic energy of a final state nucleon. Such
a large momentum transfer suggests that heavier mesons exchanges in a
baryon-baryon potential could play a significant role.
However this review is dealing with pseudoscalar meson exchanges, such
as $\pi,\ K$ and $\eta$ only. The potential pieces due to the vector
and/or axial-vector meson exchanges will be studied along the similar
lines in the following paper [1]. The same goes for the calculations of the
hypernuclear decay widths $\Gamma$, which involve detailed description
of nuclear states [2]. In hypernuclear
decays one can study parity conserving (PC) and parity violating
(PV) part of the weak interaction. In that sense the nonmesonic decay
$$
\aligned
\Lambda\ +\  N\ &\to \ N\ +\  N\qquad\text{or}\\
\Lambda\ +\  p^+\ &\to \ p^+\ +\ n^0\qquad\text{and}\\
\Lambda\ +\  n^0\ &\to \ n^0\ +\  n^0
\endaligned
\tag\uvo-1
$$
are the $\Delta S=1$ analogy of the weak $NN\to NN$ nuclear PV reaction.
However
the weak PC $\Lambda N\to
NN$ decay is also observable in experiments. Experimental hypernuclear
programs within BNL, KEK, CEBAF and DUBNA promise enough data for
exhaustive hypernuclear studies [3-7].

In this paper we will employ one pseudoscalar
meson exchange mechanism to induce the
effective baryon-baryon (hyperon) potential (Fig. \uvo.1).

\midinsert
\vbox{\medskip
\centerline{\epsfysize=4cm\epsfbox{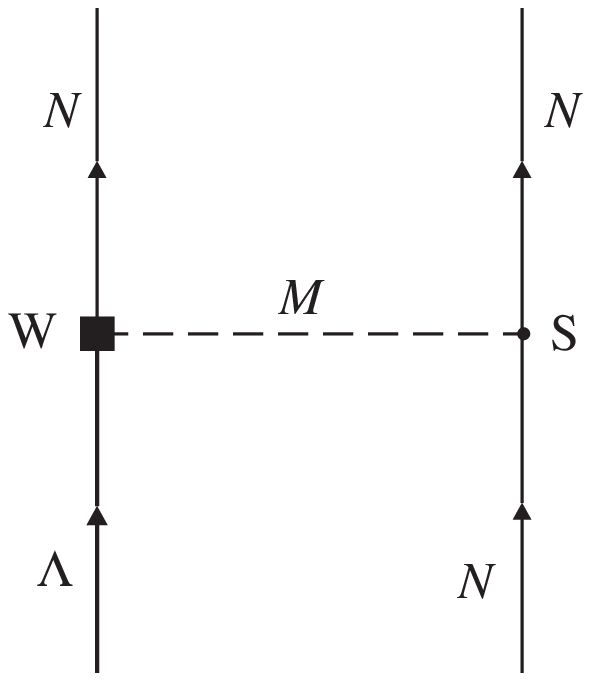}}
\smallskip
{\narrower\narrower
\noindent
\uskodevet
\it Fig. \uvo.1 - One meson baryon-baryon exchange diagram. $N$ and
$\Lambda$ are $S=0$ and $S=-1$ baryons and $M$ is a non-strange
pseudoscalar meson. $W$ and $S$ are the
strong and weak vertices.
\smallskip}
\medskip}
\endinsert

The weak vertex $W$ for pseudoscalar meson exchange
is closely connected with the theory of hyperon
nonleptonic decays [8-15].
These decays have been studied and analised in detail. Here we will
review a particular theoretical approach [8,13,14]. The related methods
have already been applied [5] in estimates of the "new" $\Delta S=1$
vertices, such as $NNK$ which appear in processes (\uvo-1) and which are
shown in Fig.\uvo.1. Here we intend to include all known and relevant
contributions, but for decuplet poles [16].

The strength of a weak vertex $W$
$$
W=\overline u(p_N)(A+B\gamma_5)u(p_{\Lambda})
\tag\uvo-2
$$
can be parametrized by $A$ and $B$ amplitudes. For certain weak vertices
such as $\Lambda\to N+\pi$ for example, the amplitudes $A$ and $B$
can be taken directly from the
experiment. In the experimentaly measured decay all particles are on the
mass shell. That is not necessarily so in the case of the exchange
interaction Fig.\uvo-1 inside the nuclear matter. However all possible
corrections will be neglected in the following. Theoretical
uncertainties, as explained in detail below, are so large, that such
niceties have no practical importance.

The theoretical analysis will be  carried out for some of  directly
measurable amplitudes also in order to check the
theoretical accuracy. One has to
rely entirely on the theory when dealing with weak vertex corresponding
to $N\to N+K$ and $\Lambda\to N+\eta$ transitions. In that
sense the investigation of the
strangeness violating nuclear
interaction is a wellcome test
for the general theoretical understanding of the hyperon nonleptonic
processes.

Use will be made of
theoretical schemes which were developed for the hyperon
nonleptonic decays [13, 14, 17, 18] and for the deduction of the weak parity
violating nuclear potential [19,20].

For the strong meson vertices  the standard $PC$ (parity coneserving) form
and the flavour SU(3) symmetry is assumed.

The study of the hyperon nonleptonic decays has in the eighties reached
certain level of success based on current algebra and pole dominance.
Some results [14, 21]  were elaborated in monographs and reviews
[8, 11, 13].

The  theoretically consistent
and complete (as far as the mentioned approximations  go)
results are obtained by combining contributions which previously were
used in either one or the other of discussed works. For
example, in describing $p-$wave ($B$) amplitudes  Ref. [21] used
baryon and meson poles. Ref.s [14, 17, 18] used baryon poles and separable
(factorizable) contributions. It turns out, as discussed in detail
below, that one has to combine all three contributions in a well defined
way.

Separable contributions associated with operators $\Cal
O_1\,\dots,\Cal O_4$ (see \drug\ below)
contain an axial vetor current matrix element
$$
\brik{B_f}{A_{\mu}^a}{B_i}\sim \overline u_{B_f}[\gamma_{\mu}\gamma_5
g_A+iq_{\mu}\gamma _5 g_P]u_{B_i}.
\tag\uvo-3
$$
Here the $g_A$ contribution contains a formfactor which is associated
with the axial vector meson exchanges [8, 22, 23] and thus it should be
included as "an axial vector meson pole contribution". The term $g_P$
was not included in the earlier estimate [14, 21]. It is dominated
by the pseudoscalar meson (i.e. kaon) pole, so that it's contribution is
included in the more general kaon pole contributions. Separable
contributions from operators $\Cal O_5$ and $\Cal O_6$ are, as shown
below, contained in the meson pole term.

The parity violating ($s-$wave) amplitudes were estimated by Ref.
[21] using the current algebra terms (CAT) and the contributions
comming from the commutators involving $\Cal O_5$ and $\Cal O_6$
operators. In the case of the $p-$wave amplitudes, the analogous terms
are included in the kaon pole pieces [13, 21]. Ref.s
[14, 17, 18] used CAT and factorizable contributions. Obviously the
complete estimate, involving leading poles should contain everything.

The weak PV
($A$) amplitude contains {\it current algebra\/}
and  {\it separable \/} parts, whereas the PC ($B$) amplitude gets
its contributions from {\it pole terms\/} and {\it separable\/}
parts, i.e.
$$
\aligned
A&=A_{\text{CA}}+A_{\text{SEP}}\\
B&=B_{\text{POLE}}+B_{\text{SEP}},
\endaligned
\tag\uvo-4a
$$
or more precisely
$$
\aligned
A&=\sum_{a=\pi,K,\eta}\left[A_{\text{CA}/a}+A_{\text{SEP}/a}\right]\\
B&=\sum_{a=\pi,K,\eta}\left[B_{\text{POLE}/a}+B_{\text{SEP}/a}\right].
\endaligned
\tag\uvo-4b
$$
The following sections will be devoted to
specific {\it pion, kaon\/} and $\eta-$contributions to PV and PC weak
$\Lambda-$decay amplitudes.
The separable pieces are given in Section 3 while the baryon pole
contributions are described in Section 4. Section 5 is devoted to so
called current algebra contributions. The connection between separable
contributions and the meson poles is discussed in Section 6. Section 7
compares the calculated weak $BBM$ amplitudes with the measured ones.
The theoretically predicted $NNK$ and $NN\eta$ amplitudes are compared
with some other theoretical predictions. The derivation of the
nonrelativistic weak potentials is described in Section 8. The $\Delta
I=1/2$ and $\Delta I=3/2$ pieces are listed separably. The effective
$\Delta S=1$ nonrelativistic potential, which can serve as an input in
nuclear calculations, are listed in Section 9. Various theoretical and
calculational details can be found in numerous appendices.

Some attention will be paid to the so called "double counting problem"
[11, 18] by comparing various contributions
to $B$ amplitudes.

Closely related theoretical methods were used in Ref. [5]. This
approach investigates the importance (or unimportance) of additional
contributions, such as separable terms and kaon poles, which are
introduced here.

The weak $NNK$ interactions were investigated using heavy baryon chiral
peturbation theory also [22]. Their results will be compared
with results obtained by methods corresponding to the scheme outlined by
formulae (\uvo-4).

\vfill
\mpikk
\eject

\def\drug{2}
%
%

\noindent{\bf \drug. Weak And Strong Hamiltonian}
\def\naspog{\drug. Weak And Strong Hamiltonian}
\vskip 1cm

The weak one pion exchange potential can be extracted from the Feynman
amplitude shown in Fig.\uvo.1. This diagram corresponds to a second
order term in the $S$-matrix expansion. The details of that expansion,
performed in an effective field theory, can be found in Section
\efe\ below. Here we want to specify the strength of the weak
(i.e. $A$, $B$) and of the strong vertex (i.e. $g_{B'BM}$). The
calculation of the weak strengths $A$ and $B$, which correspond to the
hyperon nonleptonic decay amplitudes [13,14,17,18,21] will be the main topic
of the following Sections \sep.-\meso.

The weak vertices are determined by the effective weak $\Delta S|=1$
Hamiltonian
which is in the quark basis given
by [23,24]
$$
\Cal H^{(W)}_{\text{eff.}}=\frac{G_F}{\sqrt{2}}V_{ud}V^*_{us}
\sum_i C_i\Cal O_i,
\tag\drug-1
$$
can be conveniently  written as
$$
\Cal H^{(W)}_{\text{eff.}}=-\sqrt{2}{G_F}\sin\theta_C\cos\theta_C
\sum_i C_i\Cal O_i.
\tag\drug-2
$$

Here $C_i'$s  are the QCD Wilson coefficients [23] calculated in the
six-quark Standard model environment by solving the renormalization
group equations to the one-loop QCD corrections, evaluated are the scale
$\mu=0.5\,$GeV. Their values  are given in
Table \drug.1. $G_F$ is  Fermi constant and
$\sin\theta_C$ and $\cos\theta_C$ (later denoted by $s_C$ and $c_C$) are
sine/cosine of the Cabibbo angle.

%
%
\parskip 0pt
\baselineskip 10pt
\medskip
$$
\table{}
$C_1$ ! $ C_2 $  !$ C_3$ ! $ C_4 $  ! $C_5$ ! $C_6$ \rr
   -2.358 ! 0.080 ! 0.082  ! 0.411  ! -0.080 ! -0.021
\caption{\centerline{\uskodevet\it Table \drug.1 - Wilson coefficients
evaluated are the scale $\mu=0.5\,$GeV }}
$$
\medskip
%
%
%

The four-quark $(V-A)$ operators comprising the effective
weak Hamiltonian (\drug-1) are
$$
\alignedat 3
\Cal O_1&=:(\overline d_L\gamma_{\mu}s_L)(\overline u_L\gamma^{\mu} u_L)
-(\overline d_L\gamma_{\mu}u_L)(\overline u_L\gamma^{\mu} s_L):
&\qquad{(8,\,1/2)}\\
\Cal O_2&=:(\overline d_L\gamma_{\mu}s_L)(\overline u_L\gamma^{\mu} u_L)
+(\overline d_L\gamma_{\mu}u_L)(\overline u_L\gamma^{\mu} s_L)&{}\\
&+2(\overline d_L\gamma_{\mu}s_L)(\overline d_L\gamma^{\mu} d_L)+
2(\overline d_L\gamma_{\mu}s_L)(\overline s_L\gamma^{\mu} s_L):
&\qquad{(8,\,1/2)}\\
\Cal O_3&=:(\overline d_L\gamma_{\mu}s_L)(\overline u_L\gamma^{\mu} u_L)
+(\overline d_L\gamma_{\mu}u_L)(\overline u_L\gamma^{\mu} s_L)&{}\\
&+2(\overline d_L\gamma_{\mu}s_L)(\overline d_L\gamma^{\mu} d_L)-
3(\overline d_L\gamma_{\mu}s_L)(\overline s_L\gamma^{\mu} s_L):
&\qquad{(27,\,1/2)}\\
\Cal O_4&=:(\overline d_L\gamma_{\mu}s_L)(\overline u_L\gamma^{\mu} u_L)
+(\overline d_L\gamma_{\mu}u_L)(\overline u_L\gamma^{\mu} s_L)&{}\\
&-(\overline d_L\gamma_{\mu}s_L)(\overline d_L\gamma^{\mu} d_L):
&\qquad{(27,\,3/2)}\\
\Cal O_5&=:(\overline d_L\gamma_{\mu}\lambda_As_L)
(\overline u_R\gamma^{\mu}\lambda_A u_R)
+(\overline d_L\gamma_{\mu}\lambda_As_L)(\overline d_R\gamma^{\mu}
\lambda_A d_R)&{}\\
&+(\overline d_L\gamma_{\mu}\lambda_As_L)
(\overline s_R\gamma^{\mu}\lambda_A s_R):
&\qquad{(8,\,1/2)}\\
\Cal O_6&=:(\overline d_L\gamma_{\mu} s_L)
(\overline u_R\gamma^{\mu} u_R)
+(\overline d_L\gamma_{\mu} s_L)(\overline d_R\gamma^{\mu} d_R)&{}\\
&+(\overline d_L\gamma_{\mu} s_L)(\overline s_R\gamma^{\mu} s_R):
&\qquad{(8,\,1/2)}
\endalignedat
\tag\drug-3
$$
All the operators in (\drug-3) are normal ordered and their
SU(3) ({\it flavour representation,\, isospin}) content is also given.
In the above expressions the following notation is used
$$
\aligned
u_L&\equiv \frac12(1-\gamma_5)u;\quad
u_R\equiv \frac12(1+\gamma_5)u
\quad\text{so for instance}\\
(\overline d_L\gamma_{\mu}s_L)(\overline u_L\gamma^{\mu} u_L)&
\equiv[\overline d\frac12 \gamma^{\mu}(1-\gamma_5) s]
[\overline u\frac12
\gamma_{\mu}(1-\gamma 5) u]\\
&=\frac14 [\overline d \gamma_{\mu}(1-\gamma_5) s]
          [\overline u \gamma^{\mu}(1-\gamma_5) u]\\
&=\frac14 (\overline d s)_{(V-A)}(\overline u
u)_{(V-A)},\quad\text{and}\\
(\overline d_R\gamma_{\mu}d_R)&\equiv[\overline d\frac12
(1-\gamma_5)\gamma_{\mu}\frac12(1+\gamma_5) d]\\
&=\frac12\overline d\gamma_{\mu}(1+\gamma_5)d
\endaligned
\tag\drug-4a
$$

The strong and weak vertices corresponding to effective
baryon-baryon-meson couplings are
given by
$$
\Cal H^{(S)}_{NN\pi}=ig_{NN\pi}\overline{\Psi}_N\gamma_5\Psi_N\Phi_{\pi}
\tag\drug-5
$$
and
$$
\Cal H^{(W)}_{\Lambda N\pi}=-iG_F m_{\pi}^2\overline{\Psi}_N(A-B\gamma_5)
\Psi^S_{\Lambda}\Phi_{\pi}.
\tag\drug-6
$$
Here is $G_F$ the Fermi coupling constant whose value is usually
given in the following combination
$$
G_Fm_{\pi^+}^2=2.21\times10^{-7}\qquad\text{i.e.}\qquad
G_F=1.16639\times10^{-5}\,\text{GeV}^{-2}.
\tag\drug-7
$$
$\Psi^S_{\Lambda}$ is the isospin spurion $\pmatrix 0\\ 1\endpmatrix$
included here to enforce the $\Delta I=1/2$ rule observed in the
$\Lambda$ decays.
The weak amplitudes $A$ and $B$ are calculated below, see Table \efe.1.
They obtain contributions corresponding to current algebra forms,
separable forms and various pole terms.

The strong coupling constants differ according to their isomultiplet
content which can be inferred from their SU(3) coupling. The following
results are obtained (see for instance [25-28]):
\item{-} $NN\pi$: $g_{pn\pi^+}=\sqrt{2}g_{NN\pi}$;
$g_{nn\pi^0}=-g_{NN\pi}$; $g_{pp\pi^0}=g_{NN\pi}$;
\smallskip
\item{-} $N\Sigma K$: $g_{nK^+\Sigma^-}=\sqrt{2}g_{N\Sigma K}$;
$g_{\overline p K^0\Sigma^-}=\sqrt{2}g_{N\Sigma K}$;
$g_{\overline p K^+\Sigma^0}=g_{N\Sigma K}$;
$g_{n  K^0\Sigma^0}=-g_{N\Sigma K}$;
\smallskip
\item{-} $\Xi\Xi \pi$: $g_{\Xi^-\Xi^0\pi^+}=-\sqrt{2}g_{\Xi\Xi\pi}$;
$g_{\Xi^-\Xi^-\pi^0}=-g_{\Xi\Xi\pi}$;
$g_{\Xi^0\Xi^0\pi^0}=g_{\Xi\Xi\pi}$;
\smallskip
\item{-} $NK\Lambda$: $g_{pK^+\Lambda}=g_{\overline n K^0\Lambda}=
g_{NK\Lambda}$;
\smallskip
\item{-} $\Lambda\Sigma\pi$: $g_{\Lambda\Sigma^+\pi^-}=
g_{\Lambda\Sigma^-\pi^+}=g_{\Lambda\Sigma^0\pi^0}=g_{\Lambda\Sigma\pi}$;

\midinsert
\uskoosam{
%
%
\parskip 0pt
\baselineskip 10pt
\medskip
$$
\table{}
$M\,B\,B'$ ! $ \pi NN $ !$\pi\Lambda\Sigma$ ! $\pi\Sigma\Sigma $!
$\pi\Xi\Xi$ ! $K\Lambda N$!$K\Xi\Lambda$!$K\Sigma N$ \rr
$g_{MBB'}$ !$1$ !$\displaystyle\frac{2(1-f)}{\sqrt{3}}$!$2f$ !$(2f-1)$!
$\displaystyle\frac{-(1+2f)}{\sqrt{3}}$!
$\displaystyle\frac{4f-1}{\sqrt{3}}$!$1-2f$
\caption{\centerline{\uskodevet\it Table \drug.2 - Strong coupling
constants in units of $g$; eg. $g_{\pi NN}=g$ etc. }}
$$
\medskip
}
\endinsert
%
%

\vfill
\mpikk
\eject

\def\sep{3}

%

\noindent{\bf \sep. Separable contributions}
\def\naspog{\sep- Separable Contributions}
\bigskip
In this section we calculate the separable contributions to $A$ and $B$
amplitudes.
In the quark basis they are symbolized by diagrams $(a)$ and $(b)$ in
Fig.\sep.1.

\midinsert
\vbox{\medskip
\centerline{\epsfxsize=10cm\epsfbox{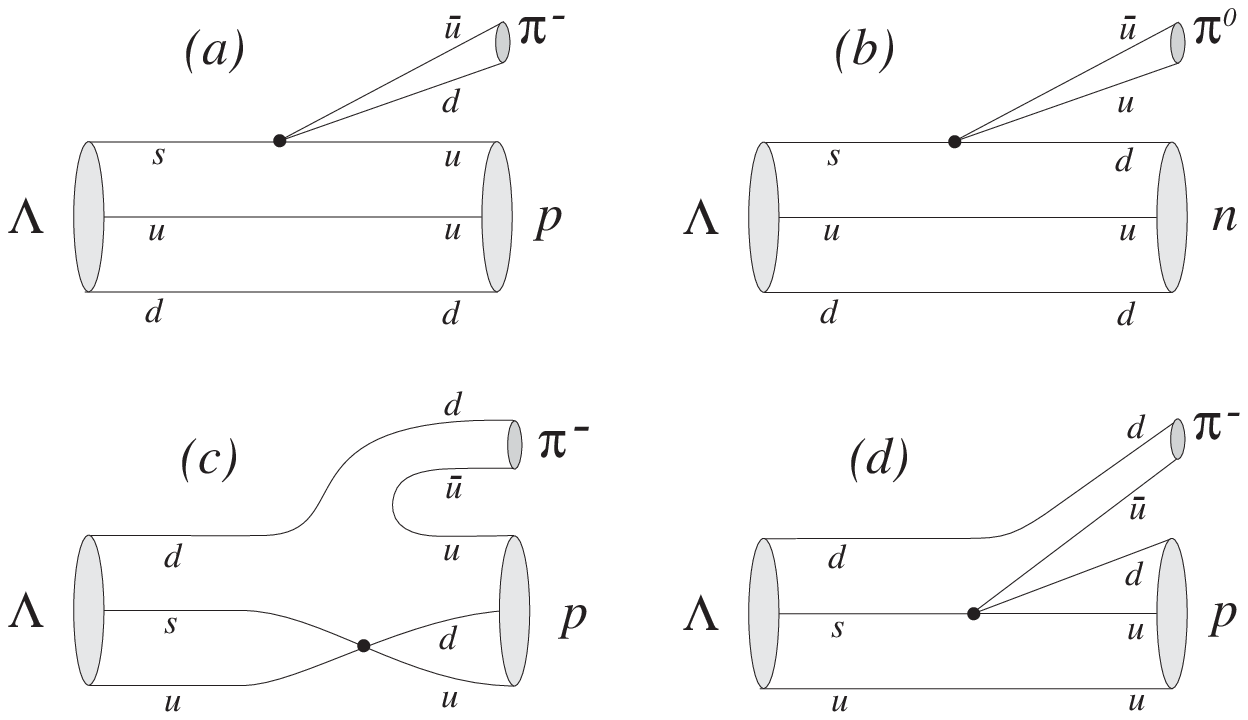}}
\medskip
\centerline{{\uskodevet\it Fig. \sep.1 - Separable ((a), (b)) and
non-separable ((c), (d)) contributions in $\Lambda$}}
\centerline{{\it decays in a proton $p$ and neutron $n$. $\Cal
H_{\text{eff}}^{(W)}$ acts at the black dot.}}
\medskip}
\endinsert

Four quarks emerging from the black dot in Fig.\sep.1 are produced by
the four-quark operators in (\drug-1) and (\drug-2).

In order to account for all the possible processes one has to include
contributions from the Fierz-transformed (FIT) operators $\Cal O_i$
averaged over colours. This average brings in a factor of 1/3.
The two relations govern all the Fierz transformations
$$
\aligned
[\overline q_1\gamma_{\mu}(1-\gamma_5)q_2]\cdot
[\overline q_3\gamma^{\mu}(1-\gamma_5)q_4]&=
[\overline q_1\gamma_{\mu}(1-\gamma_5)q_4]
\cdot [\overline q_3\gamma^{\mu}(1-\gamma_5)q_2]\\
[\overline q_1\gamma_{\mu}(1-\gamma_5)q_2]\cdot
[\overline q_3\gamma^{\mu}(1+\gamma_5)q_4]&=
-2[\overline q_1(1+\gamma_5)q_4]
\cdot [\overline q_3(1-\gamma_5)q_2].
\endaligned
\tag\sep-1
$$
In the above transformations the anticommutativity of fermion fields has
been taken into account.
The Fierz identities for eight Gell-Mann SU(3) $\lambda$-matrices
which are necessary for the calculations are
$$
\aligned
\lambda_b^a\cdot\lambda^c_d&=\frac{16}9\delta^a_d\delta^c_b-\frac13
\lambda^a_d\lambda^c_b\\
\lambda_d^a\cdot\lambda^c_b&= 2\delta^a_b\delta^c_d-
-\frac{2}3\delta^a_d\delta^c_b.
\endaligned
\tag\sep-2
$$
FIT operators could be expressed in terms of the original $\Cal
O_i$ operators:
$$
\aligned
\Cal O_1^{\text{FT}}&=-\Cal O_1,\quad\text{so that}\quad
\Cal O_1\to \Cal O_1+\frac13\Cal O_1^{\text{FT}}=
\left(1-\frac13\right)\Cal O_1\\
\Cal O_2^{\text{FT}}&=\Cal O_2,\quad\text{so that}\quad
\Cal O_2\to \Cal O_2+\frac13\Cal O_2^{\text{FT}}=
\left(1+\frac13\right)\Cal O_2\\
\Cal O_3^{\text{FT}}&=\Cal O_3,\quad\text{so that}\quad
\Cal O_3\to \Cal O_3+\frac13\Cal O_3^{\text{FT}}=
\left(1+\frac13\right)\Cal O_3\\
\Cal O_4^{\text{FT}}&=\Cal O_4,\quad\text{so that}\quad
\Cal O_4\to \Cal O_4+\frac13\Cal O_4^{\text{FT}}=
\left(1+\frac13\right)\Cal O_4\\
\Cal O_6^{\text{FT}}&=-2(\overline d_Lq_R^i)(\overline q_R^is_L),
\quad\text{so that}\quad
\Cal O_6\to \Cal O_6+\frac13\Cal O_6^{\text{FT}}\\
&=
\Cal O_6-\frac23(\overline d_Lq_R^i)(\overline q_R^is_L)\\
\Cal O_5^{\text{FT}}&=\frac{3}{16}\Cal O_6
\quad\text{so that}\quad
\Cal O_5\to \Cal O_5+\Cal O_5^{\text{FT}}=
\Cal O_5+\frac{3}{16}\Cal O_6
\endaligned
\tag\sep-3
$$
There is also a simple relation between a matrix element of the
operators $\Cal O_5$ and $\Cal O_6$ which is used frequently
$$
\brik{B'}{\Cal O_5}{B}=\frac{16}{3}\brik{B'}{\Cal O_6}{B}.
\tag\sep-4
$$
In the FIT operators $\Cal O_{5,6}$ the following quark combinations occur
$$
\aligned
(\overline d_L s_R)(\overline d_R d_L)&\equiv[\overline d\frac12
(1+\gamma_5)\frac12(1+\gamma_5) s]
[\overline d\frac12
(1-\gamma_5)\frac12(1-\gamma_5) d]\\
&=\frac14[\overline d(1+\gamma_5)s][\overline d(1-\gamma_5)d].
\endaligned
\tag\sep-5
$$

From the Dirac equation for quark fields
one gets usefull relations
which are necessary to calculate matrix elements of scalar and
pseudo-scalar densities occuring in the FIT operators $\Cal O_5$
and $\Cal O_6$  (see (\sep-7), (\sep-8) and (\sep-4a)). For instance
$$
\aligned
\overline d\gamma_5u&=(p_d-p'_u)_{\mu}\overline
d(p_d)\gamma^{\mu}\gamma_5u(p'_u)\cdot\frac1{m_d+m_u}\\
\overline u s&=(p_u-p'_s)_{\mu}\overline
u(p_u)\gamma^{\mu} s(p'_s)\cdot\frac1{m_u-m_s}.
\endaligned
\tag\sep-6
$$

The effective Hamiltonian (\drug-1) is used to calculate the weak decay
amplitudes which correspond to the weak vertex in Fig.\uvo.1.
The general procedure is briefly outlined: the invariant
decay amplitude for the process
(baryon$\to$baryon+meson)
$$
B\to B'+M
\tag\sep-7
$$
is given by the factorization (separation) assumption as
$$
\brik{M,B'}{\Cal H_{\text{eff}}}{B}=
-\sqrt{2}{G_F}\sin\theta_C\cos\theta_C
\sum_i C_i
\brik{M}{(V-A)_i}{0}\brik{B'}{(V-A)_i}{B}
\tag\sep-8
$$
where the $(V-A)_i$ are currents (operator parts) which could account
for the particular transition between the vacuum $\ket{0}$ and a meson
$\bra{M}$ and between two (octet) baryons $B$ and $B'$.
The character of
the meson matrix element $\brik{M}{(V-A)}{0}$ depends on the meson
state: if $M$ belongs to the SU(3) pseudo-scalar octet, only $A$ part of
the four-quark operator contributes. (When a vector meson is considered
only $V$ part remains.)  For the baryon matrix element both $V$ and $A$
parts occur giving PV or PC contributions. Baryon matrix elements
are expressed in terms of vector/axial-vector form factors
$v_{BB'}/a_{BB'}$ which occur
in semileptonic baryon decays. These form factors are
expressible in terms of $F$ and $D$ constants. The form factors are given
in Table \sep.1.

\midinsert
{\uskoosam{
%
%
\parskip 0pt
\baselineskip 10pt
\medskip
$$
\table{}
B\,B'! SU(3) content ! $v_{NN'}$  ! $a_{NN'}$ \jrr
$n\, p$      !\hfill $\tilde{\pi}^-$\hfill   ! $1$              ! $F+D$             \rr
$n\, n$      !\hfill$\tilde{\pi}^0$\hfill    !
$-\displaystyle\frac{1}{\sqrt{2}}$
! $-\displaystyle\frac1{\sqrt{2}}(F+D) $    \rr
$p\, p$      !\hfill$\tilde{\pi}^0$\hfill    !
$\displaystyle\frac{1}{\sqrt{2}}$
! $\displaystyle\frac1{\sqrt{2}}(F+D) $    \rr
$p\, p$      !\hfill$\tilde{\eta}_8$ \hfill   !
$\displaystyle\frac{3}{\sqrt{6}}$
! $\displaystyle\frac1{\sqrt{6}}(3F-D) $    \rr
$n\, n$      !\hfill$\tilde{\eta}_8$ \hfill   !
$\displaystyle\frac{3}{\sqrt{6}}$
! $\displaystyle\frac1{\sqrt{6}}(3F-D) $    \rr
$n\, n$      !\hfill$\overline u u$ \hfill   !
$1$  ! $F-D $\rr
$n\, n$      !\hfill$\overline d d$ \hfill   !
$2$  ! $2F$\rr
$n\, n$      !\hfill$\tilde{\eta}_1$ \hfill   !
$\displaystyle\frac{3}{\sqrt{3}}$
! $\displaystyle\frac1{\sqrt{3}}(3F-D) $\rr
$p\, p$      !\hfill$\tilde{\eta}_1$ \hfill   !
$\displaystyle\frac{3}{\sqrt{3}}$
! $\displaystyle\frac1{\sqrt{3}}(3F-D) $\rr
$\Lambda\, p$      !\hfill$K^-$ \hfill   !
$\displaystyle-\frac{3}{\sqrt{6}}$
! $\displaystyle-\frac1{\sqrt{6}}(3F+D) $\rr
$\Lambda\, n$      !\hfill$-\overline{K^0}$ \hfill   !
$\displaystyle-\frac{3}{\sqrt{6}}$
! $\displaystyle-\frac1{\sqrt{6}}(3F+D) $
\caption{\centerline{{\uskodevet\it Table \sep.1 -
Nucleon vector and axial-vector form factors}}}
$$
\medskip
}}
\endinsert
%
%

One defines
$$
\brik{B'}{V^{\mu}_a}{B}=\overline u_{B'}(p')\gamma^{\mu} v^a_{BB'}
u_B(p)
\tag\sep-9
$$
and
$$
\brik{B'}{A^{\mu}_a}{B}=\overline u_{B'}(p')\gamma^{\mu}\gamma_5 a^a_{BB'}
u_B(p)
\tag\sep-10a
$$
or
$$
a^a_{BB'}=g_A\Lambda_a.
\tag\sep-10b
$$
Here $\Lambda_a=\lambda_a/2$ is the well known SU(3) matrix [9,26].
A very important ingredient for the above outlined calculation is the
knowledge of a meson matrix element. It is calculated either by using the
partially conserved axial-vector current (PCAC) hypothesis
[8,9,11] (in the case
of pseudo scalar (PS) mesons $\pi,\ \eta, \ K$) or by using
the current-field identity (CFI) [19,29] or the  meson-nucleon
$\sigma-$term [30-32] (in the case of vector mesons).

The PCAC hypothesis is based on the divergence of the axial-vector
current
$$
A_{a}^{\mu}=\overline{\psi}\gamma^{\mu}\gamma_5\frac{\lambda^a }{2}\psi.
\tag\sep-11
$$
This is an operator relation and its divergence
is applied to a
(physical) state represented by a {\it ket\/} (or {\it bra\/}), i.e.
$$
\partial_{\mu}A_a^{\mu}\ket{0}=if_{\phi}p_{\mu}p^{\mu}
\phi^{\dagger}\ket{0}=if_{\phi}m_{\phi}^2\ket{\phi}.
\tag\sep-12
$$
Since we want to express the meson matrix element of the form given in
(\sep-3), i.e. $\brik{M}{A_\mu }{0}$, the $\phi$ operator has {\it to
create} a state $\ket{M}$
$$
\partial_{\mu}A^{\mu}_a\ket{0}\sim\ket{\phi_M}
\tag\sep-13
$$
(Recall that it is usual to define an operator to anihilate
$\ket{\hphantom{0}}$, i.e.
$\brik{0}{\partial_{\mu}A^{\mu}_a}{\phi_M}\not=0$!) For instance, if one
has to calculate the matrix element where the physical pion $\pi^+$ is
being emitted, the first step is to determine the $\lambda-$matrix
content of the axial-vector current (\sep-11), i.e. one has to find
$$
M\sim\overline q\frac{\lambda_{a+ib}}{\sqrt{2}}q
\tag\sep-14
$$
which, for $\pi^0$ gives $\pi^0=\overline q(\lambda_3/\sqrt{2}) q$, and
$q^{\text{T}}\equiv (u,d,s)$ (here "T" means {\it transposed\/}), or for
$\eta=\overline q(\lambda_8/\sqrt{2}) q$. For the positive pion one gets
that $a+ib$ from (\sep-14) equals $1-i2$, so we have
$$
\partial_{\mu}A_{1-i2}^{\mu}=Cf_{\pi}m_{\pi}^2\pi^+=Cf_{\pi}m_{\pi}^2
\deu_d^{\dagger}\beu_u^{\dagger}.
\tag\sep-15
$$
Here $\beu_u$ anihilates a quark of the flavour $u$ whereas $\deu_d$
anihilates an antiquark of the flavour $d$. The constant $C$ is to be
determined and it depends on the isospin content of the meson in
question. In the above case the calculation gives $C=1$. We can write
now all the relevant PCAC relations
$$
\gathered
{\alignedat 3
\partial_{\mu}A_{3}^{\mu}&=\frac{\sqrt{2}}{2}f_{\pi}m_{\pi}^2\pi^0&\qquad
\partial_{\mu}A_{1-i2}^{\mu}&=f_{\pi}m_{\pi}^2\pi^+\\
\partial_{\mu}A_{4-i5}^{\mu}&=f_{K}m_{K}^2K^+&\qquad
\partial_{\mu}A_{6-i7}^{\mu}&=f_{K}m_{K}^2K^0
\endalignedat}
        \\
\partial_{\mu}A_{8}^{\mu}=\frac{\sqrt{2}}{2}f_{\eta}m_{\eta}^2\eta.
\endgathered
\tag\sep-16
$$
The above relations enable us to relate the following matrix elements
$$
\brik{\pi^0}{\overline u\gamma_{\mu}\gamma_5 u}{0}=-
\brik{\pi^0}{\overline d\gamma_{\mu}\gamma_5 d}{0}=\frac{\sqrt{2}}{2}
\brik{\pi^-}{\overline d\gamma_{\mu}\gamma_5 u}{0}
\tag\sep-17
$$
etc.

\vfill
\mpikk
\eject


%
\def\pol{4}
%

\noindent{\bf \pol. The Baryon Pole Contributions}
\def\naspog{\pol. The Baryon Pole Contributions}

The parity conserving (PC) $B$ amplitude gets part of its contributions
from the {\it baryon pole terms\/} given for a particular case of the
$\Lambda$ decay in the Fig.\pol.1.

\midinsert
\vbox{\medskip
\centerline{\epsfxsize=12cm\epsfbox{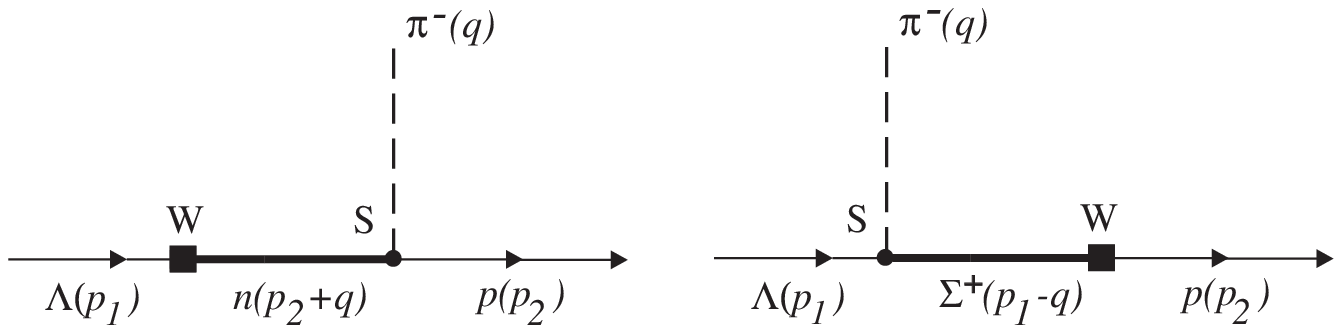}}
\medskip
\centerline{{\uskodevet\it Fig.\pol.1 - Baryon pole terms contributions:
(a) $s$-channel and (b) $u$-channel.}}
\medskip}
\endinsert

We will consider one particular example of the baryon pole calculation
in full detail.
Two possible $\Lambda-$decays (which almost 100\% saturate the $\Lambda$
width) are
$$
\alignedat 3
&A(\Lambda_0^0)\qquad \Lambda^0&\to n+\pi^0\\
&A(\Lambda_-^0)\qquad \Lambda^0&\to p+\pi^-.
\endalignedat
\tag\pol-1
$$
By asigning the momenta to incomming and outgoing particle, i.e.
$$
\Lambda (p_1)\underset S\to\longrightarrow \pi ^-(q)+\Sigma^+(p_1-q)
\underset W\to\longrightarrow p(p_2)
\tag\pol-2
$$
we calculate the Feynman amplitude, corresponding to the $u-$channel,
Fig.\pol.1, using the effective Hamiltonians
given in (\drug-5) and (\drug-6), as follows
$$
\overline u(p_2)(A-B\gamma_5)\frac{(\sc{p}_1-\sc{q})+m_{\Sigma}}
{(p_1-q)^2-m_{\Sigma}^2}\gamma_5g_{\Sigma^+\pi^-\Lambda}u(p_1).
\tag\pol-3
$$
Here $g_{\Sigma\pi\Lambda}$ is the strong coupling constant listed in
table \drug.2.

Since $p_1-q=p_2$ and $p_2^2=m_p^2$ we get
$$
\overline u(p_2)(A-B\gamma_5)\frac{m_p+m_{\Sigma}}{(m_p-m_{\Sigma})
(m_p+m_{\Sigma})}\gamma_5 g_{\Sigma\pi\Lambda}u(p_1).
\tag\pol-4
$$

The result for $B(\Lambda^0_-)$ amplitude, Fig.\pol.1, is
$$
B^{\text{POLE}}(\Lambda^0_-)=
g_{\Lambda\Sigma^+\pi^-}\frac{a_{\Sigma^+ p}}{\Sigma^+-
p}+g_{n\pi^-p}\frac{a_{\Lambda n}}{n-\Lambda}.
\tag\pol-5
$$
However, in the soft pion limit, which is used to determine the $a_{ij}$
amplitude (see Section \cur\ below), $B$ amplitude has to vanish. That
imposes the condition
that is given by a {\it subtraction\/} of the soft amplitude
$B^{\text{POLE}}(q^2=0)$, i.e.
$$
B^{\text{POLE}}(q^2)-B^{\text{POLE}}(0)\equiv
B^{\text{POLE}}\,(\text{eq.}(\pol-7,8)).
\tag\pol-6
$$
Here $q^2$ is the pion (kaon, $\eta$) four-momentum. As ref. [3]
uses $B^{\text{POLE}}(q^2)$ only,
that explains some  numerical differences
which will appear in Tables(\wbb.1, \wbb.3 and \wbb.4. In detail one finds:
$$
\align
B^{\text{POLE}}_{\pi}(\Lambda^0_-)&=
2g(\Lambda+p)\left[\frac{a_{\Sigma p}}{\sqrt{3}}\cdot\frac{d}{(\Sigma^+-
p)(\Lambda+\Sigma^+)}-
\frac{a_{\Lambda n}}{\sqrt{2}}\cdot\frac{f+d}{(\Lambda-
N)(N+P)}\right]
\tag\pol-7\\
B^{\text{POLE}}_{\pi}(\Lambda^0_0)&=-\frac1{\sqrt{2}}
B^{\text{POLE}}(\Lambda^0_-).
\tag\pol-8
\endalign
$$
In the above expression the following notation for masses is used:
$\Lambda\equiv m_{\Lambda}$, or $p\equiv m_p$, etc. Recall also that
$f+d=1$, so only $f'$s appear in  (\pol-8).

\vbox{\medskip
\centerline{\epsfxsize=12cm\epsfbox{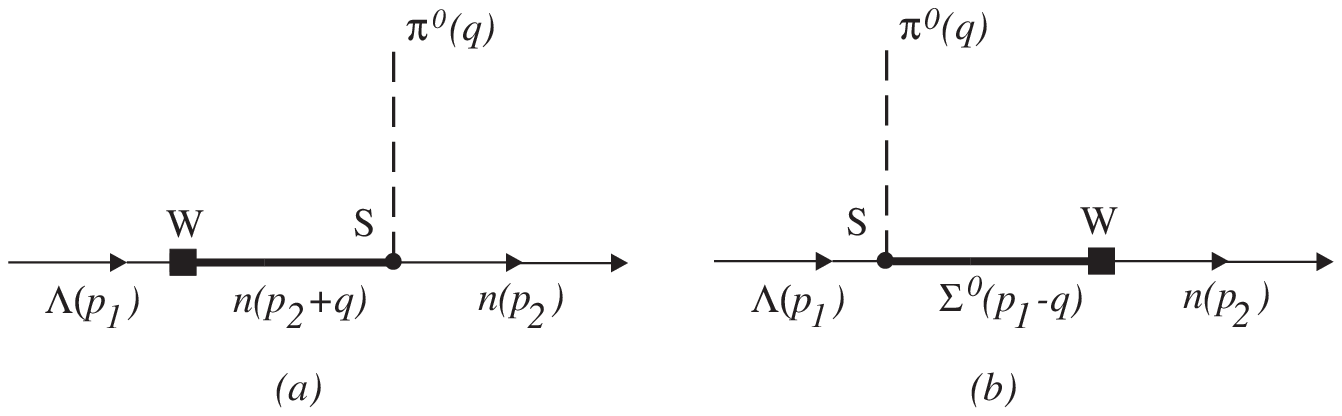}}
\medskip
\centerline{{\uskodevet\it Fig.\pol.2 - Baryon pole terms contributions:
(a) $s$-channel and (b) $u$-channel, $\pi^0$ emission.}}
\medskip}
\vfill
\eject
%
%

Kaon  exchange contributions are determined by

\midinsert
\vbox{\medskip
\centerline{\epsfxsize=10cm\epsfbox{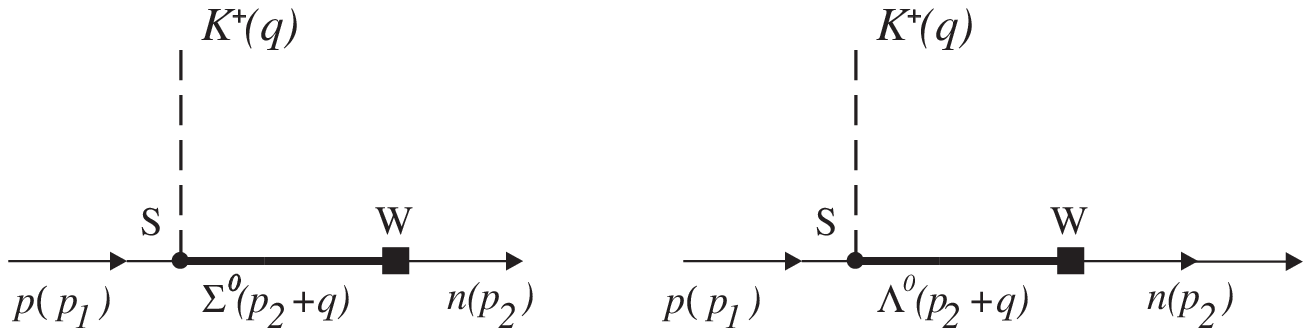}}
\medskip
\centerline{{\it Fig. \pol.3 - Pole diagrams for the proton nonleptonic
decay: $p\to n\,K^+$}}
\medskip}
\endinsert

\midinsert
\vbox{\medskip
\centerline{\epsfxsize=10cm\epsfbox{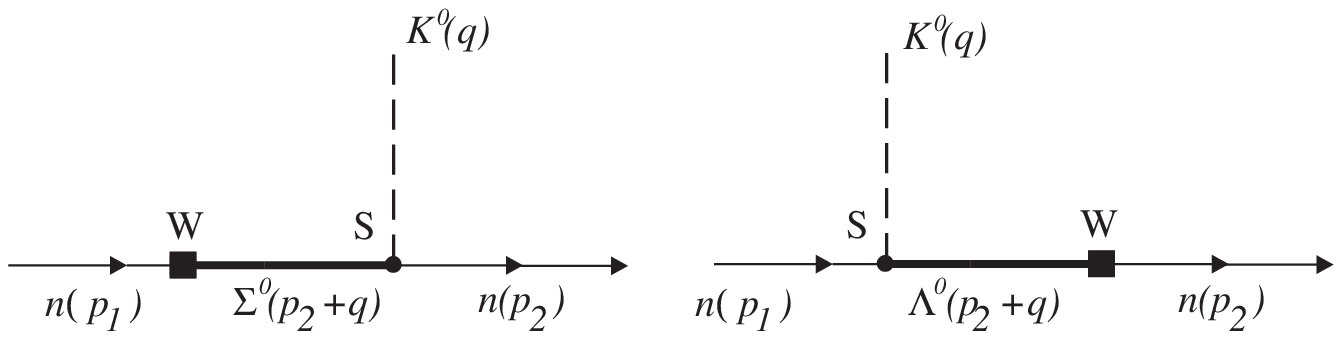}}
\medskip
\centerline{{\it Fig. \pol.4 - Pole diagrams for the neutron nonleptonic
decay: $n\to n\,K^0$}}
\medskip}
\endinsert

$$
\align
B^{\text{POLE}}_K(p^+_+)&=g(p+n)\Big[
(1-2f)\frac{a_{\Sigma^0 n}}{(\Sigma^0-n)
(\Sigma^0+p)}-\frac1{\sqrt{3}}(1+2f)\frac{a_{\Lambda^0 n}}
{(\Lambda^0-n)(\Lambda ^0+p)}\Big]
\tag\pol-9\\
B^{\text{POLE}}_K(n^0_0)
&=2gn\left\{\left[-\frac1{\sqrt{3}}(1+2f)\right]
\frac{a_{\Lambda n}}{{\Lambda^0}^2-n^2}-
(1-2f)\frac{a_{\Sigma n}}{\Sigma^2-n^2}
\right\}
\tag\pol-10\\
B^{\text{POLE}}_K(p^+_0)
&=4gp(1-2f)\sqrt{2}\frac{a_{\Sigma^+ p}}{{\Sigma^0}^2-p^2}
\tag\pol-11
\endalign
$$
Here for instance is (see Sec. \cur).
$$
a_{\Sigma^0 n}=f_{\pi}{\sqrt{2}}A(\Sigma^-_-),
\tag\pol-12
$$
etc.

\midinsert
\vbox{\medskip
\centerline{\epsfxsize=10cm\epsfbox{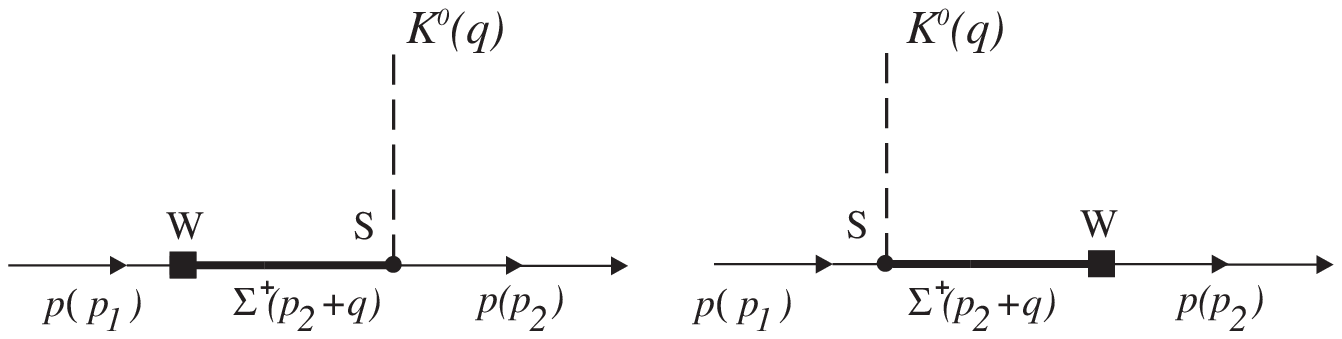}}
\medskip
\centerline{{\it Fig. \pol.5 - Pole diagrams for the proton nonleptonic
decay: $p\to p\,K^0$}}
\medskip}
\endinsert

\medskip

$\eta$\/ exchange contributions are:
$u-$channel contribution for the sub-process $\Lambda\to \eta_1+n$
evaluated by the baryon-pole model contains $\Lambda^0$ as the
intermediate hyperon whereas in the $t-$channel $n$ is the intermediate
baryon, so that strong vertices contain  $g_{\Lambda\Lambda\eta}$ and
$g_{NN\eta}$ form-factors (see Fig.\pol.6).
$$
\align
B^{\text{POLE}}_{\eta_1}(\Lambda^0_{\eta_1})&=g\sqrt{\frac23}(1-f)
\frac{\Lambda+n}{\Lambda -n}\left(\frac1{\Lambda}-\frac2{n}\right)
\left(\frac{1}{\sqrt{2}}\cdot a_{\Lambda n}\right)
\tag\pol-13\\
B^{\text{POLE}}_{\eta_8}(\Lambda^0_{\eta_8})&=\frac{a_{\Lambda^0n}}{2}
\frac{\Lambda+n}{\Lambda -n}\left(\frac{g_{\Lambda\Lambda \eta}}
{\Lambda}-\frac{g_{NN\eta}}{n}\right)
\left(\frac{1}{\sqrt{2}}\cdot a_{\Lambda n}\right)
\tag\pol-14
\endalign
$$

\midinsert
\vbox{\medskip
\centerline{\epsfxsize=11cm\epsfbox{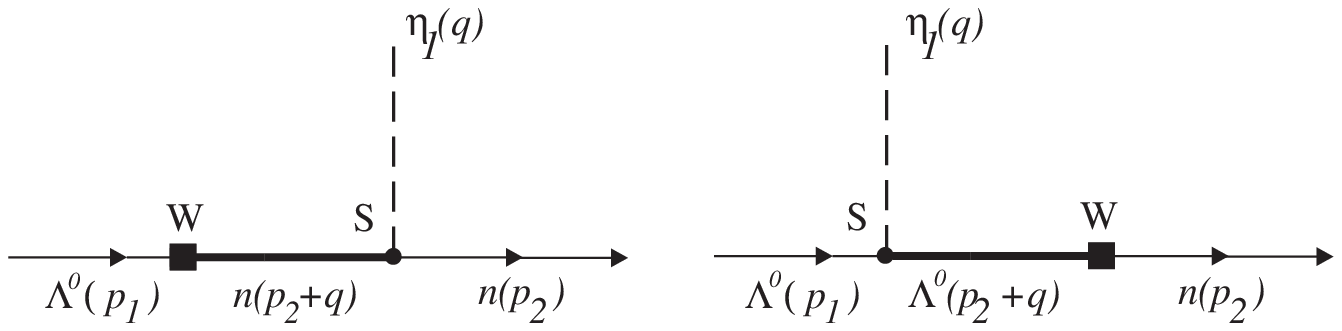}}
\medskip
\centerline{{\it Fig. \pol.6 - Baryon pole $\eta_1$ emission: $s-$channel
and $u-$channel contributions}}
\medskip}
\endinsert

In the numerical calculations (see \wbb.\ below) we will be using
quantities $\tilde a_{if}$ instead of $a_{if}$ introduced above.

\vfill
\mpikk
\eject
%
%
\def\cur{5}

\noindent{\bf \cur. Current Algebra Contributions}
\def\naspog{\cur. Current Algebra Contributions}
\vskip 1cm

Calculation of weak matrix elements by methods of current algebra (CA)
consists of several approximation procedures which lead eventually to
simple relations among transition amplitudes.

The $S$ matrix element is expressed in term of $in$ and $out$ field
operators where [33,34]
$$
\alignedat 3
\phi(\vx,t)&\simeq\phi_{in}(\vx,t) &\quad &\text{for}\quad t\to-\infty\\
\phi(\vx,t)&\simeq\phi_{out}(\vx,t) &\quad &\text{for}\quad t\to+\infty.
\endalignedat
\tag\cur-1
$$
(Here we ignore the renormalization constant appearing in front of the
$in/out$-states.)
The initial particle state is of the form
$$
\ket{k_1,\dots,k_N;in}=a_{in}^{\dagger}(k_N)\cdot\dots\cdot
a_{in}^{\dagger}(k_1)\ket{0}
\tag\cur-2
$$
and the $S$-matrix serves to connect the $in$ and $out$ states
$$
\ket{k_1,\dots,k_N;in}=S\ket{k_1,\dots,k_N;out}.
\tag\cur-3
$$
The meson field
operator $\phi_A(x)$, for which we are going to use more handy notation i.e.
$\phi(x)_A\to M_A(x)$, could be expressed by the PCAC  (\sep-16)
$$
M_{(A=a+ib)}=\frac1{m_M^2f_M}\partial_{\mu}A^{\mu}_{a+ib},
\tag\cur-4
$$
and the meson field, being the Heisenberg field operator, annihilates a
meson state with the normalization in accordance with the
creation/annihilation operator commutation relations
$$
\brik{0}{M_{a+ib}(x)}{M_{a'+ib'}(k)}=\frac{1}{(2\pi)^32\omega_k}e^{-ikx}
(\delta_{aa'}\delta_{bb'}).
\tag\cur-5
$$
Using the time-ordering operator $\Cal T$ one can calculate the matrix
element  as follows
$$
\aligned
\brik{B'M_A(k;out)}{\Cal H_W(0)}{B}&=i\int
d^4x\frac{e^{ikx}}{(2\pi)^32\omega_k}(\overrightarrow{\dalamb_x+m_M^2})
\brik{B'}{\Cal T[M_{a+ib}(x)\Cal H_W(0)]}{B}\\
&=i\frac{C_M}{f_M}\left(1-\frac{k^2}{m_M^2}\right)\int d^4x e^{ikx}
\brik{B'}{\Cal T[\partial_{\mu}A^{\mu}_{a+ib}(x)\Cal H_W(0)]}{B}.
\endaligned
\tag\cur-6
$$
Here $C_M$ depends on the particular (PS-)meson.
The time ordered product could be written in the following way
$\Cal T[A(x)B(0)]=\theta(x^0)[A(x),B(0)]+B(0)A(x)$.
From
$$
\split
e^{ikx}\theta(x_0)\partial_{\mu}A^{\mu}_{a+ib}(x)&=
\partial^{\mu}\left[e^{ikx}\theta(x_0)A^{\mu}_{a+ib}(x)\right]-
ik_{\mu}\theta(x_0)e^{ikx}A^{\mu}_{a+ib}(x)\\
&-\left[\partial_0\theta(x_0)
\right]e^{ikx}A^{\mu}_{a+ib}(x)
\endsplit
\tag\cur-7
$$
and using the derivative of the Heviside function
$$
\frac{\partial}{\partial t}\theta(t)=-\frac{\partial}{\partial
t}\theta(-t)=\delta(t)
\tag\cur-8
$$
we find
$$
\split
e^{ikx}\Cal T[\partial_{\mu}A^{\mu}_{a+ib}(x)\Cal H_W(0)]=
&-ik_{\mu}e^{ikx}\Cal T[A^{\mu}_{a+ib}(x)\Cal H_W(0)]
-\delta(x_0)
[A^0_{a+ib}(x),\Cal H_W(0)]\\
&+\partial_{\mu}\left\{e^{ikx}\Cal T
[A^{\mu}_{a+ib}(x)\Cal H_W(0)]\right\}.
\endsplit
\tag\cur-9
$$
Hence we can write the
above matrix element in the following way
$$
\split
\brik{B'M_A(k;out)}{\Cal H_W(0)}{B}=
&i\frac{C_M}{f_M}\left(1-\frac{k^2}{m_M^2}\right)\int d^4x\,e^{ikx}
\{-ik_{\lambda}\brik{B}{\Cal T[A_{a+ib}^{\lambda}(x)\Cal H_W(0)}{B}\\
&-\delta(x^0)\brik{B'}{[A^0_{a+ib}(x),\Cal H_W(0)]}{B}\}.
\endsplit
\tag\cur-10
$$
By taking the limit $k\to 0$ ({\it soft pion limit\/} for off-shell
pions) we get a typical current algebra (CA) relation
$$
\Cal M(q\to 0)=-i\frac{C_M}{f_M}\brik{B'}{[F^5_{a+ib}(0),
\Cal H_W(0)]}{B}.
\tag\cur-11
$$
Here the SU(3) (axial) charge is defined by
$$
F^5_{a+ib}(t)=\int d^3x\, A^0_{a+ib}(t,\vx).
\tag\cur-12
$$
The following CA relations hold [9,13,24]
$$
\aligned
[F_{a+ib}^5,\Cal H_W]&=[F_{a+ib},\Cal H_W^{PV}]\\
[F_{a+ib}^5,\Cal H_W^{PC}]&=[F_{a+ib},\Cal H_W^{PV}]\\
[F_{a+ib}^5,\Cal H_W^{PV}]&=[F_{a+ib},\Cal H_W^{PC}].
\endaligned
\tag\cur-13
$$
Here $F_{a+ib}$'s are (vector) charges defined in  the similar way as
$F_{a+ib}^5$'s, (\cur-12) i.e.
$$
F_{a+ib}=\int d^3x\, V_{a+ib}^0(t,\vx).
\tag\cur-14
$$
In order to evaluate the above commutator we can use the SU(3) relations
connecting the axial charges and field operators
$$
\aligned
M_{a+ib}&=\overline q\frac12(\lambda_a+i\lambda_b\gamma_5)q=
\sqrt{2}F^5_{a+ib}
\quad\text{and}\\
M_A\ket{B}&=if_{ABC}\ket{C}
\endaligned
\tag\cur-15
$$
where $f_{ABC}$ are the SU(3) structure constants and $q^T=(u,d,s)$ ($T$
stands for {\sl transposed\/}!).
For instance for the
$K^0$ field we have
$$
M_{a+ib}\to \phi_{K^0}=\sqrt{2}F_{"K^0"}=\sqrt{2}F_{6-i7}=
\overline {q}\frac12(\lambda_6-i\lambda_7)\gamma_5 q.
\tag\cur-16
$$
The above field anihilates a $K^0$ in $\ket{\hphantom{K}}$ or creates a
$K^0$ in $\bra{\hphantom{K}}$. Also $\phi_{K^+}=\sqrt{2}F_{4-i5}$ etc.
So by here displayed procedure we have extracted the {\it current
algebra\/} contribution to the matix element $\Cal M$.

The above relations can be neatly expressed by the SU(3) baryon
(antibaryon) $\Cal B$ ($\overline{\Cal B}$) and
meson $\Cal P$ octet matrices
{\uskoosam{
$$
\Cal B^a_b=\pmatrix
\displaystyle \frac{\Sigma^0}{\sqrt{ 2}}+\frac{\Lambda^0}{\sqrt{6}} &
\Sigma^+ & p\\
\Sigma^- &\displaystyle  -\frac{\Sigma^0}{\sqrt{2}}+
\displaystyle \frac{\Lambda^0}{\sqrt{6}}& n\\
\Xi^- & \Xi ^0 &\displaystyle  -\frac{2\Lambda ^0}{\sqrt{6}}
\endpmatrix\qquad
\overline{\Cal B^a_b}=\pmatrix
\displaystyle \frac{\overline{\Sigma^0}}
{\sqrt{ 2}}+\frac{\overline{\Lambda^0}}{\sqrt{6}} &
\overline{\Sigma^+} & \overline{\Xi^-}\\
\overline{\Sigma^+} &\displaystyle  -\frac{\overline{\Sigma^0}}{\sqrt{2}}+
\displaystyle \frac{\overline{\Lambda^0}}{\sqrt{6}}&\overline{\Xi^0}\\
\overline{p} &\overline{n} &\displaystyle
-\frac{2\overline{\Lambda ^0}}{\sqrt{6}}
\endpmatrix
\tag\cur-17a
$$
}}
and
{\uskoosam{
$$
\Cal P^a_b=\pmatrix
\displaystyle \frac{\pi^0}{\sqrt{ 2}}+\frac{\eta^0}{\sqrt{6}} &
\pi^+ & K^+\\
\pi^- &\displaystyle  -\frac{\pi^0}{\sqrt{2}}+
\displaystyle \frac{\eta^0}{\sqrt{6}}& K^0\\
K^- & \overline{K ^0} &\displaystyle  -\frac{2\eta ^0}{\sqrt{6}}
\endpmatrix.
\tag\cur-17b
$$
}}
Alternatively one can write
$$
\alignedat 3
\Sigma ^+&=\frac{1}{\sqrt{2}}(B_1-iB_2) &\qquad
p&=\frac1{\sqrt{2}}(B_4-iB_5)\\
\Sigma ^-&=\frac{1}{\sqrt{2}}(B_1+iB_2) &\qquad
n&=\frac1{\sqrt{2}}(B_6-iB_7)\\
\Sigma ^0&=B_3 &\qquad
\Xi^-&=\frac1{\sqrt{2}}(B_4+iB_5)\\
\Lambda ^0&=B_8 &\qquad
\Xi^0&=\frac1{\sqrt{2}}(B_6+iB_7)
\endalignedat
\tag\cur-18a
$$
and
$$
\alignedat 3
\pi ^+&=\frac{1}{\sqrt{2}}(P_1-iP_2) &\qquad
K^+&=\frac1{\sqrt{2}}(P_4-iP_5)\\
\pi ^-&=\frac{1}{\sqrt{2}}(P_1+iP_2) &\qquad
K^0&=\frac1{\sqrt{2}}(P_6-iP_7)\\
\pi ^0&=P_3 &\qquad
K^-&=\frac1{\sqrt{2}}(P_4+iP_5)\\
\eta_8 &=P_8 &\qquad
\overline{K^0}&=\frac1{\sqrt{2}}(P_6+iP_7)
\endalignedat
\tag\cur-18b
$$
The quark
SU(3)${}_{\text{flavour}}$ contents of the baryon and meson octet states
are:
$$
\alignedat 3
\Sigma ^+&\sim\ u\,u\,s  &\qquad
p&\sim\ u\,u\,d \\
\Sigma ^-&\sim\ d\,d\,s &\qquad
n&\sim\ u\,d\,d \\
\Sigma ^0&\sim\ u\,d\,s &\qquad
\Xi^-&\sim\ d\,s\,s \\
\Lambda ^0&\sim\ u\,d\,s  &\qquad
\Xi^0&\sim\ u\,s\,s
\endalignedat
\tag\cur-18c
$$
and
$$
\alignedat 3
\pi ^+&=\overline d\, u &\qquad
K^+&=\overline s\, u\\
\pi ^-&=\overline u \, d &\qquad
K^0&=\overline s\, d\\
\pi ^0&=\frac{1}{\sqrt{2}}(\overline u\, u-\overline d\, d) &\qquad
K^-&=\overline u\, s\\
\eta_8 &=\frac{1}{\sqrt{6}}(\overline u\,u+\overline d\, d-
2\overline s\, s) &\qquad
\overline{K^0}&=\overline d s
\endalignedat
\tag\cur-18d
$$
(The U(3) isosinglet field $\eta_1$ is given in terms of the U(3)
$\lambda_0$ matrix: $\lambda_0=\sqrt{\frac23}\hat 1$, i.e.
$\eta_1=(\overline u\,u+\overline d\,d+\overline s\,s)/\sqrt{3}$.)
Quark-antiquark combinations could be inverted as to be expressed in
terms of corresponding meson states, i.e.
$$
\alignedat 3
\overline u u&=\overline q\left(\frac{\sqrt{3}}{6}\lambda^8+
\frac{\sqrt{6}}{6}\lambda^0+\frac12\lambda^3\right) q&\qquad&\text{corresponding
to}\\
\overline u u&\to\frac{\sqrt{6}}{6}\eta_8+
\frac{\sqrt{3}}{3}\eta_1+\frac{\sqrt{2}}2\pi^0&\qquad&{}\\
\overline d d&=\overline q\left(\frac{\sqrt{3}}{6}\lambda^8+
\frac{\sqrt{6}}{6}\lambda^0-\frac12\lambda^3\right) q&\qquad&\text{corresponding
to}\\
\overline d d&\to\frac{\sqrt{6}}{6}\eta_8+
\frac{\sqrt{3}}{3}\eta_1-\frac{\sqrt{2}}2\pi^0&\qquad&{}\\
\overline s s&=\overline q\left(-\frac{\sqrt{3}}{3}\lambda^8+
\frac{\sqrt{6}}{6}\lambda^0\right) q&\qquad&\text{corresponding
to}\\
\overline s s&\to-\frac{\sqrt{6}}{3}\eta_8+
\frac{\sqrt{3}}{3}\eta_1&\qquad&{}
\endalignedat
\tag\cur-18e
$$

The SU(3) meson states states could be written in the standard basis
[9,35,36]
as $\ket{\text{\bf M}}=\ket{\text{\bf 8};\ Y,I,I_3}$, with
$Y=B+S$. So
$$
\alignedat 4
\ket{\pi^+}&=\Cal P_1^2\ket{0}&\qquad&=-\ket{\text{\bf 8};\ 0,1,+1}\\
\ket{\pi^-}&=\Cal P_2^1\ket{0}&\qquad&=\hphantom{-}
\ket{\text{\bf 8};\ 0,1,-1}\\
\ket{\pi^0}&=\frac{1}{\sqrt{2}}(\Cal P_1^1-
\Cal P_2^2)\ket{0}&\qquad&=\hphantom{-}\ket{\text{\bf 8};\ 0,1,0}\\
\ket{K^+}&=\Cal P_1^3\ket{0}&\qquad&=\hphantom{-}\ket{\text{\bf 8};\ 1,1/2,1/2}\\
\ket{K^0}&=\Cal P_2^3\ket{0}&\qquad&=\hphantom{-}\ket{\text{\bf 8};\ 1,1/2,-1/2}\\
\ket{\overline K^0}&=\Cal P_3^2\ket{0}&\qquad&=
\hphantom{-}\ket{\text{\bf 8};\ -1,1/2,1/2}\\
\ket{K^-}&=\Cal P_3^1\ket{0}&\qquad&=-\ket{\text{\bf 8};\ -1,1/2,-1/2}\\
\ket{\eta^0}&=-\frac{3}{\sqrt{6}}\Cal P_3^3\ket{0}
        &\qquad&=\hphantom{-}\ket{\text{\bf 8};\ 0,0,0}
\endalignedat
\tag\cur-18f
$$
The corresponding SU(2) (iso)multiplets are
$$
\alignedat 5
K^a&=\pmatrix K^+\\
     K^0\endpmatrix
&\qquad \overline K^a&=\pmatrix -\overline K^0\\ K^-\endpmatrix
&\qquad
\pi_a^b&=\pmatrix \frac{\pi^0}{\sqrt{2}} & \pi^+\\
                  \pi^- & -\frac{\pi^0}{\sqrt{2}}
         \endpmatrix \\
N^a&=\pmatrix p^+\\
     n^0\endpmatrix
&\qquad \overline \Xi^a&=\pmatrix -\overline \Xi^0\\ \Xi^-\endpmatrix
&\qquad
\Sigma_a^b&=\pmatrix \frac{\Sigma^0}{\sqrt{2}} & \Sigma^+\\
                  \Sigma^- & -\frac{\Sigma^0}{\sqrt{2}}
         \endpmatrix
\endalignedat
\tag\cur-18g
$$

When in (\cur-18a,b) defined operators act on a initial {\it ket\/} or final
{\it bra\/} they annihilate or create corresponding states. From the
SU(3) charges one can construct the well known operators $F_+$ or $F_-$
for instance (together with $F_3$) which have the standard SU(2) group
properties \footnote{$^{\dag}$}{\uskoosam
One has to be very carefull with the definitions of the initial and
final states and with operators acting on the states. The usual
definition which is adopted here as well is that operators {\it
annihilate\/} the corresponding particle state in {\it the initial
state\/}. Careless application of this convention could lead to much
confussion in a calculation to which many authors contribute by loosely
following a given convention!}

$$
\alignedat 3
\bra{p}F_+&=\bra{n} &\qquad  \bra{n}F_3&=-\frac12\bra{n}\\
F_+\ket{\Xi^-}&=-\ket{\Xi^0} &\qquad
F_3\ket{\Xi^0}&=\frac12\ket{\Xi^0}\\
F_+\ket{\Sigma^-}&=\sqrt{2}\ket{\Sigma^0} &\qquad
F_-\ket{\Sigma^+}&=-\sqrt{2}\ket{\Sigma^0}
\endalignedat
\tag\cur-19
$$

When the commutators appearing in the matrix elements $\Cal M$
(\cur-15) are
evaluated one gets the transition amplitudes which are the {\it current
algebra\/} contributions to $A$.\footnote{$^{\ddag}$}{\uskoosam
The PC amplitude $B$ does not have the
contribution comming from the above commutators, but only the pole
diagrams contribute as we
have shown in the preceding section.} Therefore
$$
\align
A(\Lambda^0_-)&=-\frac{\sqrt{2}}{f_{\pi}}\frac1{\sqrt{2}}\brik{p}
{[F_+,\Cal H_W^{PC}(0)]}{\Lambda}=-\frac{1}{f_{\pi}}\brik{n}
{\Cal H^{PC}_W(0)}{\Lambda}\\
&=-\frac{1}{f_{\pi}}a_{\Lambda n}
\tag\cur-20\\
A(\Lambda^0_0)&=-\frac{\sqrt{2}}{f_{\pi}}\frac1{\sqrt{2}}\brik{n}
{[F_3,\Cal H_W^{PC}(0)]}{\Lambda}\\
&=\frac{1}{\sqrt{2}f_{\pi}}a_{\Lambda n}
\tag\cur-21\\
A(\Xi^-_-)&=-\frac{\sqrt{2}}{f_{\pi}}\frac1{\sqrt{2}}\brik{\Lambda}
{[F_+,\Cal H_W^{PC}(0)]}{\Xi^-}=-\frac1{f_{\pi}}\brik{\Lambda}
{\Cal H_W^{PC}}{\Xi^0}\\
&=-\frac{1}{f_{\pi}}a_{\Xi^0\Lambda}
\tag\cur-22\\
A(\Xi^0_0)&=-\frac{1}{\sqrt{2}f_{\pi}}a_{\Xi^0\Lambda}
\tag\cur-23\\
A(\Sigma^-_-)&=\frac{\sqrt{2}}{f_{\pi}}a_{\Sigma^0n}
\tag\cur-24\\
A(\Sigma^+_+)&=-\frac{1}{f_{\pi}}\left[a_{\Sigma^+p}+\sqrt{2}a_{\Sigma^0
n}\right]
\tag\cur-25\\
A(\Sigma^+_0)&=\frac{1}{\sqrt{2}f_{\pi}}a_{\Sigma^+p}
\tag\cur-26
\endalign
$$

\vfill
\mpikk

\eject
%

%
%
\def\meso{6}

\noindent{\bf \meso. Meson Poles and Separable Contributions}
\def\naspog{\meso. Meson Poles and Separable Contributions}
\vskip 1cm

Some terms in separable contributions (Sec. \sep\,) are
also included in the meson
pole contributions, Fig. \meso.1. However, with
some care, one can avoid the double
counting. The separable contributions due to
operators $\Cal O_1,$  $\Cal O_2,$
$\Cal O_3,$ and $\Cal O_4$ as calculated in Section \sep\.
are not included in the meson (for example kaon)
contributions. The meson (kaon) pole contributes to the parity
conserving $B$ amplitudes only.

The separable $PC$ contributions (\sep-8) by
operators $\Cal O_1,$  $\Cal O_2,$ $\Cal O_3,$ and $\Cal O_4$ were in
our case calculated using the product of axial vector currents
$A_a^{\mu}(x)$:
$$
\brik{B_{\beta}}{A_a^{\mu}}{B_{\alpha}}\cdot \brik{L_a}
{A_{\mu}^a}{0}=
B(\text{sep.})
\tag\meso-1
$$
Here, in accordancs with (\sep-10),
$$
\aligned
\brik{B_{\beta}}{A_a^{\mu}(0)}{B_{\alpha}}&=\overline u_{\beta}(p')
\left[\gamma^{\mu}\gamma_5 g_A(q^2)\right.\\
&\left.+\frac{1}{M_{\alpha}+M_{\beta}}g_P(q^2) q^{\mu}\gamma_5\right]
\Lambda_a u_{\alpha}(p)\\
q&=p-p';\qquad g_A(0)\simeq 1.25
\endaligned
\tag\meso-2
$$
and $L_a$ is some
meson.\footnote{$^{\dag}$}{\uskoosam
For $B(\Lambda_-^0)$ $L_a=\pi^-$.}
The results presented in  Sec. \sep\  were obtained by assuming that
$g_P$ term (induced pseudoscalar form factor) can be neglected. {\it The
contribution due to the induced pseudoscalar\/} $g_P$ is contained only in
the kaon-pole term as described below. Some needed definitions and
conventions are given in Appendix A.

The meson pole contribution to the process in which a meson $L$ is
emitted
$$
B_{\alpha}\quad\to\quad B_{\beta}\quad+\quad L_a
\tag\meso-3
$$
is shown in Fig.\meso.1.

\vbox{\medskip
\centerline{\epsfysize=4cm\epsfbox{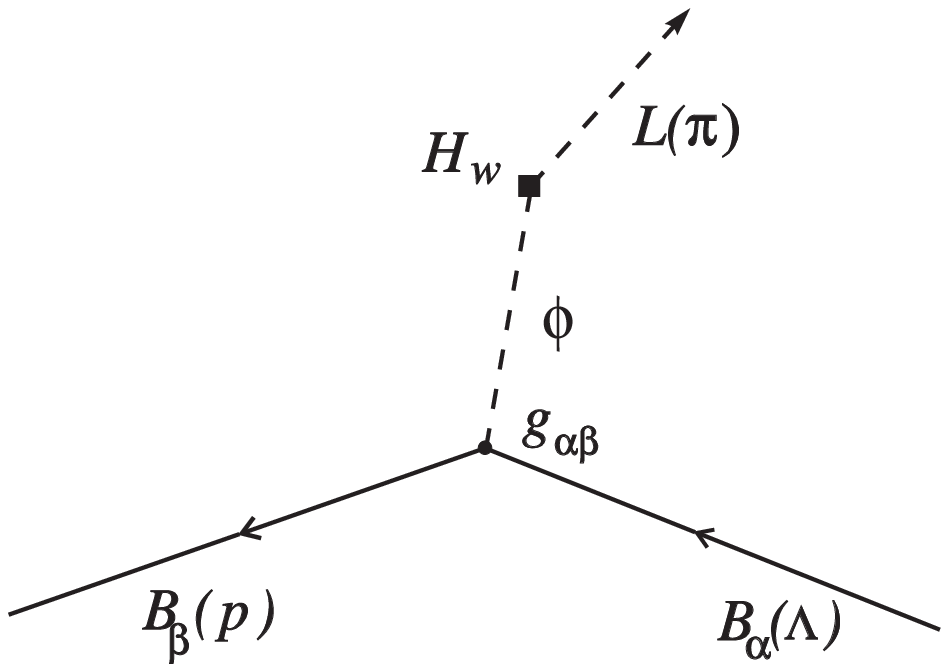}}
\medskip
{\uskodevet{
\centerline{{\it Fig. \meso.1 - The strong vertex is $g_{\alpha\beta}$ and
the weak Hamiltonian $H_w$}}
\centerline{{\it contains the current product (\meso-1). The particles in
the }}
\centerline{{\it process $\Lambda\to p+\pi$ are indicated in the
parentheses.}}}
}
\medskip}

The amplitude shown in Fig.\meso.1 which corresponds to $\phi$ meson
pole, is\footnote{$^{\dag}$}{\uskoosam Here $B^{MB}$ is a
contribution due to a meson pole, while $B^{\text{POLE}}$ in Ch.\pol\
denotes a contribution connected with a baryon pole.}
$$
B^{MP}=\brik{L}{H_w}{\phi}\frac{i}{q^2-m_{\phi}^2+i\epsilon}
\overline u_{\beta}\gamma_5 \Lambda_a u_{\alpha}g_{\alpha
\beta}.
\tag\meso-4
$$
When one approximates
$$
\brik{L}{H_w}{\phi}\simeq\brik{L}{A_{\mu}^b}{0}
\brik{0}{A^{\mu}_a}{\phi},
\tag\meso-5
$$
then by using the relation from Appendix A
$$
\frac{g_P(q^2)}{M_{\alpha}+M_{\beta}}\simeq(-)
\frac{g_{\alpha\,\beta} f_{\phi}(q^2)}{q^2-m_{\phi}^2},
\tag\meso-6a
$$
and $if_{\phi}q^{\mu}=\brik{0}{A^{\mu}}{\phi}$, one can show that
$$
B^{MP}\simeq \brik{L}{A_{\mu}^b}{0}\cdot \overline u_{\beta}
\frac{1}{M_{\alpha}+M_{\beta}} g_P q_{\mu}\gamma_5 \Lambda_a u_{\alpha}.
\tag\meso-6b
$$
Thus the induced pseudoscalar ($g_P$) contribution to the
$B(\text{sep})$ is included in the $M(\text{pole})$ term. When
calculating the separable contribution from operators
$\Cal O_1,\, \dots,\, \Cal O_4$, one should include only the $g_A$ form
factors as it has been done in Sec. \sep.

The operators $\Cal O_5$ and $\Cal O_6$ contribute pieces
which look like
$$
\aligned
\Cal O_x\sim &(\overline \psi_1\gamma_5\psi_2)(\overline
\psi_3\gamma_5\psi_4)\\
&\brik{B_{\beta}}{\overline \psi_1\gamma_5\psi_2}{B_{\alpha}}\not=0\\
&\brik{L}{\overline \psi_3\gamma_5\psi_4}{0}\not=0
\endaligned
\tag\meso-7
$$
Here $\psi_i$ symbolize quark fields appearing in $H_w$. The
corresponding $B^{MP}$ contribution is obtained from (\meso-4)
by the replacement
$$
\brik{L}{H_w}{\phi}\quad\to\quad \brik{L}{\Cal O_x}{\phi}.
\tag\meso-8
$$
For meson pseudoscalar densities one uses identities
$$
\aligned
\overline \psi_i\gamma_5\psi_j&=\frac{1}{m_i+m_j}(-i)(\partial_{\mu}
\overline \psi_i\gamma^{\mu}\gamma_5\psi_j)\\
&=\frac{1}{m_i+m_j}(-i)(i)(p_i^{\mu}-p_j^{\mu})\overline
\psi_i\gamma_{\mu}\gamma_5\psi_j.
\endaligned
\tag\meso-9
$$
Thus
$$
\aligned
\brik{B_{\beta}}{\overline \psi_1\gamma_5\psi_2}{B_{\alpha}}&=
(-i)\brik{B_{\beta}}{\partial_{\mu}A^{\mu}_a}{B_{\alpha}}
\frac 1{(m_1+m_2)}\\
&=\frac{f_{\phi}\tilde q^2}{q^2-m_{\phi}^2}g_{\alpha \beta}\overline
u_{\beta}\gamma_5 \Lambda_a u_{\alpha}\cdot
\frac{1}{m_{1}+m_{2}}.
\endaligned
\tag\meso-10
$$
Here the notation $\tilde q^2$ was used instead $m_M^2$.

A separable contribution corresponding to (\sep-8) thus have a generic form
$$
B_x({\text{sep}})=\frac{f_{\phi}\tilde
q^2}{(m_1+m_2)(q^2-m_{\phi}^2)}g_{\alpha\beta}\overline u_{\beta}\gamma_5
\Lambda_a
u_{\alpha}\brik{L}{\overline\psi_3\gamma_5\psi_4}{0}.
\tag\meso-11
$$
It openly displays $\phi-$meson pole so one expects that it must be
included in the $B^{MP}$ contribution. In order to test that
one again approximates (\meso-5) and writes (\meso-4)
$$
B_x^{MP}\simeq \brik{L}{\overline \psi_3\gamma_5\psi_4}{0}
\brik{0}{\overline \psi_1\gamma_5\psi_2}{\phi}
\frac{i}{q^2-m_{\phi}^2}
g_{\alpha\beta}\overline u_{\beta}\gamma_5
\Lambda_au_{\alpha}.
\tag\meso-12a
$$
The first mesonic matrix element is
$$
\brik{0}{\overline\psi_1\gamma_5\psi_2}{\phi}=\frac1{m_1+m_2}(-i)
\brik{0}{\partial_{\mu}A^{\mu}_a}{\phi_a}=\frac{1}{m_1+m_2}(-i)q^2
f_{\phi}.
\tag\meso-13
$$
Thus
$$
B_x^{MP}\simeq \frac{f_{\phi}q^2}{(m_1+m_2)(q^2-m_{\phi}^2)}
g_{\alpha\beta}\overline u_{\beta}\gamma_5 \Lambda_a u_{\alpha}
\brik{L}{\overline \psi_3\gamma_5\psi_4}{0}.
\tag\meso-12b
$$
The approximation (\meso-12) is exactly equal to $B_x(\text{sep})$
(\meso-11) if one puts
$$
\tilde q^2\equiv q^2.
\tag\meso-14
$$

Alternatively and formally one can compare the meson pole approximation
(\meso-12a)
with the expression (\meso-11) converting
$$
\frac{f_{\phi} m_{\phi}^2}{m_1+m_2}=
\frac{\brik{0}{\partial_{\mu}A^{\mu}_a}{\phi}}
{m_1+m_2}=i\brik{0}{\overline \psi_1\gamma_5\psi_2}{\phi}\qquad
(\tilde q^2=m_{\phi}^2).
\tag\meso-15
$$
The insertion of (\meso-15) in (\meso-11) converts it into (\meso-12a).
However in that formal procedure one has neglected the question about
meson $\phi$ being off-mass-shell. Some additional details can be found in
Appendix \apa.

In the derivation of PCAC constant  one has assumed that the
meson $\phi$ was on the mass shell. As that
meson is {\it not\/} on the mass shell neither in the diagram in
Fig.\meso.1, nor in the expression (\meso-11), the reading
(\meso-14) seems to be amply justified. The separable contributions from
operators $\Cal O_5$ and $\Cal O_6$ are included in the meson (for
example kaon) pole.

For the process
$$
B_{\alpha}\quad\to\quad B_{\beta}\quad+\quad M_a.
\tag\meso-16
$$
one can now formulate {\it the rule for the calculation of $B$
amplitudes\/}:
$$
B=B(\text{baryon pole})+B(\text{sep})+B^{MP}.
\tag\meso-17
$$
Here $B(\text{sep})$ contains only the contribution from the $g_A$ terms
associated with the operators $\Cal O_1$,  $\Cal O_2$,  $\Cal O_3$ and
$\Cal O_4$. The $g_P$ terms for these operators and
the separable contributions from $\Cal O_5$ and $\Cal O_6$
are included in $B^{MP}$ piece. The contributions
$B(\text{baryon-pole})=B^{\text{POLE}}$ were calculated in Section \pol\
while $B(\text{sep})$ can be found in Sec. \sep.

\vfill
\mpikk
\eject
%

%
%
%
\def\wbb{7}

\centerline{\bf \wbb. Weak Baryon-Baryon-Meson Amplitudes}
\def\naspog{\wbb. Weak Baryon-Baryon-Meson Amplitudes}
\vskip 1cm

It is very important to test the quality of the approximate procedure
which was used to calculate $A$ and $B$
amplitudes. That can be done by:
\item{(\it i\,)} Comparison whith other theoretical calculations,
\item{(\it ii\,)} Comparison with experimental results.

The first task can be easily accomplished by studying the amplitudes
corresponding to $\eta$ and $K$ exchanges
given in Tables \wbb.1-\wbb.4. Ref. [5] did not
distinguished $\eta_1$ (singlet) from $\eta_8$ (octet) exchanges. In
their approach $\eta$ was supposed to belong to an SU(3) octet. Our
results for the octet $\eta$ are fairly close to Ref. [5].
$A_{\eta_8}$ differs by about 8\,\%, while the discrepancy for
$B_{\eta_8}$ is 18\,\%. The origin of those discrepancies is obviously
connected with the separable  and pole  terms
which appear in our calculational sheme. As discussed in Sections
\sep\ and \pol, they were not used or they were used in
different forms by ref. [5]. In view of that, the agreement within
18\,\% suggests that used calculational shemes are relatively stable
and, hopefully, reliable to within about 20\,\%. That is the accuracy
that was hoped for in the theoretical descriptions of the hyperon
nonleptonic decays [8-15].

The amplitudes appearing with the kaon strangeness violation vertices
are compared in Table \wbb.3. The differences are small, below 29\,\%,
for $A(p_0^+)$, $A(p_+^+)$ and $A(n_0^0)$. They are much larger for the
corresponding $B$ amplitudes. There the absolute value of $B(p^+_0)$
amplitudes, as calculated by [22], is about 3 times smaller then
ours. However in all cases signs are the same. If $B_K(p^+_0)$ is
excluded, the largest discrepancy is about 21\,\%. Our result for
$B_K(p^+_0)$ amplitude differs from Ref. [5] by about 35\,\% only.

Comparison with experimental reults is, and can be, performed in a
limitted sense only. The experimental $A_{\pi}$ ampitudes are used to
find $\tilde a(B'B)$ matrix elements via a substraction procedure
described in Appendix \aph.
Those $\tilde a_{BB'}$ quantites are
then used in baryon pole terms which were defined in Section \pol.
Together with separable contributions and kaon poles theay lead to the
theoretical prediction of $B_{\pi}$ amplitudes, which are shown in
Tables \wbb.1 and \wbb.4. Reference [5] has used simple,
unsubstracted pole terms as was discussed in Section \pol. More
complicated and hopefully better, spproximation leads to somewhat better
agreement with the experiment in most cases. According to Table \wbb.4
one finds that
discrepancies with experiments for various amplitudes are as follows:
$B_{\pi}(\Lambda_-^0)$, 6\%; $B_{\pi}(\Lambda_0^0)$, 2\%;
$B_{\pi}(\Sigma_+^+)$, 14\%; $B_{\pi}(\Sigma_0^+)$, 48\%;
$B_{\pi}(\Xi_0^0)$, 4\%; and $B_{\pi}(\Xi_-^-)$, 6\%. While
$B_{\pi}(\Lambda)$ and $B_{\pi}(\Xi)$ amplitudes are reproduced within
6\%, the theoretical prediction for $B_{\pi}(\Sigma)$ is poor. Our
theoretical $B_{\pi}(\Sigma_-^-)$ amplitude is off by factor 5.2.
However the corresponding numbers, which were found using a simplified
approximation, analogous to the one applied by Ref. [5] are:
$B_{\pi}(\Lambda^0_-)$ 44\%; $B_{\pi}(\Lambda^0_0)$ 42\%;
$B_{\pi}(\Sigma^+_+)$ 35\%; $B_{\pi}(\Sigma^+_0)$ 3\,times too small;
$B_{\pi}(\Xi^0_0)$ 1\%;
$B_{\pi}(\Xi^-_-)$ 1\%, and $B_{\pi}(\Sigma^-_-)$ disagrees by
factor 6.3. Thus for most
amplitudes simplified approximation gives poorer agreement with the
experimental values.

There were always some difficulties associated with the theoretical
description of the $\Sigma$-hyperon nonleptonic decays. Attempted
explanations involved additional $1/2^-(1/2^+)$ baryon resonance poles
[37-40], instantons [41] etc.

Overall one feels that the first column in Table \wbb.3 might constitute
a better approximation of the real physical results, than given by the
last column. No theoretical comparison with Ref. [22] is feasible,
as that reference uses a quite different method of evaluation.
However it
is encouraging that all methods give similar relative magnitudes and
relative sign. The worst disagreement is for $B_K(p^+_0)$. However,
$A_K(p_0^+)$ agrees within 14\%. If the agreement among various
theoretical results is an acceptable indicator, one could say that
$A_K(N)$ amplitudes are determined with better accuracy than the
$B_K(N)$ ones. A sceptical observer would claim that the results
presented in Table \wbb.3 are the order of magnitude estimates at best.

Our method has produced both $\Delta I=1/2$ and $\Delta I=3/2$ pieces in
the effective potential (\efp-6). In the formula (\efp-11) those
pieces are written separately, with isospin operators ${\bold
1}\cdot{\bold 1}$ and $\vec\tau_i\cdot\vec\tau_j$ for $\Delta I=1/2$ and
$\vec T_i\tau_j$ for $\Delta I=3/2$. However, $\Delta I=3/2$ pieces are
small and their magnitude is comparable with the theoretical errors
which were discussed above. This can be starkly illustrated if one
calculates the $\Delta I=3/2$ piece, associated with $\Lambda\to N\pi$
weak amplitudes. One finds
$$
B_{\pi}(\Lambda^0_-)_{\text{exp}} +\sqrt{2}
B_{\pi}(\Lambda^0_0)_{\text{exp}}=0.32\times10^{-7}.
\tag\wbb-1
$$
This experimental $\Delta I=3/2$ piece has a different sign than the
predicted one:
$$
B_{\pi}(\Lambda^0_-)_{\text{BHNT}}+\sqrt{2}
B_{\pi}(\Lambda^0_-)_{\text{BHNT}}=-0.26\times10^{-7}.
\tag\wbb-2
$$
Effect is due to subtraction of large (relatively speaking) numbers.
Yet $B(\Lambda)_{\text{exp}}$ agree within 6\% with
$B_{\pi}(\Lambda)_{\text{BHNT}}$. The same degree of agreement is shown
by $B(\Xi)$ amplitudes. But in that case $\Delta I=3/2$ pieces
($B\Xi_-^-) +\sqrt{2}B\Xi^0_0)$ are $0.013\times10^{-7}$ (exp.) and
$0.018\times10^{-7}$ (theor.).

In the deduction of the nuclear potential (\efp-11) we have used,
naturally, the experimental $B_{\pi}(\Lambda)$ value. Thus the whole
discussion, involving the theoretical $B_{\pi}(\Lambda)$ and
$B_{\pi}(\Xi)$ values, serves as an indication for theoretical
uncertainities connected with the predicted $A_K(N)$ and $B_K(N)$
(Table \wbb.3) values. It is hard to quantify those uncetainties. While
the theoretical $\Delta I=3/2$ piece, connected with
$B_{\pi}(\Lambda)$'s, has wrong sign (but its magnitude is comaparable
with the experimental value) the $B_{\pi}(\Xi)$ based $\Delta I=3/2$
piece agrees with the experimental value within 40\%.

{\uskoosam{
%
%
\parskip 0pt
\baselineskip 10pt
\medskip
$$
\table{}
Amplitude                    ! Exp.$\times 10^{7}$!
BHKNT(a)$\times 10^{7}$ ! BHKNT(b)$\times 10^{7}$!Ref. [5]$\times 10^{7}$
 \jrr
$A_{\pi}(\Lambda^0_-)$       ! 3.25  !  $3.25^{(*)}$  ! 2.97
!$3.25^{\dagger}$  \rr
$B_{\pi}(\Lambda^0_-)$       ! 22.40 !  $22.27$  !19.38 ! $22.25^{\dagger}$
       \rr
$ A_{\pi}(\Lambda^0_0)$      ! -2.36 !  $-2.36^{(*)}$ !-2.10 ! $-2.37^{\dagger}$
       \rr
$ B_{\pi}(\Lambda^0_0)$      ! -15.61 !  $-15.93$ ! -13.70! $-15.8^{\dagger}$
       \rr
$ A_{\pi}(\Sigma^+_0)$       ! -3.27 ! $ -3.27^{(*)}$ ! -3.71 ! $-3.27^{\dagger}$
        \rr
$ B_{\pi}(\Sigma^+_0)$       ! 26.74 ! 18.09  !18.04 ! $26.6^{\dagger}$
        \rr
$ A_{\pi}(\Sigma^-_-)$        !4.27   ! $4.27^{(*)}$ !4.60 ! $4.27^{\dagger}$
       \rr
$ B_{\pi}(\Sigma^-_-)$        ! -1.44  ! $-7.46$ !-5.39 ! $-1.44^{\dagger}$
       \jrr
$ A_{K}(p^+_ +)  $            !  --   ! 1.31    ! 1.02  ! 1.69
       \rr
$ B_{K}(p^+_ +)  $            !  --   ! 42.38  ! 34.38 ! 41.30
        \rr
$ A_K(n^0_0)  $               !  --   ! 6.25    ! 6.25  !  6.33
       \rr
$ B_K(n^0_0)  $               !  --   ! 17.99  ! 21.56 !  26.58
        \rr
$ A_K(p^+_0) $                !  --   ! 4.08    !  5.25  ! 4.64
       \rr
$ B_K(p^+_0) $                !  --   ! -21.87   !  -14.59 ! 14.72
       \rr
$A_{\eta}(\Lambda^0_{\eta_1})$! --    ! 5.60    ! 5.54 ! --
       \rr
$B_{\eta}(\Lambda^0_{\eta_1})$! --     ! 41.10 ! 26.09 ! --
       \rr
$A_{\eta}(\Lambda^0_{\eta_8})$! --     ! 5.19  ! 5.54 ! 5.63
       \rr
$B_{\eta}(\Lambda^0_{\eta_8})$! --     ! 38.41 ! 28.82  ! 31.60
\caption{\uskodevet{\it Table \wbb.1 - Transition amplitudes:
[BHKNT]$\equiv$ present work: (a) a complete
amplitude; (b) an amplitude without the separable
contribution (this work). The asterisk ${}^{(*)}$ denotes
that the experimental amplitude was used instead.
${}^{\dagger}\equiv$ experimental value assumed.}}
$$
\medskip
}}
%
%

{\uskoosam{
%
%
\parskip 0pt
\baselineskip 10pt
\medskip
$$
\table{}
 Ref. [5]                !  Total Ampl.! Ampl. without SEP\jrr
$A_{\pi}: \hfill 3.25 $  !\hfill    3.25       !\hfill  2.97        \rr
$B_{\pi}: \hfill 22.35$ !\hfill    22.27     !\hfill  19.38         \rr
$A_{\eta_1}: \hfill -- $ !\hfill 5.60          !\hfill   5.54        \rr
$A_{\eta_8}: \hfill 5.63 $ !\hfill 5.19          !\hfill   5.54        \rr
$B_{\eta_1}: \hfill -- $!\hfill   41.10   !\hfill 26.09  \rr
$B_{\eta_8}: \hfill 31.60 $!\hfill   38.41($m_{\eta}^2$)  !\hfill 28.82 \jrr
$C_K^{PV}: \hfill 2.38 $   !$ A_{K}(p^+_ +): \hfill 1.31 $!\hfill  1.02  \r
$\hphantom{C_K^{PV}}\hfill 1.25$ !         !         \rr
$C_K^{PC}: \hfill 41.30 $   !$ B_{K}(p^+_ +):   \hfill 42.38 $!\hfill
34.38 \r
$\hphantom{C_K^{PC}}\hfill 43.76 $!         !         \rr
$D_K^{PV}: \hfill 6.53 $   !$ A_K(p^+_0): \hfill 4.08 $!\hfill    5.25  \r
$\hphantom{D_K^{PV}}\hfill 4.69$ !         !         \rr
$D_K^{PC}: \hfill -14.72 $   !$ B_K(p^+_0): \hfill -21.87 $
!\hfill  -14.59  \r
$\hphantom{C_K^{PV}}\hfill -10.00$ !         !         \rr
$C_K^{PV}+D_K^{PV}: \hfill 6.33 $   !$ A_K(n^0_0):  \hfill 6.25 $
!\hfill    6.25  \r
$\hphantom{C_K^{PV}+C_K^{PV}}\hfill 5.94$ !              !  \rr
$C_K^{PC}+D_K^{PC}: \hfill 26.58 $   !$ B_K(n^0_0): \hfill 17.99 $
!\hfill 21.56  \r
$\hphantom{C_K^{PV}+D_K^{PC}}\hfill 33.75 $ !             !
%
\caption{\uskodevet\it Table \wbb.2 - Comparison between
transition amplitudes (in $10^{-7}$ units, w.o. dimension) as given
in [5] and  this work. A complete amplitude
includes  a separable contribution. Two different results in the Ref.
[5] coloumn correspond to two different sets of results as quoted
there.}
$$
\medskip
}}
%
%

\midinsert
%
%
\parskip 0pt
\baselineskip 10pt
\medskip
$$
\table{}
Amplitude ! Tot. ampl.${}^{\dagger}$ ! Ref.[19] ! Ref.[3] \rr
$A(p^+_0)$ ! 4.08    ! 4.09 ! 4.64           \r
$B(p^+_0)$ ! -21.87    ! -7.6 ! -14.72           \r
$A(p^+_+)$ ! 1.31    ! 1.09  ! 1.69     \r
$B(p^+_+)$ ! 42.38   ! 33.40 ! 41.30     \r
$A(n^0_0)$ ! 6.25    ! 5.19  ! 6.33         \r
$B(n^0_0)$ ! 17.99    ! 26.16 ! 26.58
\caption{{\it Table \wbb.3 - Nonleptonic amplitudes ($\times
10^7$); ${}^{\dagger}$ this work, compared with rescaled values of
Ref. [22]
and [5].
}}
$$
\medskip
\endinsert
%
%

\midinsert
{\uskoosam{
%
%
\parskip 0pt
\baselineskip 10pt
\medskip
$$
\table{}
Ampl.! $B_{\text{exp.}}$!$B^{\text{(POLE)}}$!
Sep.! Mes. ! Tot.${}^{\dagger}$ !Pole! Sep.! Ref. \r
$\times 10^7$!   !$(\tilde A)$! $(1-4)^{*}$ ! pole !amp. !   !
$(5-6)^{\flat}$ ![5] \jrr
$B_{\pi}(\Lambda_-^0)$! 22.4! 19.38 !4.79 ! -1.9 ! 22.27 !
23.03 ! 6.98 ! 15.6 \rr
$B_{\pi}(\Lambda_0^0)$! -15.61! -13.70 ! -1.03 ! -1.2 ! -15.93 !
-16.28 ! -3.68 ! -11.03 \rr
$B_{\pi}(\Sigma_+^+)$! 41.83! 48.73 !-- ! -- ! 48.73 !
37.68 ! -- ! 30.97 \rr
$B_{\pi}(\Sigma_0^+)$! 26.74 ! 18.04 ! -0.45 !  0.5 ! 18.09 !
15.85 ! -1.62 ! 9.13 \rr
$B_{\pi}(\Sigma_-^-)$! -1.44! -5.39 ! -2.17 ! 0.1 ! -7.46 !
-8.83 ! -2.79 ! -9.08 \rr
$B_{\pi}(\Xi_0^0)$! -12.13 ! -12.67 ! -0.29  ! 0.3 ! -12.66 !
-11.28 ! -1.05 ! -12.21 \rr
$B_{\pi}(\Xi_-^-)$! 17.45 ! 17.45  ! -1.40 ! 0.4 ! 16.45 !
15.97 ! -1.94 ! 17.26 \jrr
$B_{K}(p_+^+)$! -- ! 34.38 ! -7.5 ! 15.5  ! 42.38 !
47.60 ! -138.6 ! 41.3 \rr
$B_{K}(p_0^+)$! -- ! -14.59 ! 0.12 ! -7.4  ! -21.87 !
-12.99 ! -11.76 ! -6.63 \rr
$B_{K}(n_0^0)$! -- ! 21.56 !-10.97  ! 7.4  ! 17.99 !
24.58 ! -57.03 ! 26.71 \rr
$B_{\eta_1}(\Lambda_0^0)$! -- ! 26.09 ! 1.16 !13.85  ! 41.10 !
28.16 ! -130.8 ! -- \rr
$B_{\eta_8}(\Lambda_0^0)$! -- ! 28.82 ! -6.76 ! 16.35  ! 38.41 !
31.10 ! 762.36 ! 31.60
\caption{{\it Table \wbb.4 - Nonleptonic $p-$wave (parity conserving)
amplitudes ($\times 10^7$); ${}^{\dagger}$ this work.
${}^{*}$Contributions from operators $\Cal O_1-\Cal O_4$;
${}^{\flat}$Contributions from operators $\Cal O_{5,6}$.
}}
$$
\medskip
}}
\endinsert
%
%

To facilitate a comparison between our results  and those
of  ref. [5] we
translate the ref. [5] formulae to a more transparent form, i.e. we
write
$$
\aligned
C_K^{PV}\vert_{[5]}&=A_K(p_+^+)\,\vert_{[\text{BHNT}]}\\
C_K^{PC}\vert_{[5]}&=B_K(p_+^+)\,\vert_{[\text{BHNT}]}\\
D_K^{PV}\vert_{[5]}&=A_K(p_0^+)\,\vert_{[\text{BHNT}]}\\
D_K^{PC}\vert_{[5]}&=B_K(p_0^+)\,\vert_{[\text{BHNT}]}\\
C_K^{PV}+D_M^{PV}\vert_{[5]}&=A_K(n_0^0)\,\vert_{[\text{BHNT}]}\\
C_K^{PC}+D_M^{PC}\vert_{[5]}&=B_K(n_0^0)\,\vert_{[\text{BHNT}]}
\endaligned
\tag\wbb-3
$$
The above equalities should be fulfilled only for "bare" amplitudes i.e.
the amplitudes without the separable contributions!

In some future study one has to consider the contribution of the SU(3)
decuplet poles, which are succesfully employed by ref. [37].

\vfill
\mpikk
\eject
%

%
%
%
\def\efe{8}

\centerline{\bf \efe. Effective Field Theory and The Weak
$|\Delta S|=1$ Potential}
\def\naspog{\efe. Effective Field Theory and The Weak
$|\Delta S|=1$ Potential}
\vskip 1cm

Instead of starting with quarks $(u,d,s)$ we formulate formalism (an
effective one) in which baryons $(N,\ \Lambda)$ and mesons $(\pi,\ K,\
\eta)$ appear.
Such an approach facilitates the deduction of the effective potential
presented in the next section.
We will start with $\pi-$exchange contribution.

Process (transition, potential) under consideration is
$$
\Lambda + N\to N+N;\qquad
N=\pmatrix p \\ n \endpmatrix
\tag\efe-1
$$
In a typical diagram one has one weak and one strong vertex, as shown in
Fig.\uvo.1. In the perturbation calculation this diagram can be deduced
from the effective interaction Hamiltonian (\efe-2)
$$
\aligned
\Cal H_{\text{eff}}&=\int d^4x\,
\overline\psi_n(x)G_Fm_{\pi}^2
(A+\gamma_5 B)
\psi_{\Lambda}(x)\phi _{\pi^0}(x)\\
&+\int d^4x\,
\overline\psi_p(x)G_Fm_{\pi}^2
(A+\gamma_5 B)
\psi_{\Lambda}(x)\phi _{\pi^+}(x)\\
&+g_{\pi NN}\int d^4x\,
\overline\psi_N(x)\gamma_5 \tau_i \psi_{N}(x)\phi _{\pi_i}(x)\\
&=\Cal H_{W\,\pi^0}+\Cal H_{W\,\pi^+}+\Cal H_s\\
&=\int d^4x\, (h_0(x)+h_+(x)+h_S(x)).
\endaligned
\tag\efe-2a
$$
The last term in (\efe-2a) is the strong nucleon-pion interaction, which
can be written as
$$
\aligned
\Cal H_S&=g_{\pi NN}\int d^4x\,
\overline\psi_N(x)\gamma_5 \tau_i \psi_{N}(x)\phi _{\pi_i}(x)\\
&=g_{\pi NN}\int d^4x\,
\left\{\left[\overline\psi_p(x)\gamma_5  \psi_{p}(x)-
\overline\psi_n(x)\gamma_5  \psi_{n}(x)\right]\phi _{\pi^0}(x)+
\sqrt{2}\overline\psi_p(x)\gamma_5  \psi_{n}(x)\phi
_{\pi^+}(x)\right.\\
&\hphantom{g_{\pi NN}}\left.+\sqrt{2}\overline\psi_n(x)\gamma_5
\psi_{p}(x)\phi _{\pi^-}(x)
\right\}.
\endaligned
\tag\efe-2b
$$
Sometimes one also writes
$$
\alignedat 4
g_{pp\pi^0}&=&\quad  -&g_{nn\pi^0}&\quad &=g_{\pi NN}\\
g_{pn\pi^+}&=&\quad  &g_{np\pi^-}&\quad &=\sqrt{2}g_{\pi NN}.
\endalignedat
\tag\efe-2c
$$
The diagram Fig.\uvo.1 is due to the second order contribution
$$
\aligned
\Cal H_{\text{eff}} \Cal H_{\text{eff}}&=\int d^4x_1\,d^4x_2
[h_0(x_1)+h_+(x_1)+h_S(x_1)] [h_0(x_1)+h_+(x_1)+h_S(x_1)]\\
&=\int d^4x_1\,d^4x_2
[h_0(x_1)h_S(x_2)+h_S(x_1)h_0(x_2)+
h_+(x_1)h_S(x_2)\\
&\hphantom{+}+h_S(x_1)h_+(x_2)]+R
\endaligned
\tag\efe-3
$$
Here only $|\Delta S|=1$ terms are shown. The part $R$ contains the
strong pion exchange such as
$$
\Cal H_{\text{eff}} \Cal H_{\text{eff}}=\int d^4x_1\,d^4x_2
h_S(x_1)h_S(x_2).
\tag\efe-4
$$
In the terms shown in (\efe-3) the pion field contraction must be carried
out. That leaves us with an effective baryon-baryon (i.e. $\Lambda+N\to
N+N$) interaction. For example (see Appendix E for more details)
$$
\aligned
V_{\text{op}}&=\int d^4x_1\,d^4x_2
[\pkditrk{\phi_{\pi^0}(x_1)}{\phi_{\pi^0}(x_2)}\overline
\psi_n(x_1)G_Fm_{\pi}^2
A(\Lambda_0^0)\psi_{\Lambda}(x_1)\cdot
g_{\pi NN}\overline\psi_n(x_2)\gamma_5\psi_n(x_2)\\
\hphantom{\int d^4x_1}&+\text{similar terms of the same type}
\endaligned
\tag\efe-5a
$$
One can replace:
$$
\int d^4x_1\,d^4x_2\pkditrk{\phi_{\pi^0}(x_1)}{\phi_{\pi^0}(x_2)}\quad
\longrightarrow\quad
\int d^3x_1\,d^3x_2\,\Delta(1,2).
\tag\efe-5b
$$
Here $\Delta(1,2)$ is the Yukawa function. Integration over time components
goes into overall energy conservation. In order to do that one assumes
that all baryon statyes are stationary states, i.e. that baryon
operator is of the form
$$
\psi(x)=\sum_i a_i\,e^{-iE_it/\hbar}\phi_i(\vec r).
\tag\efe-5c
$$
Time dependence of the pion propagator can be also included explicitly
or replaced by $\delta(t_1-t_2)$. More about that can be found in the
Appendix \apg\,.

The final $V_{\text{op}}$ is
$$
\aligned
V_{\text{op}}&=\int d^3 x_1\,d^3 x_2\Big\{
\left[-g_{\pi NN}{\bold:}\overline\psi_n(x_1)G_Fm_{\pi}^2
[A(\Lambda_0^0)+
B(\Lambda_0^0)\gamma_5]
\psi_{\Lambda}(x_1)\right.\\
&\hphantom{++++}\cdot
\overline\psi_n(x_2)\gamma_5\psi_n(x_2){\bold:}\Delta (1,2)\\
&\hphantom{++++}+\left.(1\leftrightarrow 2)\right]\\
&+g_{\pi NN}{\bold:}\overline\psi_n(x_1)G_Fm_{\pi}^2
[A(\Lambda_0^0)+
B(\Lambda_0^0)\gamma_5]
\psi_{\Lambda}(x_1)\cdot
\overline\psi_p(x_2)\gamma_5\psi_p(x_2){\bold :}\Delta(1,2)
+(1\leftrightarrow
2)\\
&+g_{\pi NN}\sqrt{2}{\bold:}\overline\psi_p(x_1)G_Fm_{\pi}^2
[A(\Lambda_-^0)+
B(\Lambda_-^0)\gamma_5]
\psi_{\Lambda}(x_1)\cdot
\overline\psi_n(x_2)\gamma_5\psi_p(x_2){\bold :}\Delta(1,2)\\
&\hphantom{++++}+(1\leftrightarrow
2)]\Big\}
\endaligned
\tag\efe-6
$$
Here the first two terms contribute to $\Lambda+n\to n+n$ while the rest
contributes to $\Lambda +p\to n+p$. The notation $\bold :\quad \bold :$
indicates the normal ordering.

In the nonrelativistic limit we have:
$$
\psi_a\simeq\pmatrix
        \chi_a\\
        \displaystyle\frac{\vsi\cdot\vp_a}{2m_a}\chi_a
\endpmatrix
\tag\efe-7
$$
where $\chi_a$ a solution of a Schr\"odinger equation
is a two-component spinor.
From the
expression (\efe-6) we can see that there are two contributions to the
weak-strong mixing amplitude whose space-time properties are determined
by the Lorentz structure of vertices. First amplitude (see Fig.\efe-1)
has $[\gamma_5(S)]\otimes [1(W)]$ structure and the second one has
$[\gamma_5(S)]\otimes [\gamma_5(W)]$ structure.

\midinsert
\vbox{\medskip
\centerline{\epsfysize=5cm\epsfbox{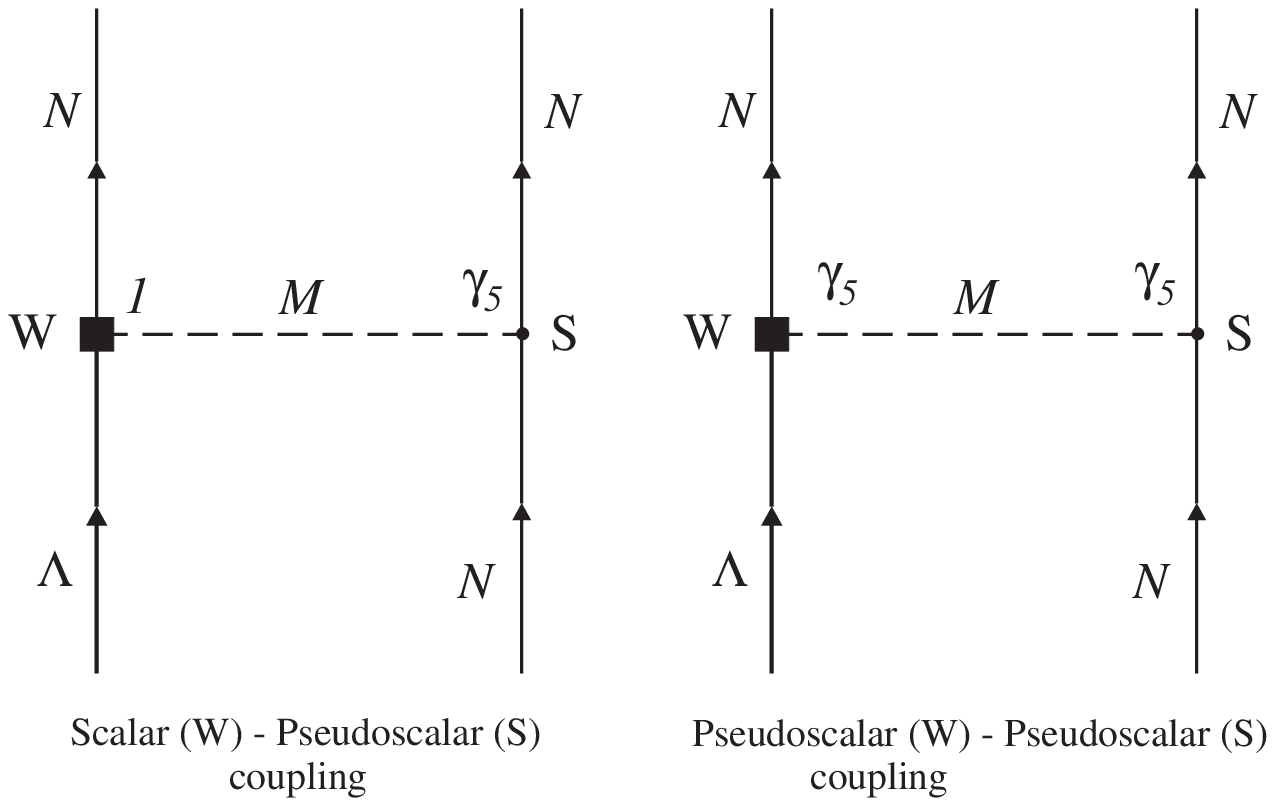}}
\medskip
\centerline{{\uskodevet \it Fig.\efe.1 - Two different weak (W) - strong (S)
combinations for the OME amplitude}}
\medskip}
\endinsert

Two different contributions are treated on the same footing: in the
nonrelativistic limes of Dirac spinors given in (\efe-7)
i.e.
for slowly moving particles we can write for different combinations of
spinors (to the leading terms in $1/m_N$) ($\vec q=\vec p_1-\vec p_3,$)
$$
\alignat 3
\overline \psi_3 \psi_1&\simeq (\chi_3^{\dagger}\, \chi_1),
&\ \text{scalar}\quad\tag\efe-8a\\
\overline \psi_3\gamma_5 \psi_1&\simeq \left(\frac1{2m_N}\right)
\chi_3^{\dagger}(\vsi\cdot \vec q)\chi_1;
&\ \text{pseudoscalar}\quad\tag\efe-8b\\
\overline \psi_3\gamma^i\gamma_5 \psi_1&\simeq
\chi_3^{\dagger}(\sigma^i)\chi_1
&\ \text{space comp. of a axial vector}\quad\tag\efe-8c\\
\overline \psi_3\gamma^0\gamma_5 \psi_1&\simeq \left(\frac1{2m_N}\right)
\chi_3^{\dagger}(\vsi\cdot\vec p_1+\vsi\cdot\vec p_3)\chi_1
&\ \text{time comp. of a axial vector}\quad\tag\efe-8d\\
\overline \psi_3\gamma^i \psi_1&\simeq \left(\frac1{2m_N}\right)
\chi_3^{\dagger}(\sigma^i\vsi\cdot\vec p_1+\vsi\cdot\vec p_3\sigma^i)\chi_1
&\ \text{space comp. of a vector}\quad\tag\efe-8e\\
\overline \psi_3\gamma^0 \psi_1&\simeq
\chi_3^{\dagger}\chi_1
&\ \text{time comp. of a vector}\quad\tag\efe-8f\\
\endalignat
$$
From
$$
(2\pi)^{-3}
(\q^2+m_M^2)^{-1}=\frac1{4\pi}\int
e^{i\q\cdot\bold r}\frac{e^{-m_Mr}}{r}\,d^3\bold r
\tag\efe-9a
$$
one gets
$$
\Delta(|\vec r_1-\vec r_2|)=V(r)=-\frac1{4\pi}\frac{e^{-m_Mr}}{r}
\tag\efe-9b
$$
i.e. the known {\it Yukawa potential\/}. As discussed in Appendix \apg\,
the mass dependence can be more complicated than given in (\efe-9).

For the vertices of the form $[\gamma_5(S)]\otimes [\gamma_5(W)]$ we get
a baryon-baryon potential
$$
V(r)\sim (\vsi_1\nabla_1)(\vsi_2\nabla_2)\frac{e^{-m_Mr}}{r},
\tag\efe-10
$$
where $r=|\vr_1-\vr_2|$.
By using some of the identities
given in Appendix \apg\,  we obtain
four different contributions (in the
coordinate space) for the effective potential
$$
\aligned
V_S&=\frac13\left[\frac{m_M^2}{4\pi r}e^{-m_Mr}-\delta(r)\right]\\
V_T&=\frac13\frac{m_M^2}{4\pi r}e^{-m_Mr}\left[1+\frac{3}{m_Mr}+
\frac{3}{(m_Mr)^2}\right]\\
V_{PV}&=\frac{m_M}{4\pi r}e^{-m_Mr}\left(1+\frac{1}{m_Mr}\right)
\endaligned
\tag\efe-11
$$

If the weak decay $\Lambda\to N+\pi$ obeys the isospin selection rule
$|\Delta I|=1/2$
then it is possible to establish the connection
$$
\aligned
\frac{F(\Lambda_-^0)}{F(\Lambda_0^0)}&=-\sqrt{2};\qquad (F=A,B)\\
\Lambda^0_- &\quad\sim \Lambda ^0\to p+\pi^-\\
\Lambda^0_0 &\quad\sim\quad \Lambda ^0\to p+\pi^-.
\endaligned
\tag\efe-12
$$
With $f=G_Fm_{\pi}^2(A+B\gamma_5)$ one can write
$$
\aligned
V_{\text{op}}&=\int d^3 x_1\,d^3 x_2\Big\{
g_{\pi NN}{\bold:}\overline\psi_n(x_1)f
\psi_{\Lambda}(x_1)\cdot
\left[-\overline\psi_n(x_2)\gamma_5\psi_n(x_2)\right.\\
&\hphantom{++++}
+\left.\overline\psi_p(x_2)\gamma_5\psi_p(x_2)\big]{\bold:}\Delta(1,2)
+(1\leftrightarrow2)\right.\\
&-2g_{\pi NN}{\bold:}\overline\psi_p(x_1)f
\psi_{\Lambda}(x_1)\cdot
\overline\psi_n(x_2)\gamma_5\psi_p(x_2){\bold:}\Delta(1,2)
+
(1\leftrightarrow 2) \Big\}
\endaligned
\tag\efe-13
$$
The bilinear combinations which contain $\gamma_5$ and which are
associated with the strong vertex in Fig.\uvo.1, can be written as
$$
\overline \psi_N\tau_3\psi_N=\overline
\psi_p\psi_p-\overline\psi_n\psi_n;
\qquad
\overline \psi_N\tau_+\psi_N=\overline \psi_p\psi_n
\tag\efe-14
$$
The $|\Delta I|=1/2$ weak Hamiltonian transforms as the Pauli spinor
$$
\chi^{-1/2}=\pmatrix 0 \\ 1 \endpmatrix.
\tag\efe-15
$$
Such spinor is sometime known as {\it spurion\/}. The bilinear
combination containing $\Lambda$ field can be written as
$$
\aligned
\overline\psi _N\tau^3 \pmatrix 0 \\ 1 \endpmatrix \psi_{\Lambda}&=
(\overline\psi_p,\overline\psi_n)\pmatrix 1 & 0 \\ 0 & -1 \endpmatrix
\pmatrix 0 \\ 1 \endpmatrix\psi_{\Lambda}\\
&=(-)\overline\psi_n\psi_{\Lambda}.\\
\overline\psi _N\tau^+ \pmatrix 0 \\ 1 \endpmatrix \psi_{\Lambda}&=
(\overline\psi_p,\overline\psi_n)\pmatrix 0 & 1 \\ 0 & 0 \endpmatrix
\pmatrix 0 \\ 1 \endpmatrix\psi_{\Lambda}\\
&=\overline\psi_p\psi_{\Lambda}.\\
\endaligned
\tag\efe-16a
$$
Formally one can assume that $\Lambda-$field is a quasi-spinor
$$
\psi_{\Lambda\,n}=\pmatrix 0 \\ \Psi_{\Lambda}\endpmatrix.
\tag\efe-16b
$$
For future purpose we will also introduce
$$
\psi _{\Lambda\,p}=\tau^+\psi_{\Lambda\,n}=\pmatrix \psi_{\Lambda}\\
0 \endpmatrix
\tag\efe-16c
$$
With conventions (\sep-7) one can put (\sep-5) into the form (only the
part is kept in which $\Lambda$ is in $x_1$!)
$$
\aligned
V_{\text{op}}&=-\sum_{i}\int d^3 x_1\,d^3 x_2
g_{\pi NN}\overline\psi_N(x_1)\tau_i\,f
\psi_{\Lambda\,n}(x_1)\cdot
\overline\psi_N(x_2)\tau_i\gamma_5\psi_N(x_2)
\endaligned
\tag\efe-17
$$
which has been used by [3].

With (\efe-14) and (\efe-16) one  obtains
$$
\aligned
(-)\left(\overline\psi_N\tau^3\psi_{\Lambda\,n}\right)
\left(\overline\psi_N\tau_3\psi_N\right)&=+
\left(\overline\psi\psi_{\Lambda}\right)[\overline\psi_p\psi_p-\overline
\psi_n\psi_n]\\
(-)2\left(\overline\psi_N\tau^+\psi_{\Lambda\,n}\right)
\left(\overline\psi_N\tau_-\psi_N\right)&=-2
\left(\overline\psi\psi_{\Lambda}\right)(\overline\psi_n\psi_p)
\endaligned
\tag\efe-18
$$
We have also used
$$
\sum_i\tau_1^i\tau_2^i=\vec\tau_1\cdot\vec\tau_2=
\tau_1^3\tau_2^3+2(\tau_1^+\tau_1^-+\tau_1^-\tau_2^+).
\tag\efe-19
$$
Thus the $|\Delta I|=1/2$ $V_{\text{op}}$
is given by (\efe-17).

One can add the $|\Delta I|=3/2$ piece (see Appendix \cpc\,)
$$
\aligned
V_{\text{op}}(3/2)&=\sum_{h}\int d^3x_1\,d^3x_2\, g_{\pi NN}\overline\psi
_N(x_1)\cdot T_i^m\chi^m_{1/2}\,h\,\psi_{\Lambda}(x_1)\cdot
\overline\psi_N(x_2)\tau_i\gamma_5\psi_N(x_2)\\
h&=\alpha\text{ or }\beta\gamma_5.
\endaligned
\tag\efe-20a
$$
The magnitudes of $\alpha$ (or $\beta$), as well as their relative signs
with respect to $A$ (or $B$) (\efe-2) are calculated theoretically in
Sec. \wbb. Here $\psi_N$ are isospinors
$$
\psi_N(x)=\pmatrix \psi_p(x) \\ \psi_n(x) \endpmatrix,
\tag\efe-20b
$$
and
$$
\chi_{1/2}^m\quad\to\quad \pmatrix 1 \\ 0 \endpmatrix;\qquad \pmatrix 0
\\ 1 \endpmatrix
\tag\efe-20c
$$
is an isospinor while
$$
\psi_{\Lambda}(x)
\tag\efe-20d
$$
is an isoscalar.

If one calculates with nuclear wave functions then the potential can be
obtained from
$$
V_{\text{op}}=V_{\text{op}}(1/2)+V_{\text{op}}(3/2),
\tag\efe-21
$$
by omitting the baryon fields $\psi_N$ and $\psi_{\Lambda}$. Thus
$V_{\text{op}}\to V$
$$
\aligned
V=(-)&\sum_f g_{\pi NN}(f)_1(\gamma_5)_2\vec\tau_1\cdot\vec\tau_2+\\
     &\sum_h g_{\pi NN} (h)_1 (\gamma_5)_2 \left(\vec T_1^m\cdot
     \chi_{1/2}^m(1)\right)\cdot\vec\tau_2.
\endaligned
\tag\efe-22a
$$
The notation
$$
\vec T_1^m\cdot \chi_{1/2}^m(1)
\tag\efe-22b
$$
means that in the actual calculation of that term one has to associate
an isospinor $\chi_{1/2}^m(1)$ with the $\Lambda$ particle, which is
also in the position $x_1$. (Thus (\efe-22) is {\it not\/}
symmetrized.) In the first term in (\efe-22a) the $\Lambda$ particle,
which is neutral like neutron, automatically has the spinor
$\chi_{1/2}^{-1/2}(1)$. The second term in (\efe-22a) must be read as
$$
N^{\dagger}\vec T_1^m{\chi_{1/2}}^m\vec\tau_2\to\cases
(\text{proton:})\ {\chi^{1/2}}^{\dagger}\vec
T_1^{1/2}\chi^{1/2}\cdot\vec\tau_2\quad\text{or}\\
(\text{neutron:})\ {\chi^{-1/2}}^{\dagger}\vec T_1^{-1/2}
\chi^{-1/2}\cdot\vec\tau_2
\endcases
\tag\efe-23a
$$
with
$$
\aligned
{\chi^{1/2}}^{\dagger}\vec T_1^{1/2}\chi^{1/2}\cdot\vec\tau_2&=
\kleb{1}{-1}{1/2}{1/2}{3/2}{-1/2} \vec t_1^{-1}\cdot \vec\tau_2\\
{\chi^{-1/2}}^{\dagger}\vec T_1^{-1/2}\chi^{1/2}\cdot\vec\tau_2&=
\kleb{1}{0}{1/2}{-1/2}{3/2}{-1/2} \vec t_1^{0}\cdot \vec\tau_2\\
\vec t_1^{-1}\cdot \vec\tau_2=\frac{1}{\sqrt{2}}(1,-i,0)(\tau_1,
\tau_2,\tau_3)&=\sqrt{2}\tau_-\\
\vec t_1^0\cdot\vec\tau_2&=\tau_2^3.
\endaligned
\tag\efe-23b
$$
The notation in (\efe-1a) and (\efe-3b) can be understood without the
explicit reference to isospinot $\chi_{1/2}^m(1)$ (see also Ref.
[42,43], formulae (2.29) and (2.43): their $T=T^{\dagger}$(ours).) One
can write
$$
\overline \psi_N(x_1)T_i\psi_{\Lambda}^{-1/2}(x_2)
\tag\efe-24
$$
In isospin formalism that means
$$
\chi^{m_s}(1/2)^{\dagger} T_i \Delta ^M(3/2).
\tag\efe-25a
$$
Here $N$ and $\Delta$ are $ I=1/2$ and $I=3/2$ spinors, with the
projections $m_s$ and $M$. The actual values of these projections are
$$
\alignedat 3
m_s&=\pm1/2 &\qquad &\text{for proton, neutron}\\
M&=-1/2     &\qquad &\text{for weak transitions}.
\endalignedat
$$
The transition isospin $\vec T$ is defined by its matrix
elements\footnote{$^{\dag}$}{\uskoosam
Other details are given in Appendix \cpc.
}
$$
\left(T_i\right)_{m_s\,M}=\sum_r \kleb{1}{r}{1/2}{m_s}{3/2}{M} t_i^r.
\tag\efe-25b
$$

The form (\efe-3), and analogous forms for $\eta$ and $K$ exchanges,
automatically lead to a symmetric potential. Our final results,
in the isospin dependent formalism, are displayed in Section \efp. Some
formal details are explained in Appendices \cpc\ and \apd.

\vfill
\mpikk
\eject
%
%
%
%
%
%
\def\efp{9}

\centerline{\bf  \efp. The  Effective Potential}
\def\naspog{\efp. The Effective Potential }

\medskip

Methods described in Section \efe.\ and Apendices \apa\ and \apg\ allow
construction of the strangeness violating potential which corresponds to
the exchange of the pseudoscalar mesons $\pi,\ K$ and $\eta$.
Its radial dependence is described by
$$
\align
V_C(M)&=\frac {e^{-m_Mr}}{4\pi r}\tag\efp-1\\
V_S(M)&=\frac13\left[\frac{m_M^2}{4\pi
r}e^{-m_Mr}-\delta(r)\right]\tag\efp-2\\
V_T(M)&=\frac13\frac{m_M^2}{4\pi r}e^{-m_Mr}\left[1+\frac{3}{m_Mr}+
\frac{3}{(m_Mr)^2}\right]\tag\efp-3\\
V_{PV}(M)&=\frac{m_M}{4\pi r}e^{-m_Mr}\left(1+\frac{1}{m_Mr}\right)\\
&(M=\{\pi,K,\eta\}).
\tag\efp-4a
\endalign
$$
According to Appendix \apg\ one can replace $m_{M}$ by
$\epsilon_M=[m_M^2-(m_{\Lambda}-m_N)^2/4]^{1/2}.$
There is a simple derivative connection between different radial
functions, i.e.:
$$
\gathered
-\frac{\partial}{\partial r}V_C=V_{PV}\\
\frac{\partial}{\partial r}\left[\frac{1}{r}
\frac{\partial}{\partial r}V_C\right]=3V_{T}
=-3\frac{\partial}{\partial r}\left[\frac{1}{r}
V_{PV}\right]
\endgathered
\tag\efp-4b
$$

The above radial parts get multiplied by spin-isospin components of the
weak-strong vertices and corresponding amplitudes given in the former
sections whose numerical values are given in Sec. \wbb.
Therefore, written formally
$$
\align
V_C&\to V_C\cdot 0
\tag\efp-5a\\
V_S&\to V_S(\{\pi,K,\eta\})
\cdot(\vsi_1\cdot\vsi_2)\cdot \{B_{\pi},\ B_{K},\ B_{\eta}\}
\tag\efp-5b\\
V_T&\to V_T(\{\pi,K,\eta\})\cdot(\hat S_{12})\cdot \{B_{\pi},\  B_{K},\  B_{\eta}\}
\tag\efp-5c\\
V_{PV}&\to V_{PV}(\{\pi,K,\eta\})\cdot(\hr\cdot\vsi_2)\cdot
\{A_{\pi},\  A_{K},\  A_{\eta}\}.
\tag\efp-5d
\endalign
$$
Here we introduced $\hat S_{12}=3\cdot(\vsi_1\cdot\hr\vsi_2\cdot\hr
-\vsi_1\cdot\vsi_2/3)$.

The effective potential could be written in two different (but
equivalent) ways: the first version displays the {\it particle
content\/} of the different part of the potential, whereas the other
version shows more explicitly its {\it isospin character\/}.

The {\it particle content version\/} of the effective potential has to
be written in such a way that the particular channel of the
$\Lambda-Nucleon$ interaction is specified.
The {\it isospin character\/} version
unifies both channels and enables us to write the effective potential
as an operator in the isospace.

The {\it particle content\/} version is given first, for the
$\Lambda+p\to p+n$ channel:
\vfill
\eject

{\uskoosam{
$$
\aligned
V(\vr)&_{[\Lambda\ p\to p\,n]}=V_C(r)\cdot 0\\
  &+(\vsi_1\cdot\vsi_2)
      \cdot\Big\{V_S(\pi)\cdot
      \frac{g_{NN\pi}}{2m_N(m_{\Lambda}+m_N)}
      \left(2\tilde a+\sqrt{\frac23}\tilde b\right)\frac12
      [(\overline n
      p)_2(\overline p \Lambda)_1+(\overline n
      p)_1(\overline p \Lambda)_2]\\
     &\hphantom{+xx(\vsi_1\cdot\vsi_2)\cdot}
      +V_{S}(\pi)\cdot\frac{g_{NN\pi}}{2m_N(m_{\Lambda}+m_N)}
      \left(-\tilde a+\sqrt{\frac23}\tilde b\right)\frac12
      [(\overline p p)_2(\overline n \Lambda)_1 +(\overline p p)_1
      (\overline n \Lambda)_2]\\
      &\hphantom{+xx(\vsi_1\cdot\vsi_2)\cdot}
        +\frac{g_{K\Lambda N}}{2m_N(m_{\Lambda}+m_N)}
      \big(V_S(K^+)(\tilde c+\tilde e)\frac12[(\overline n
      p)_1(\overline p \Lambda)_2+(\overline n p)_2(\overline p
      \Lambda)_1]\\
      &\hphantom{+xx(\vsi_1\cdot\vsi_2)\cdot}
      +V_S(K^0)(\tilde d+\tilde e)\frac12[(\overline p
      p)_1(\overline n \Lambda)_2+(\overline p p)_2(\overline n
      \Lambda)_1]\big)\\
      &\hphantom{+xx(\vsi_1\cdot\vsi_2)\cdot}
        +\frac{1}{2m_N(m_{\Lambda}+m_N)}
      \Big(V_S(\eta_1)\cdot
      g_{\eta_1 NN}\tilde f\frac12[(\overline p
      p)_2(\overline n \Lambda)_1+(\overline p p)_1(\overline n
      \Lambda)_2]\\
      &\hphantom{+xx(\vsi_1\cdot\vsi_2)\cdot}
      +V_S(\eta_8)\cdot g_{\eta_8 NN}
                \tilde g\frac12[(\overline p
      p)_2(\overline n \Lambda)_1+(\overline p p)_1(\overline n
      \Lambda)_2]\Big)\Big\}\\
   &+\hat S_{12}\ \cdot\Big\{
          V_T(\pi)\cdot
      \frac{g_{NN\pi}}{2m_N(m_{\Lambda}+m_N)}
      \left(2\tilde a+\sqrt{\frac23}\tilde b\right)\frac12[(\overline n
      p)_2(\overline p \Lambda)_1+(\overline n p)_1(\overline p
      \Lambda)_2 ]\\
      &\hphantom{+xx(\vsi_1\cdot\vsi_2)\cdot}
      +V_{T}(\pi)\cdot\frac{g_{NN\pi}}{2m_N(m_{\Lambda}+m_N)}
      \left(-\tilde a+\sqrt{\frac23}\tilde b\right)\frac12
      [(\overline p p)_2(\overline n \Lambda)_1 +(\overline p p)_1
      (\overline n \Lambda)_2]\\
      &\hphantom{+xx(\vsi_1\cdot\vsi_2)\cdot}
        +\frac{g_{K\Lambda N}}{2m_N(m_{\Lambda}+m_N)}
      [V_T(K^+)(\tilde c+\tilde e)\frac12[(\overline n
      p)_1(\overline p \Lambda)_2+
      (\overline n p)_2(\overline p \Lambda)_1]\\
      &\hphantom{+xx(\vsi_1\cdot\vsi_2)\cdot}
      +V_T(K^0)(\tilde d+\tilde e)\frac12[(\overline p
      p)_1(\overline n \Lambda)_2+
      (\overline p p)_2(\overline n \Lambda)_1]\\
      &\hphantom{+xx(\vsi_1\cdot\vsi_2)\cdot}
        +\frac{1}{2m_N(m_{\Lambda}+m_N)}
      \Big([V_T(\eta_1)\cdot
      g_{\eta_1 NN}\tilde f\frac12[(\overline p
      p)_2(\overline n \Lambda)_1+
      (\overline p p)_1(\overline n \Lambda)_2]\\
      &\hphantom{+xx(\vsi_1\cdot\vsi_2)\cdot}
      +V_T(\eta_8)\cdot g_{\eta_8 NN}\tilde g\frac12[(\overline p
      p)_2(\overline n \Lambda)_1+(\overline p p)_1(\overline n
      \Lambda)_2]\Big)
      \Big\}\\
      &\hphantom{++}+\Big\{V_{PV}(\pi)\cdot\frac{g_{NN\pi}}{2m_N}
      \left(2a+\sqrt{\frac23}b\right)\frac12
      [-(\vsi_2\hr)(\overline n
      p)_2(\overline p \Lambda)_1+(\vsi_1\hr)(\overline n
      p)_1(\overline p \Lambda)_2]\\
      &\hphantom{++}+V_{PV}(\pi)\cdot\frac{g_{NN\pi}}{2m_N}
      \left(-a+\sqrt{\frac23}b\right)\frac12
      [-(\vsi_2\hr)(\overline p
      p)_2(\overline n \Lambda)_1 +(\vsi_1\hr)(\overline p p)_1
      (\overline n \Lambda)_2]\\
      &\hphantom{+xx(\vsi_1\cdot\vsi_2)\cdot}
        +\frac{g_{K\Lambda N}}{2m_N}
      \big(V_{PV}(K^+)(c+e)\frac12[-(\vsi_2\hr)(\overline n
      p)_1(\overline p \Lambda)_2\\
      &\hphantom{+xx(\vsi_1\cdot\vsi_2)\cdot}
      +(\vsi_1\hr)(\overline n p)_2(\overline p
      \Lambda)_1]\\
      &\hphantom{+xx(\vsi_1\cdot\vsi_2)\cdot}
      +V_{PV}(K^0)( d+e)\frac12[-(\vsi_2\hr)(\overline p
      p)_1(\overline n \Lambda)_2+(\vsi_1\hr)(\overline p p)_2(\overline n
      \Lambda)_1]\big)\\
      &\hphantom{+xx(\vsi_1\cdot\vsi_2)\cdot}
        +\frac{1}{2m_N}
      \Big(V_{PV}(\eta_1)\cdot
      g_{\eta_1 NN} f\frac12[-(\vsi_2\hr)(\overline p
      p)_2(\overline n \Lambda)_1+\\
      &\hphantom{+xx(\vsi_1\cdot\vsi_2)\cdot}
      (\vsi_1\hr)(\overline p p)_1(\overline n
      \Lambda)_2]\Big)\\
      &\hphantom{+xx(\vsi_1\cdot\vsi_2)\cdot}
      +V_{PV}(\eta_8)\cdot g_{\eta_8 NN}
                 g\frac12[-(\vsi_2\hr)(\overline p
      p)_2(\overline n \Lambda)_1\\
      &\hphantom{+xx(\vsi_1\cdot\vsi_2)\cdot}
      +(\vsi_1\hr)(\overline p p)_1(\overline n
      \Lambda)_2]\Big\}\\
\endaligned
\tag\efp-6
$$
}}
\vfill

Here the quantities $\tilde a$, $\tilde b$,  $\tilde c$ etc. are
vertices - weak (nonleptonic) amplitudes for the isoscalar and
isotensor parts
of the potential, and "bare" quantities like $a$, $b$, $c$ etc. are the
corresponding parity-violating amplitudes.
These quantities connect two different ways (versions, see above) in
which the effective potential could be written. A particle exchanged
could be inferred from the radial function ($V_k(\pi)$ for instance, for
$k=T,S,PV$) multiplying the corresponding terms.

For the other  channel the particle version is
given
{\uskodevet{
$$
\aligned
V(\vr)&_{[\Lambda\ n\to n\,n]}=V_C(r)\cdot 0\\
      &+(\vsi_1\cdot\vsi_2)
      \cdot\Big\{
      \frac{g_{\pi
      NN}}{2m_N(m_{\Lambda}+m_N)}\big(V_S(\pi)\left(\tilde a-\sqrt{\frac23}
      \tilde b\right)\frac12[(\overline n n)_2(\overline n\Lambda)_1
      +(\overline n n)_1(\overline n\Lambda)_2]
      \big)\\
       &\hphantom{(\vsi_1\cdot\vsi_2)\cdot\cdot}
        +\frac{g_{KN\Lambda}}{2m_N(m_{\Lambda}+m_N)}
       V_S(K^0)(\tilde c+\tilde d-\tilde e)
      \frac12[(\overline n n)_2(\overline n\Lambda)_1
      +(\overline n n)_1(\overline n\Lambda)_2] \\
       &\hphantom{(\vsi_1\cdot\vsi_2)\cdot\cdot}
        +\frac{1}{2m_N(m_{\Lambda}+m_N)}
      \Big(g_{\eta_1 NN}V_S(\eta_1)\tilde f+g_{\eta_8 NN}
      V_S(\eta_8)\tilde g\Big)\\
      &\hphantom{xxV_S(r)\cdot (\vsi_1\cdot\vsi_2)\cdot}\cdot
      \frac12[(\overline n n)_2(\overline n\Lambda)_1
      +(\overline n n)_1(\overline n\Lambda)_2]\Big\}
\\
           &+\hat S_{12}
      \cdot\Big\{
      \frac{g_{\pi NN}}{2m_N(m_{\Lambda}+m_N)}
      \big(V_T(\pi)\left(\tilde a-\sqrt{\frac23}
      \tilde b\right)\frac12[(\overline n n)_2(\overline n\Lambda)_1
      +(\overline n n)_1(\overline n\Lambda)_2]
      \big)\\
       &\hphantom{(\vsi_1\cdot\vsi_2)\cdot\cdot}
        +\frac{g_{KN\Lambda}}{2m_N(m_{\Lambda}+m_N)}
      V_T(K^0)(\tilde c+\tilde d-\tilde e)
      \frac12[(\overline n n)_2(\overline n\Lambda)_1
      +(\overline n n)_1(\overline n\Lambda)_2] \\
       &\hphantom{(\vsi_1\cdot\vsi_2)\cdot\cdot}
        +\frac{1}{2m_N(m_{\Lambda}+m_N)}
      \Big(g_{\eta_1 NN}V_T(\eta_1)\tilde f+g_{\eta_8 NN}
      V_T(\eta_8)\tilde g\Big)\\
      &\hphantom{xxV_S(r)\cdot (\vsi_1\cdot\vsi_2)\cdot}\cdot
      \frac12[(\overline n n)_2(\overline n\Lambda)_1
      +(\overline n n)_1(\overline n\Lambda)_2]
      \Big\}\\
      &+\Big\{
      \frac{g_{\pi NN}}{2m_N}
      \big(V_{PV}(\pi)\left( a-\sqrt{\frac23}
       b\right)\frac12[-(\vsi_2\hr)(\overline n n)_2(\overline n\Lambda)_1
      +(\vsi_1\hr)(\overline n n)_1(\overline n\Lambda)_2]
      \big)\\
       &\hphantom{(\vsi_1\cdot\vsi_2)\cdot\cdot}
        +\frac{g_{KN\Lambda}}{2m_N}
      V_{PV}(K^0)(\tilde c+\tilde d-\tilde e)
      \frac12[(\vsi_1\hr)(\overline n n)_2(\overline n\Lambda)_1
      -(\vsi_2\hr)(\overline n n)_1(\overline n\Lambda)_2] \\
       &\hphantom{(\vsi_1\cdot\vsi_2)\cdot\cdot}
        +\frac{1}{2m_N}
      \Big(g_{\eta_1 NN}V_{PV}(\eta_1)\tilde f+g_{\eta_8 NN}
      V_{PV}(\eta_8)\tilde g\Big)\\
      &\hphantom{xxV_S(r)\cdot (\vsi_1\cdot\vsi_2)\cdot}\cdot
      \frac12[-(\vsi_2\hr)(\overline n n)_2(\overline n\Lambda)_1
      +(\vsi_1\hr)(\overline n n)_1(\overline n\Lambda)_2]
      \Big\}
\endaligned
\tag\efp-7
$$
}}

{\uskoosam{
%
%
\parskip 0pt
\baselineskip 10pt
\medskip
$$
\table{}
 {}             !       {}              !          \r
Weak vertices     ! Analytic expression  ! Numerical value  \r
 {}             !       {}              !          \jrr
\hfill$a$\hfill !$\frac{\sqrt{2}}3 A(\Lambda^0_-)-\frac13 A(\Lambda ^0_0)$! 2.3187
\rr
\hfill$b$\hfill !$\sqrt{\frac32}\left[
	\frac{\sqrt{2}}3 A(\Lambda^0_-)+\frac23 A(\Lambda ^0_0)\right]$!
   -0.0505        \rr
\hfill$\tilde a$\hfill !$\frac{\sqrt{2}}3 B(\Lambda^0_-)-\frac13 B(\Lambda ^0_0)$!15.862
\rr
\hfill$\tilde b$\hfill !$\sqrt{\frac32}\left[
	\frac{\sqrt{2}}3 B(\Lambda^0_-)+\frac23 B(\Lambda ^0_0)\right]$!
  0.032       \rr
\hfill$c$\hfill !$\frac13\left[ A_K(n^0_0)-A_K(p^+_0)+2A_K(p^+_+)\right]$! 1.5967
\rr
\hfill$d$\hfill !$\frac13\left[ A_K(n^0_0)+2A_K(p^+_0)-A_K(p^+_+)\right]$! 4.3667
\rr
\hfill$e$\hfill !$\frac13\left[ -A_K(n^0_0)+A_K(p^+_0)+A_K(p^+_+)\right]$! -0.2867
\rr
\hfill$\tilde c$\hfill !$\frac13\left[ B_K(n^0_0)-B_K(p^+_0)+2B_K(p^+_+)\right]$!
20.8733
\rr
\hfill$\tilde d$\hfill !$\frac13\left[ B_K(n^0_0)+2B_K(p^+_0)-B_K(p^+_+)\right]$!
-12.3767
\rr
\hfill$\tilde e$\hfill !$\frac13\left[ -B_K(n^0_0)+B_K(p^+_0)+B_K(p^+_+)\right]$!
-9.4933
\rr
\hfill$ f$\hfill !$ A_{\eta_1}(\Lambda^0_{\eta_1})$!5.60
\rr
\hfill$ \tilde f$\hfill !$ B_{\eta_1}(\Lambda^0_{\eta_1})$!27.53
\rr
\hfill$ g$\hfill !$ A_{\eta_8}(\Lambda^0_{\eta_8})$!5.19
\rr
\hfill$ \tilde g$\hfill !$ B_{\eta_8}(\Lambda^0_{\eta_8})$!22.97
\caption{\uskodevet\it Table \efp.1 - Weak vertices and their connection
with the weak nonleptonic amplitudes}
$$
\medskip
}}
%
%

To introduce the isospin formalism, as done already before in Sec.\efe,
one recalls the three different combinations which occur in the
effective weak-strong interference Hamiltonian, i.e.
$$
\alignedat 5
&\left(\overline N\,{\bold 1}\,\Lambda\pmatrix 0 \\ 1 \endpmatrix\right)_1
\left(\overline N\,{\bold 1}\, N\right)_2 &\quad &=\beta_1 &\qquad
&\Delta I=1/2\\
&\left(\overline N\,{\vec\tau}\,\Lambda\pmatrix 0 \\ 1 \endpmatrix\right)_1
\left(\overline N\,{\vec\tau}\, N\right)_2 &\quad &=\beta_{\tau} &\qquad
&\Delta I=1/2\\
&\left(\overline N\,[{\vec T\,\chi]}\,\Lambda\right)_1
\left(\overline N\,{\vec\tau}\, N\right)_2 &\quad &=\beta_T &\qquad
&\Delta I=3/2
\endalignedat
\tag\efp-8
$$
Their particle content is
$$
\aligned
\beta_1&=(\overline n\Lambda)_1[(\overline p p)_2+(\overline n n)_2]\\
\beta_{\tau}&=(\overline n\Lambda)_1[(\overline n n)_2-(\overline p p)_2]
+2(\overline p\Lambda)_1(\overline n p)_2\\
\beta_T&=\sqrt{\frac23}\left\{(\overline p\Lambda)_1(\overline n p)_2+
(\overline n\Lambda)_1[(\overline p p)_2-(\overline n n)_2]\right\}.
\endaligned
\tag\efp-9
$$

The  {\it isospin explicit version\/} of the
effective potential could now be written in such a way (for the PV
exchange of the pions, for instance) by connecting both versions
$$
\aligned
V_{\pi}&=a\beta_{\tau}+b\beta_T\\
&=a(\overline n\Lambda)_1[(\overline p p)_2+(\overline n n)_2]\\
&+b\sqrt{\frac23}\left\{(\overline p\Lambda)_1(\overline n p)_2+
(\overline n\Lambda)_1[(\overline p p)_2-(\overline n n)_2]\right\}\\
&=(\overline n\Lambda)_1(\overline n
n)_2\left(a-\sqrt{\frac23}b\right)\\
&+  (\overline n\Lambda)_1(\overline p
  p)_2\left(-a+\sqrt{\frac23}b\right)\\
&+(\overline p\Lambda)_1(\overline n p)_2\left(2a+\sqrt{\frac23}b\right)
\endaligned
\tag\efp-10
$$

Finally the {\it isospin explicit version\/}  is given by
\vfill
\eject

{\uskoosam{
$$
\aligned
V(\vr)&_{}=V_C(r)\cdot 0\\
  &+(\vsi_1\cdot\vsi_2)
      \cdot\Big\{
      V_S(\pi)\cdot
      \frac{g_{NN\pi}}{2m_N(m_{\Lambda}+m_N)}\cdot
      \tilde a \cdot
      \frac12(\vec\tau_{1\,\Lambda}\cdot\vec\tau_2+
        \vec\tau_1\cdot\vec\tau_{2\,\Lambda})\\
     &\hphantom{+xx(\vsi_1\cdot\vsi_2)\cdot}
      +V_{S}(\pi)\cdot\frac{g_{NN\pi}}{2m_N(m_{\Lambda}+m_N)}\cdot
      \tilde b \cdot
      \frac12(\vec T_{1\,\Lambda}\cdot\vec\tau_2+
        \vec\tau_1\cdot\vec T_{2\,\Lambda})\\
      &\hphantom{+xx(\vsi_1\cdot\vsi_2)\cdot}
        +\frac{g_{K\Lambda N}}{2m_N(m_{\Lambda}+m_N)}
      \big(V_S(K)\left(\frac12\tilde c+\tilde d\right)\frac12
      [({\bold 1}_{1\,\Lambda}\cdot {\bold 1}_2)+
       ({\bold 1}_1\cdot{\bold 1}_{2\,\Lambda})] \\
      &\hphantom{+xx(\vsi_1\cdot\vsi_2)\cdot}
      +V_S(K)\frac12\cdot\tilde c\cdot
      \frac12(\vec\tau_{1\,\Lambda}\cdot\vec\tau_2+
        \vec\tau_1\cdot\vec\tau_{2\,\Lambda})\\
      &\hphantom{+xx(\vsi_1\cdot\vsi_2)\cdot}
      +V_S(K)\cdot\tilde e\cdot
      \frac12(\vec T_{1\,\Lambda}\cdot\vec\tau_2+
        \vec\tau_1\cdot\vec T_{2\,\Lambda})\Big)\\
      &\hphantom{+xx(\vsi_1\cdot\vsi_2)\cdot}
        +\frac{1}{2m_N(m_{\Lambda}+m_N)}
      \Big(V_S(\eta_1)\cdot
      g_{\eta_1 NN}\cdot\tilde f
      +V_S(\eta_8)\cdot g_{\eta_8 NN}
                \cdot\tilde g\Big)\\
      &\hphantom{+xx(\vsi_1\cdot\vsi_2)\cdot========}
        \frac12
      [({\bold 1}_{1\,\Lambda}\cdot {\bold 1}_2)+
       ({\bold 1}_1\cdot{\bold 1}_{2\,\Lambda})]
      \Big\}\\
   &+\hat S_{12}\ \cdot\Big\{
      V_T(\pi)\cdot
      \frac{g_{NN\pi}}{2m_N(m_{\Lambda}+m_N)}\cdot
      \tilde a \cdot
      \frac12(\vec\tau_{1\,\Lambda}\cdot\vec\tau_2+
        \vec\tau_1\cdot\vec\tau_{2\,\Lambda})\\
     &\hphantom{+xx(\vsi_1\cdot\vsi_2)\cdot}
      +V_{T}(\pi)\cdot\frac{g_{NN\pi}}{2m_N(m_{\Lambda}+m_N)}\cdot
      \tilde b \cdot
      \frac12(\vec T_{1\,\Lambda}\cdot\vec\tau_2+
        \vec\tau_1\cdot\vec T_{2\,\Lambda})\\
      &\hphantom{+xx(\vsi_1\cdot\vsi_2)\cdot}
        +\frac{g_{K\Lambda N}}{2m_N(m_{\Lambda}+m_N)}
      \big(V_T(K)\left(\frac12\tilde c+\tilde d\right)\frac12
      [({\bold 1}_{1\,\Lambda}\cdot {\bold 1}_2)+
       ({\bold 1}_1\cdot{\bold 1}_{2\,\Lambda})] \\
      &\hphantom{+xx(\vsi_1\cdot\vsi_2)\cdot}
      +V_T(K)\frac12\cdot\tilde c\cdot
      \frac12(\vec\tau_{1\,\Lambda}\cdot\vec\tau_2+
        \vec\tau_1\cdot\vec\tau_{2\,\Lambda})\\
      &\hphantom{+xx(\vsi_1\cdot\vsi_2)\cdot}
      +V_T(K)\cdot\tilde e\cdot
      \frac12(\vec T_{1\,\Lambda}\cdot\vec\tau_2+
        \vec\tau_1\cdot\vec T_{2\,\Lambda})\Big)\\
      &\hphantom{+xx(\vsi_1\cdot\vsi_2)\cdot}
        +\frac{1}{2m_N(m_{\Lambda}+m_N)}
      \Big(V_T(\eta_1)\cdot
      g_{\eta_1 NN}\cdot\tilde f\\
      &\hphantom{+xx(\vsi_1\cdot\vsi_2)+\cdot}
      +V_T(\eta_8)\cdot g_{\eta_8 NN}
                \cdot\tilde g\Big)
        \frac12
      [({\bold 1}_{1\,\Lambda}\cdot {\bold 1}_2)+
       ({\bold 1}_1\cdot{\bold 1}_{2\,\Lambda})]
      \Big\}\\
      &\hphantom{++}+
      \Big\{
      V_{PV}(\pi)\cdot
      \frac{g_{NN\pi}}{2m_N}\cdot
      \tilde a \cdot
      \frac12(-(\vsi_2\hr)\vec\tau_{1\,\Lambda}\cdot\vec\tau_2+
        (\vsi_1\hr)\vec\tau_1\cdot\vec\tau_{2\,\Lambda})\\
     &\hphantom{+xx(\vsi_1\cdot\vsi_2)\cdot}
      +V_{PV}(\pi)\cdot\frac{g_{NN\pi}}{2m_N}\cdot
      \tilde b \cdot
      \frac12(-(\vsi_2\hr)\vec T_{1\,\Lambda}\cdot\vec\tau_2+
        (\vsi_1\hr)\vec\tau_1\cdot\vec T_{2\,\Lambda})\\
      &\hphantom{+xx(\vsi_1\cdot\vsi_2)\cdot}
        +\frac{g_{K\Lambda N}}{2m_N}
      \big(V_S(K)\left(\frac12\tilde c+\tilde d\right)\frac12
      [(\vsi_1\hr)({\bold 1}_{1\,\Lambda}\cdot {\bold 1}_2)
       -(\vsi_2\hr)({\bold 1}_1\cdot{\bold 1}_{2\,\Lambda})] \\
      &\hphantom{+xx(\vsi_1\cdot\vsi_2)\cdot}
      +V_S(K)\frac12\cdot\tilde c\cdot
      \frac12((\vsi_1\hr)\vec\tau_{1\,\Lambda}\cdot\vec\tau_2-
        (\vsi_2\hr)\vec\tau_1\cdot\vec\tau_{2\,\Lambda})\\
      &\hphantom{+xx(\vsi_1\cdot\vsi_2)\cdot}
      +V_S(K)\cdot\tilde e\cdot
      \frac12(-(\vsi_2\hr)\vec T_{1\,\Lambda}\cdot\vec\tau_2+
        (\vsi_1\hr)\vec\tau_1\cdot\vec T_{2\,\Lambda})\\
      &\hphantom{+xx(\vsi_1\cdot\vsi_2)\cdot}
        +\frac{1}{2m_N}
      \Big(V_S(\eta_1)\cdot
      g_{\eta_1 NN}\cdot\tilde f\\
        &\hphantom{+xx(\vsi_1\cdot\vsi_2)\cdot}
      +V_S(\eta_8)\cdot g_{\eta_8 NN}
                \cdot\tilde g\Big)
        \frac12
      [-(\vsi_2\hr)({\bold 1}_{1\,\Lambda}\cdot {\bold 1}_2)+
      (\vsi_1\hr)({\bold 1}_1\cdot{\bold 1}_{2\,\Lambda})]
        \Big\}
\endaligned
\tag\efp-11
$$
}}

\vfill
\mpik
\eject
%
\def\apa{A}
%
%
\baselineskip=20pt
\parskip 10pt

\def\doda{\text{A}}

\centerline{\bf Appendix\ \doda: Meson Poles  and Double Counting}

\def\naspog{Appendix \doda: Meson Poles  and Separable Contributions}
\vskip 1cm

Baryonic matrix element of a general axial vector current has the form
(\meso-2), i.e.
$$
\brik{B_{\beta}}{A_a^{\mu}(0)}{B_{\alpha}}=
\overline{u}_{\beta}(p')\left[\gamma^{\mu}\gamma_5 g_A(q^2)+
\frac{1}{M_{\alpha}+M_{\beta}}g_P(q^2)q^{\mu}\gamma_5\right]
\Lambda_a u_{\alpha}(p).
\tag\doda-1
$$
Here $\Lambda_a$ is some matrix (for example $\lambda_a/2$) which describes
the internal (for example SU(3) flavour) hadron suymmetry. Its precise
form is not needed here.

Meson matrix element is
$$
\brik{0}{A_a^{\mu}(0)}{M_b}=i\delta_{ab}f_M q^{\mu}.
\tag\doda-2
$$
If the meson is on the mass shell, then
$$
q^2=m_M^2.
\tag\doda-3a
$$
The PCAC relation is
$$
\partial_{\mu} A^{\mu}_a =C\Phi_a.
\tag\doda-4
$$
It leads to
$$
\aligned
\brik{0}{\partial _{\mu}A^{\mu}_a(x)}{M_a}&=\partial_{\mu}
\left[e^{-iqx}\brik{0}{A^{\mu}_a(0)}{M_a}\right]=
(-i) e^{-iqx} q_{\mu}\brik{0}{A^{\mu}_a(0)}{M}\\
&=q^2 f_M(q^2) e^{-iqx}=C(q^2) e^{-iqx}.
\endaligned
\tag\doda-5
$$
If the meson is on the mass shell, one should select
$$
C(m_M^2)=m_M^2 f_M.
\tag\doda-3b
$$
Also
$$
\aligned
\partial_{\mu}\brik{B_{\beta}}{A^{\mu}_a(x)}{B_{\alpha}}&=
\partial_{\mu}
\left[e^{i(p'-p)x}\brik{B_{\beta}}
{A^{\mu}_a(0)}{B_{\alpha}}\right]\\
&=i e^{i(p'-p)x}\overline u_{\beta}(p')\left[
g_A (\sc{p}-\sc{p}')-\frac{q^2}{M_{\alpha}+M_{\beta}}
g_P\right]\gamma_5 \Lambda_a u_{\alpha}(p)\\
&=(-i) e^{i(p'-p)x}\overline
u_{\beta}(p')\left[(M_{\alpha}+M_{\beta})g_A+
\frac{q^2}{M_{\alpha}+M_{\beta}}g_P\right]\gamma_5 \Lambda_a u_{\alpha}(p)\\
&=C\brik{B_{\beta}}{\Phi_a(x)}{B_{\alpha}}\\
&=C e^{i(p'-p)x}\brik{B_{\beta}}{\Phi_a(0)}{B_{\alpha}}\\
&=C e^{-iqx}\brik{B_{\beta}}{\Phi_a(0)}{B_{\alpha}}.
\endaligned
\tag\doda-6
$$
The matrix element $\brik{B_{\beta}}{\Phi_a}{B_{\alpha}}$ can be
calculated by using the equation of motion [33]
$$
(\dalamb +m_M^2)\Phi_a(x)=-ig_{\alpha\beta}\overline{\psi}_{\beta}(x)
\gamma_5 \Lambda_a\psi_{\beta}(x),
\tag\doda-7
$$
which leads to
$$
(-q^2+m_M^2)\brik{B_{\beta}}{\Phi}{B_{\alpha}}=-ig_{\alpha \beta}\overline
u_{\beta}\gamma_5 \Lambda_a u_{\alpha}.
\tag\doda-8
$$
Here $g_A,\ g_P,\ g_{\alpha \beta}$ and $C$ are in general some
functions of $q^2$.

By inserting (\doda-8) in (\doda-6) and dropping unimportant factors,
one finds a relation
$$
(M_{\alpha}+M_{\beta})g_A(q^2)+\frac{q^2}{M_{\alpha}+M_{\beta}}
g_P(q^2)= m_M^2 f_M \frac{g_{\alpha \beta}(q^2)}{m_M^2-q^2}.
\tag\doda-9
$$
In the limit $q^2\to 0$, assuming that
$$
\aligned
g_{\alpha\beta}(0)&=g_{\alpha\beta}(m_M^2)=g_{\alpha\beta}\\
g_A(0)&=g_A=1.25
\endaligned
\tag\doda-10
$$
one finds the famous Goldberger-Treiman (GT) relation (generalized):
$$
\frac{g_A}{f_M}=\frac{g_{\alpha\beta}}{M_{\alpha}+M_{\beta}}.
\tag\doda-11
$$
The induced pseudoscalar formfactor $g_P$ is dominated by the
pseudoscalar meson (for example pion) pole. This is described by the
diagram shown in Fig.\doda.1

\midinsert
\vbox{\medskip
\centerline{\epsfysize=6cm\epsfbox{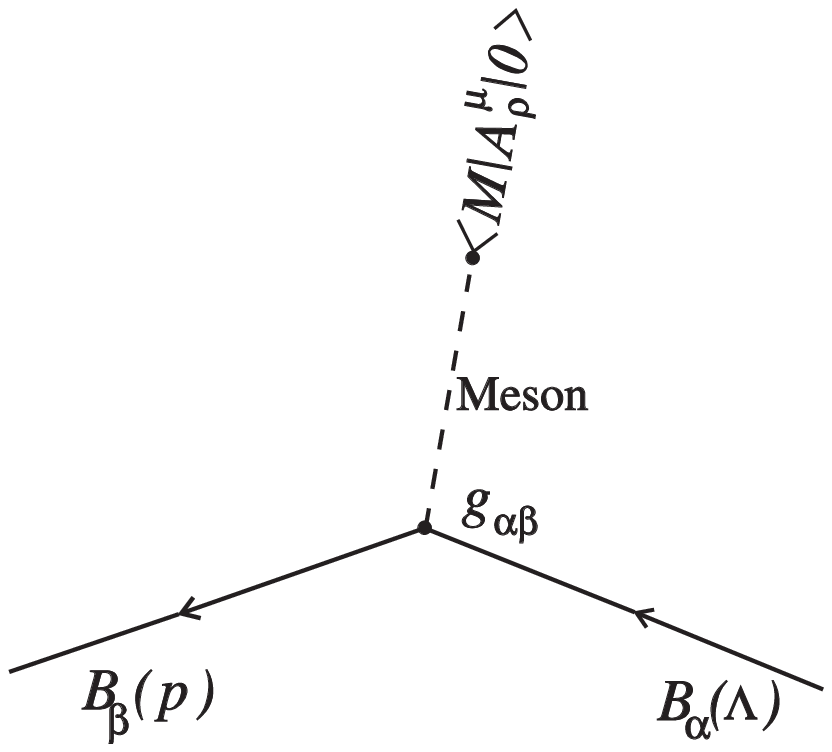}}
\medskip
\subnormal{
\centerline{{\it Fig.\doda.1 - Here $g_{\alpha\beta}$ appears in the
strong vertex. The axial}}
\centerline{{\it vector current acts on the intermediate meson $M$,
as indicated.}}
\medskip}}
\endinsert

The diagram in Fig.\doda.1 corresponds to
$$
(\text{Fig.\doda.1})^{\mu}=\overline u_{\beta}(p')\gamma_5  \Lambda_a
u_{\alpha}(p) g_{\alpha\beta}(q^2)\frac{i}{q^2-m_M^2+i\epsilon}
\brik{M_b}{A_b^{\mu}}{0}.
\tag\doda-12
$$
Inserting (\doda-2) and compairing the result with (\doda-1) one finds
$$
\frac{g_P(q^2)}{M_{\alpha}+M_{\beta}}\simeq (-)\frac{g_{\alpha\beta}(q^2)
f_M(q^2)}{q^2-m_M^2}.
\tag\doda-13a
$$
Inclusion of vertex and other corrections, such as multipion exchanges
[6], turns that into
$$
\frac{g_P(q^2)}{M_{\alpha}+M_{\beta}}=H(q^2)+\frac{g_{\alpha\beta}(q^2)
f_M(q^2)}{m_M^2-q^2}F(q^2).
\tag\doda-13b
$$
If one neglects $H$ and assumes
$$
\aligned
g_{\alpha\beta}(q^2)&\simeq g_{\alpha\beta}(m_M^2)\\
f_M(q^2)&\simeq f_M(m_M^2)=f_M\\
F(q^2)&\simeq 1,
\endaligned
\tag\doda-13c
$$
one finds the well-known estimate
$$
\frac{g_P(q^2)}{M_{\alpha}+M_{\beta}}\simeq \frac{g_{\alpha\beta}
f_M}{m_M^2-q^2}.
\tag\doda-14
$$
For the nucleon\,($N$)-pion\,($\pi$) system one finds
$$
g_P\sim\frac{4M_N^2 g_A}{m_{\pi}^2-q^2}.
\tag\doda-15
$$
In the case of the muon capture the effective pseudoscalar constant
becomes
$$
\aligned
\frac{g_P}{2M_N}\cdot m_{\mu}&=\frac{2M_N
m_{\mu}g_A}{m_{\pi}^2-q^2(m_{\mu})}\\
&\simeq
\frac{2M_Nm_{\mu}g_A}{m_{\pi}^2+0.88\cdot m_{\mu}^2}.
\endaligned
\tag\doda-16
$$
Here the mion mass  $m_{\mu}$ comes from the lepton current
$$
q_{\rho}\overline\psi_{\nu}\gamma^{\rho}\gamma_5\psi_{\mu}=
(-)m_{\mu}\overline{\psi}_{\nu}\gamma_5\psi_{\mu}.
\tag\doda-17
$$
The expression (\doda-16) corresponds to the formula (4.37) in [13].
(Their $g_P$ is not equal to our $g_P$ (\doda-15).)

The GT relation (\doda-11) and most of the following formulae were
derived by using (\doda-3) convention. Some particular questions that
might arise when the meson $\Phi_a$ is not no the mass shell will be
mentioned in the main text.

\vfill
\mpikk
\eject
%
%
  \def\bili{\text{B}}
%
%
\parskip 5pt
\baselineskip 15pt
\parindent 20pt

\centerline{\bf Appendix\ \bili: Commutators Involving Bilinear Quark Field
Combinations}
\def\naspog{Appendix \bili: Commutators Involving Bilinear Quark Field
Combinations}

\vskip 1cm

The canonical anti-commutation relations for Dirac (i.e. quark) fields
are [34]
$$
\{\psi_{\alpha}(t,\x),\psi^{\dagger}_{\beta}(t,\y)\}=\delta_{\alpha\beta}
\delta(\x-\y).
\tag\bili-1
$$

When applying current algebra (CA) on PCAC one encounters the
commutators of the bilinear combinations of quark fields such as
$$
[\psi_{\phi}^{\dagger}(z)\Delta_{\phi\epsilon}\psi_{\epsilon}(z),
\psi_{\alpha}^{\dagger}(x)\Gamma_{\alpha\beta}\psi_{\beta}(x)]
\big\vert_{z_0=x_0} =C.
\tag\bili-2
$$
Here $\Delta$ and $\Gamma$ are some products of Dirac $\gamma-$matrices
and the inner group operators corresponding to SU(3) flavor and color.
By repeatedly using (\bili-1) one can write
$$
\aligned
\psi^{\dagger}_{\phi} \psi_{\epsilon} \psi^{\dagger}_{\alpha} \psi_{\beta}
&=\psi^{\dagger}_{\alpha} \psi_{\beta} \psi^{\dagger}_{\phi} \psi_{\epsilon}+
\delta_{\epsilon\alpha}
\psi^{\dagger}_{\phi} \psi_{\beta}\delta(\x-\z)\\
&-\delta_{\phi\beta}
\psi^{\dagger}_{\alpha} \psi_{\epsilon}\delta(\x-\z).
\endaligned
\tag\bili-3
$$
This leads to
$$
\aligned
C&=[\psi_{\phi}^+\Delta_{\phi\epsilon}\Gamma_{\epsilon\beta}
\psi_{\epsilon}-
\psi_{\alpha}^+\Gamma_{\alpha\beta}\Delta_{\beta\epsilon}
\psi_{\beta}]\delta(\x-\z)\\
&=\psi^+(x)[\Delta,\Gamma]\psi(x)\delta(\x-\z).
\endaligned
\tag\bili-4
$$

Let us take $\Delta$ and $\Gamma$ of the form
$$
\aligned
\Delta&=D\,t_i\\
\Gamma&=G\,t_j;\qquad t_i=\lambda_i/2
\endaligned
\tag\bili-5a
$$
Here $\lambda_i$ are SU(3)-flavour matrices while $D$ and $G$ are some
combinations of Dirac matrices. One can write
$$
[t_i\,D,t_j\,G]=\frac12[t_i,t_j]\{D,G\}+\frac12\{t_i,t_j\}[D,G].
\tag\bili-5b
$$

It is useful to specify that for some cases:
\item{(1)}
{\narrower\narrower
\noindent
$$
\aligned
D=\gamma_5,\qquad G&=\gamma_0\Gamma^{\mu}_{L,R}\\
\Gamma^{\mu}_{L,R}&=\gamma_{\mu}(1\mp \gamma_5).
\endaligned
\tag\bili-5c
$$
This gives
$$
\aligned
[D,G]&=\gamma_5\gamma_0\Gamma^{\mu}-\gamma_0\Gamma^{\mu}\gamma_5=0\\
\{D,G\}&=2\gamma_5\gamma_0\Gamma^{\mu}\\
[t_i,t_j]&=if_{ijk}t_k
\endaligned
\tag\bili-5d
$$
\smallskip}

\item{(2)}
{\narrower\narrower
\noindent
$$
\aligned
D&=1,\qquad G=\gamma_0\Gamma^{\mu}_{L,R}\\
{}\\
[D,G]&=\gamma_0\Gamma^{\mu}-\gamma_0\Gamma^{\mu}=0\\
\{D,G\}&=2\gamma_0\Gamma^{\mu}
\endaligned
\tag\bili-5e
$$
\smallskip}

\item{(3)}
{\narrower\narrower
\noindent
$$
\aligned
D&=\gamma_5,\qquad G_{L,R}=\gamma_0(1\mp\gamma_5)\\
{}\\
[D,G_{L,R}]&=\gamma_5\gamma_0(1\mp\gamma_5)-
        \gamma_0(1\mp\gamma_5)\\
        &=2\gamma_5\gamma^0(1\mp\gamma_5)\\
\{D,G\}&=\gamma_5\gamma_0(1\mp\gamma_5)+\gamma_0(1\mp\gamma_5)\gamma_5=0.
\endaligned
\tag\bili-6
$$
\smallskip}

The relations (\bili-4) and (\bili-5) can be used to derive CA [44,
45]. With
$$
\aligned
j^{\mu}_i=\overline\psi\gamma^{\mu}t_i\psi\\
j^{\mu\,5}_i=\overline\psi\gamma^{\mu}\gamma_5t_i\psi,
\endaligned
\tag\bili-7a
$$
one finds
$$
\aligned
[j^0_i(x),j^{\mu}_j(y)]_{x_0=y_0}&=if_{ijk}j^{\mu}_k(x)\delta(\x-\y)\\
[j^0_i(x),j^{\mu\, 5}_j(y)]_{x_0=y_0}&=if_{ijk}j^{\mu\, 5}_k(x)
        \delta(\x-\y)\\
[j^{0\,5}_i(x),j^{\mu}_j(y)]_{x_0=y_0}&=if_{ijk}j^{\mu\, 5}_k(x)
        \delta(\x-\y)\\
[j^{0\,5}_i(x),j^{\mu\, 5}_j(y)]_{x_0=y_0}&=if_{ijk}j^{\mu}_k(x)
        \delta(\x-\y)
\endaligned
\tag\bili-7b
$$
This can be easily transformed into the commutation rules for the SU(3)
generators
$$
F_i=\int d^x\,j^0_i(x)
\tag\bili-8a
$$
and the axial vector charges
$$
F^5_i=\int d^x\,j^{0\,5}_i(x).
\tag\bili-8b
$$
Integration over $x$ simply removes delta-functions in (\bili-7).

General structure of the operators $\Cal O_1$, $\Cal O_2$, $\Cal O_3$
and  $\Cal O_4$ contains terms as
$$
\aligned
j_{\mu\, k}(x) j^{\mu}_{\ell}(x)+j_{\mu\,5\,k}(x) j^{\mu\,5}_{\ell}(x)
&=b(\text{PC})\\
j_{\mu\,5\,k}(x) j^{\mu\, \ell}(x)+j_{\mu\,k}(x) j^{\mu\,5}_{\ell}(x)
&=a(\text{PV}).
\endaligned
\tag\bili-9
$$
Quark fields $\psi$ in those terms are not normally oredered.
(Some details about that statement can be found in
Appendix \dode.) Thus one can simply
apply (\bili-7) and (\bili-8) in order to shaw
$$
[F^5_i,\Cal O_{A}(\text{PV})]=-[F_i,\Cal O_A(\text{PC})]\qquad
(A=1,2,3,4).
\tag\bili-10a
$$
Operators $\Cal O_5$ and $\Cal O_6$ as shown in Appendix \dode\
contain VEV's when written without normal ordering. Obviously
$$
[F^5_i,\tilde{\Cal O}_{B}(\text{PV})]=-[F_i,\tilde{\Cal O}_B(\text{PC})]
\qquad (B=1,2,3,4).\qquad
(\text{$\tilde{\Cal O}$ is not NOP})
\tag\bili-10b
$$
In addition one has to calculate the commutator with the term $\overline
d(1\pm\gamma_5)s$. One finds (with $\psi=(u,d,s)$):
$$
\aligned
[F_3^5,\overline d(1-\gamma_5) s]&=-\frac12 \overline d(1-\gamma_5)s\\
[F_3^5,\overline d(1+\gamma_5) s]&=\frac12 \overline d(1+\gamma_5)s\\
[F_3,\overline d(1-\gamma_5) s]&=-\frac12 \overline d(1-\gamma_5)s\\
[F_3,\overline d(1+\gamma_5) s]&=-\frac12 \overline d(1+\gamma_5)s
\endaligned
\tag\bili-11
$$
Thus one can write\footnote{$^{\dag}$}{\uskoosam
See  Appendix \dode\,  for the deffinition of $\hat{\Cal O}$.
}
$$
\aligned
[F^5_3,\hat{\Cal O}_6]&=[F^5_3,\tilde{\Cal O}_6]-
 \frac26\brik{0}{\overline d d}{0} \overline d(1-\gamma_5) s
+\frac26\brik{0}{\overline s s}{0} \overline d(1+\gamma_5) s\\
[F_3,\hat{\Cal O}_6]&=[F_3,\tilde{\Cal O}_6]-
 \frac26\brik{0}{\overline d d}{0} \overline d(1-\gamma_5) s
-\frac26\brik{0}{\overline s s}{0}\overline d(1+\gamma_5) s.
\endaligned
\tag\bili-12a
$$
With
$$
[F^5_3,\tilde{\Cal O}_6]=-[F_3,\Tilde{\Cal O}_6],\quad
L_d=\brik{0}{\overline d d}{0}\overline d(1-\gamma_5) s,\quad
L_s=\brik{0}{\overline s s}{0}\overline d(1+\gamma_5) s,
\tag\bili-12b
$$
one can write
$$
\aligned
[F^5_3,\hat{\Cal O}_6]&=[F_3^5,\tilde{\Cal O}_6]-\frac26 L_d
+\frac26 L_s\\
        &=-[F_3,\hat{\Cal O}_6]-\frac26 L_d-\frac26 L_s-\frac26 L_d+
                \frac26 L_s\\
        &=-[F_3,\hat{\Cal O}_6]-\frac23 \brik{0}{\overline d d}{0}
	\overline d(1-\gamma_5)s.
\endaligned
\tag\bili-12c
$$
This is the formula (17) in ref. [21] (up to the different sign
convention). The operator $\hat{\Cal O}_5$
satisfies the same relation with the replacement
$$
L_d\to \frac{16}{3} L_d.
\tag\bili-12d
$$
Everithing can be generalized for kaon emmision by selecting appropriate
indices in (\bili-11).

\vfill
\mpikk
\eject
%
%
%
%
%

\def\cpc{\text{C}}

\centerline{\bf Appendix\ \cpc: Transition Isospin $\vec T$}
\def\naspog{Appendix \cpc: Transition Isospin $\vec T$}
\vskip 1cm

This formalism, suitable for isospin $I=3/2$ (or spin $S=3/2$) particles
is described in Ref.s [42] and [43].

An isospin $I=3/2$ object can be constructed by combining the
isovectors\footnote{$^{\dag}$}{\uskoosam For spin Ref. [42] uses
$\epsilon^1=-(1,i,0)/\sqrt{2}$. The $+$ sign here leads to the usual
pion field definition $\pi^{\pm}=(\pi_1\mp i\pi_2)/\sqrt{2}$.
}
$$
{\vec t \,}^1=\frac1{\sqrt{2}}(1,i,0),\qquad
{\vec t\,}^{-1}=\frac1{\sqrt{2}}(1,-i,0),\qquad
{\vec t\,}^3=(0,0,1),
\tag\cpc-1
$$
with the isospin function $\chi_{1/2}^m$. One finds
$$
Z_i^M=\sum_{r,m}\kleb{1}{r}{1/2}{m}{3/2}{M} t_i^r \chi_{1/2}^m.\qquad
(i=1,2,3)
\tag\cpc-2
$$
A coupling with isospin $I=1$ field (for example pion $\pi$) is now
$$
{N^{m_s}}^{\dagger}\pi_i Z_i^M={\chi_{1/2}^{m_s}}^{\dagger}\sum_{r,m}
\kleb{1}{r}{1/2}{m}{3/2}{M} \chi_{1/2}^m \vec t^r\vec\pi.
\tag\cpc-3
$$
Here $N^{m_s}$ is either proton $(m_s=1/2)$ or neutron $(m_s=-1/2)$, and
$\vec\pi=(\pi_1,\pi_2,\pi_3)$. One can introduce {\it transition
isospin\/} $\vec T$ which is defined by its matrix
element\footnote{$^{\ddag}$}{\uskoosam Here again we have small
difference with the Ref. [42] convention: our $\vec T=\vec T^{*}$Ref.
[42]. In
our case $I=3/2$ object will always be on the R.H.S. in bilinear
combination (\cpc-3).}:
$$
\big(\vec T\big)_{m_s M}=\kleb{1}{r}{1/2}{m_s}{3/2}{M}\cdot \vec t^r.
\tag\cpc-4
$$
Then (\cpc-3) can be written as
$$
\aligned
{N^{m_s}}^{\dagger}\,(\vec T\cdot \vec\pi)\, Z^M(3/2)&=F(m_sM)=
{\chi_{1/2}^{m_s}}^{\dagger}\sum_{r,m}\kleb{1}{r}{1/2}{m}{3/2}{M}
\chi^{m}_{1/2}\vec t\,{}^r\vec\pi\\
&=\sum_r \kleb{1}{r}{1/2}{m_s}{3/2}{M}\vec t{}\,^r\cdot \vec\pi.
\endaligned
\tag\cpc-5a
$$
(Here ${\chi^{m_s}}^{\dagger}\chi^m=\delta_{m_s\,m}$.) As an example let
us take an object with $M=-1/2$. Then for $m_s=-1/2$ one gets
$$
\gathered
\sum_r\kleb{1}{r}{1/2}{-1/2}{3/2}{-1/2}=
\kleb{1}{0}{1/2}{-1/2}{3/2}{-1/2}=\sqrt{\frac23}\\
F(-1/2,-1/2)=\sqrt{\frac23}\vec
t^0\cdot\vec\pi=\sqrt{\frac23}\pi^0
\endgathered
\tag\cpc-5b
$$
and
$$
\overline n\, Z^{-1/2}(3/2)\cdot \pi^0=\Cal F(-1/2,-1/2)
\tag\cpc-6
$$
The expression (\cpc-6) is the isospin conserving coupling between
$I=1/2$ $\overline N(\overline p,\overline n)$ field, $I=3/2$ $\Cal E$
field and $\pi$ (pion) field.

One also finds for $m_2=1/2$:
$$
\gathered
\sum_r\kleb{1}{r}{1/2}{1/2}{3/2}{-1/2}=
\kleb{1}{-1}{1/2}{-/2}{3/2}{-1/2}=\frac{1}{\sqrt{3}}\\
F(1/2,-1/2)=\frac1{\sqrt{3}}\vec
t^{-1}\cdot\vec\pi=\frac{1}{\sqrt{3}}\frac1{\sqrt{2}}(\pi_1-i\pi_2)
=\frac1{\sqrt{3}}\pi^+,
\endgathered
\tag\cpc-7a
$$
and
$$
\aligned
\overline p Z^{-1/2}(3/2)\cdot \pi^+&=\Cal F(1/2,-1/2)\\
\Cal F(m_s,M)&=\chi^{m_s\, \dagger}\,Z^M\cdot F(m_s,M)\\
\alignedat 3
m_s&=1/2&\quad&\text{proton}\\
m_s&=-1/2&\quad&\text{neutron}
\endalignedat
\endaligned
\tag\cpc-7b
$$

Remark: the same formalism will be used for weak vertices $\overline N
\Lambda \pi$ where $\Lambda$ will be given the quasi isposipn $I=1/2$
and $I=3/2$.

\vfill
\mpikk
\eject
%
%
\def\apd{\text{D}}
%
%
\baselineskip=20pt
\parskip 10pt

\centerline{\bf Appendix\ \apd: $\Delta I=3/2$ Isospin Change}
\def\naspog{Appendix \apd: $\Delta I=3/2$ Isospin Change}
\vskip 1cm

The pion exchange contributions  contain very small $\Delta
I=3/2$ pieces. This reflects the fact that the experimental $\Lambda\to
N\pi$ decay amplitudes contain very very small $\Delta I=3/2$
contributions. For example, from $B(\Lambda_-^0)_{\text{exp}}=22.4$ and
$B(\Lambda_0^0)_{\text{exp}}=-15.61$ (in units of $10^{-7}$) one can
deduce
$$
\aligned
-B(\Lambda_0^0)/\sqrt{2}&=-15.84\\
\Big\vert
\frac{B(\Lambda_-^0)/\sqrt{2}-B(\Lambda_0^0)}{
B(\Lambda_0^0)}\Big\vert&=0.02
\endaligned
$$
Somewhat different picture emerges in the case of weak $NNK$ couplings,
whose values are shown in Table \apd.1. Our results were obtained
by including separable contributions (see Table \wbb.4) which
contain $\Delta I=3/2$ piece. Experimental values were used in pole
terms, and they had also $\Delta I=3/2$ pieces. The relative impotrance
of the $\Delta I=3/2$ terms can be directly  seen from the general weak
potential (\efp-11). One can obtain some interesting information from
Table \apd.1 also by checking how well is the $\Delta I=1/2$ sum rule
(\apd-1) satisfied
$$
\aligned
F(p^+_0)+F(p^+_+)&=\Sigma(F)\\
\Sigma(F)&=F(n^0_0)\\
F&=A,B.
\endaligned
\tag\apd-1
$$
A useful measure of the discrepancy is
$$
D(F)=\Big\vert
\frac{\Sigma(F)-F(n^0_0)}{\Sigma(F)}
\Big\vert
$$

\midinsert
%
%
\parskip 0pt
\baselineskip 10pt
\medskip
$$
\table{}
Amplitude ! Tot.amp.${}^{\dagger}$ ! Ref.[22] ! Ref.[5] \rr
$A(p^+_0)$ ! 4.08    ! 4.09 ! 4.64           \r
$B(p^+_0)$ ! -36.66    ! -7.6 ! -14.72           \r
$A(p^+_+)$ ! 1.31    ! 1.09  ! 1.69     \r
$B(p^+_+)$ ! 42.38   ! 33.40 ! 41.30     \r
$A(n^0_0)$ ! 6.25    ! 5.19  ! 6.33         \r
$B(n^0_0)$ ! 17.99    ! 26.16 ! 26.58
\caption{{\it Table \apd.1 - Nonleptonic amplitudes ($\times
10^7$); ${}^{\dagger}$ this work, compared with Ref. [22] and [5].
}}
$$
\medskip
\endinsert
%
%

In our case  one finds (in $10^{-7}$ units)
$$
\aligned
\Sigma(A)&=5.39\\
A(n_0^0)&=6.25\\
&D(A)=0.16\\
\Sigma(B)&=25.73\\
B(n^0_0)&=17.99\\
&D(B)=0.30
\endaligned
\tag\apd-3
$$
The amplitudes of Ref. [5] were obtained with the assumption that
$\Cal H_W$ contains only $\Delta I=1/2$ piece. Thus from Table \apd.1 one
finds
$$
D(A)=0;\qquad D(B)=0.
\tag\apd-4
$$
In the chiral Lagrangian approach of ref. [22] the $\Delta I=3/2$
contributions are very small
$$
D(A)=0.002,\qquad D(B)=0.014.
\tag\apd-5
$$
They have also used the $\Delta I=1/2$ Hamiltonian.

\vfill
\mpikk
\eject
%
%
%
%
%
\def\dode{\text{E}}

\centerline{\bf Appendix\ \dode: Commutators and Normal-ordered Operators}
\def\naspog{Appendix \dode: Commutators And Normal-ordered Operators}
\vskip 1cm

The product of fields appearing in the four-quark operators $\Cal O_1,\
\Cal O_2,\ \dots\ ,\Cal O_6$ comprising the effective weak Hamiltonian
are normal-ordered. Thus one has to be careful when calculating the
commutators which appear in evaluation of the current algebra
contributions (CAC). The current commutators, or more general, the
commutators of bilinear quark-field forms, are best used if the
operators $\Cal O_i$ are written in the form which is no longer normally
ordered. (Some details about commutators can be found in Section \cur.)

The task of unscrambling the normally ordered product (NOP) can be
achieved by using
\item{(a)} Wick's theorem for normal-ordered products (WT) [34],
\item{(b)} Fierz transformation (FIT) which has been described in
Sec.\sep.

The WT for a product of four quark fields $\psi_{\alpha}$ (here $\alpha$
denotes all indices: Dirac's components, SU(3) flavour, SU(3) colour,
etc.) is:
$$
\aligned
\overline\psi_{\alpha}\psi_{\beta}\,\overline\psi_{\gamma}
\psi_{\delta}
&=\,:\,\overline\psi_{\alpha}\psi_{\beta}\, \overline\psi_{\gamma}
\psi_{\delta}\, :+
\kktrk{\psi_{\alpha}}{\psi_{\beta}}\,
:\overline\psi_{\gamma}\psi_{\delta}:\\
&+\,:\, \overline\psi_{\alpha}\psi_{\beta}\,:
\kktrk{\psi_{\gamma}}{\psi_{\delta}}+
\kktrk{\psi_{\alpha}}{\psi_{\delta}}\,
\kkitrk{\psi_{\beta}}{\psi_{\gamma}}+\dots+\\
&+\kktrk{\psi_{\alpha}}{\psi_{\delta}}\,:\psi_{\beta}
        \overline\psi_{\gamma}:+
\kkitrk{\psi_{\beta}}{\psi_{\gamma}}\,
:\overline\psi_{\alpha}\psi_{\delta}:
\endaligned
\tag\dode-1a
$$
Here $:\Cal O:$ symbolizes the normal-ordering and the lower sign\
$\qquad\ktrk{}{}$\ the vacuum expectation value (VEV).
It means the following
$$
\kktrk{\psi_{\alpha}}{\psi_{\beta}}=\brik{0}{\overline\psi_{\alpha}\psi_{\beta}}{0}.
\tag\dode-1b
$$

Only the first row is important for the operators $\Cal O_1$,
$\Cal O_2$,  $\Cal O_3$ and  $\Cal O_4$. For the operators $\Cal O_5$
and $\Cal O_6$ one has to consider the first and the last row only.

This follows from a physical fact that VEV's for non-scalar (be it in
the Lorentz-space, or in the "inner"-space, i.e. SU(3) etc.) quantities
have to vanish.

This means
$$
\gathered
\brik{0}{\overline \psi_{\ell}\Gamma_{L,R}^{\mu}\psi_k}{0}=0\\
\Gamma^{\mu}_{L,R}=\gamma^{\mu}(1\mp \gamma_5)
\endgathered
\tag\dode-2a
$$
and
$$
\brik{0}{\overline \psi_m\Lambda^m \psi_p}{0}=0
\tag\dode-2b
$$
Any quark combination appearing in the operators $\Cal O_1\,-\,\Cal O_4$
of the form
$$
\overline\psi_{\alpha}\left(\Gamma^{\mu}_L\right)_{\alpha\beta}\psi_{\beta}\,
\overline\psi_{\gamma}\left(\Gamma_{\mu\, L}\right)_{\gamma\delta}
\psi_{\delta}
\tag\dode-3a
$$
FIT produces an equality
$$
\left(\overline\psi_{\alpha}\left(\Gamma^{\mu}_L\right)_{\alpha\beta}
\psi_{\beta}\right)
\,
\left(\overline\psi_{\gamma}\left(\Gamma_{\mu\, L}\right)_{\gamma\delta}
\psi_{\delta}\right)
=
\left(\overline\psi_{\alpha}\left(\Gamma^{\mu}_L\right)_{\alpha\delta}
\psi_{\delta}\right)\,
\left(\overline\psi_{\gamma}\left(\Gamma_{\mu\, L}\right)_{\gamma\beta}
\psi_{\beta}\right).
\tag\dode-3b
$$
Thus the last term in (\dode-1a), for example, can be written as
$$
\brik{0}{\overline\psi_{\alpha}(\Gamma_L^{\mu})_{\alpha\delta}
\psi_{\delta}}{0}\cdot\,:
\overline\psi_{\gamma}(\Gamma_{\mu\,L})_{\gamma\beta}
\psi_{\beta}:=0.
\tag\dode-4
$$
The same conclusion can be drawn for all other terms which contain
VEV's. When one calculates commutators involving the operators $\Cal
O_{1,2,3,4}$, NOP associated complications do not appear.

However FIT $\Cal O_5$ and $\Cal O_6$ contain scalar and pseudoscalar
quantities so that their NOP is not trivial. From
$$
(\overline\psi_{\alpha}(\Gamma_{\mu\,L})_{\alpha\beta}\psi_{\beta})
(\overline\psi_{\gamma}(\Gamma^{\mu}_{L})_{\gamma\delta}\psi_{\delta})=
(-2)
(\overline\psi_{\alpha}(1+\gamma_5)_{\alpha\delta}\psi_{\delta})
(\overline\psi_{\alpha}(1-\gamma_5)_{\gamma\beta}\psi_{\beta})
\tag\dode-5
$$
and from (\dode-1) one obtains
$$
\aligned
:(\overline\psi_{\alpha}(\Gamma_{\mu\,L})_{\alpha\beta}\psi_{\beta})
(\overline\psi_{\gamma}(\Gamma^{\mu}_{L})_{\gamma\delta}\psi_{\delta}):
&=
(\overline\psi_{\alpha}(\Gamma_{\mu\,L})_{\alpha\beta}\psi_{\beta})
(\overline\psi_{\gamma}(\Gamma^{\mu}_{L})_{\gamma\delta}\psi_{\delta})\\
&+
2\,:(\overline\psi_{\alpha}(1+\gamma_5)_{\alpha\delta}\psi_{\delta}):
\brik{0}{(\overline\psi_{\gamma}\psi_{\beta}}{0}\\
&+
2\,:(\overline\psi_{\gamma}(1-\gamma_5)_{\gamma\beta}\psi_{\beta})
\brik{0}{(\overline\psi_{\alpha}\psi_{\delta}}{0}.
\endaligned
\tag\dode-6
$$
Once this is specified for $\hat{\Cal O}_6$\footnote{$^{\sharp}$}{\uskoosam
It is defined by $\hat{\Cal O_6}=:(\overline d\Gamma^{\mu}_L s)
[\overline u \Gamma_{\mu\, R} u+
\overline d \Gamma_{\mu\, R} d + \overline s \Gamma_{\mu\, R} s]:.$
}
one has to take care of color indices too. The spinors were actually
coupled in color sectors as follows
$$
(\overline
\psi^i_{\alpha}\psi_{\beta\,i})(\overline\psi^j_{\gamma}\psi_{\delta\,j}).
\tag\dode-7a
$$
The rearangement (\dode-6) turns that into
$$
(\overline
\psi^i_{\alpha}\psi_{\delta\,j})\brik{0}{
\overline\psi^j_{\gamma}\psi_{\beta\,i}}{0}
\tag\dode-7b
$$
as only the color singlet can have VEV, one obtains
$$
(\overline
\psi^i_{\alpha}\psi_{\delta\,j})\frac13\delta_{ij}\brik{0}{
\overline\psi^k_{\gamma}\psi_{\beta\,k}}{0}.
\tag\dode-7c
$$
Finally this gives
$$
\aligned
\hat{\Cal O}_6=\tilde{\Cal O}_6&+\frac23\brik{0}{\overline d d}{0}
\overline d(1-\gamma_5)s\\
&+\frac23\brik{0}{\overline s s}{0}\overline
(1+\gamma_5)s.
\endaligned
\tag\dode-8
$$
Here $\tilde{\Cal O}_6$\footnote{$^{\flat}$}{\uskoosam
Here we are using the definition of $\Cal O_{5,6}$ which differs, with
respect to Ref. [11] for instance, by a factor 4, i.e. $\hat{\Cal
O}_{5,6}(\text{here})=4\cdot\Cal O_{5,6}(\text{Ref.[11]})$.
}\, is the $\hat{\Cal O}_6$ operator which is not
normal-ordered. This result is in full qualitative agreement with Ref.
[21], formula (33).
(One has to take into account that
$\gamma_5(\text{here})=-\gamma_5(\text{[21]})$.
The analogous result  can be obtained for
$\hat{\Cal O}_5$ which contains SU(3) color matrices $\lambda_A$. They
satisfy the equality (\sep-6)
$$
\sum_A(\lambda_A)_{a\,b}(\lambda_A)_{c\,d}=\frac{16}9\delta_{ad}\delta_{cb}
-\frac13 (\lambda)_{ad}(\lambda)_{cb}.
\tag\dode-9
$$

FIT of $\hat{\Cal O}_5$ in the Lorentz-space has to be combined with
(\dode-9). It means that in the Lorentz-space $\hat{\Cal O}_5$ must have
the same form as $\hat{\Cal O}_6$ (\dode-7), but one has to introduce
(\dode-9) into the second and third term on the right-hand-side (RHS) of
(\dode-8). The equality (\dode-2b) means that only the first RHS term in
(\dode-9) contributes. Thus VEV terms in (\dode-8) are multiplied by
16/3.
$$
\aligned
\hat{\Cal O}_5=\tilde{\Cal O}_5&+\frac{32}{9}\brik{0}{\overline d
d}{0}\overline d(1-\gamma_5)s\\
&+\frac{32}{9}\brik{0}{\overline s s}{0}\overline d(1+\gamma_5)s.
\endaligned
\tag\dode-10
$$
The expression (\dode-10) agrees fully with the formula (33) of ref.
[21].

Some additional formal manipulations are shown in Appendix \bili.

\vfill
\mpikk
\eject
%
%
%
\def\apf{\text{F}}

\baselineskip=20pt
\centerline{\bf Appendix\ \apf: Isospin and/or Baryon Decomposition
 of The Weak Potential}
\def\naspog{Appendix\, \apf: Isospin And/or Baryon Decomposition
 of The Weak Potential}

\medskip

The isospin decomposition of an effective weak two particle potential
acting among baryons depends on the isospin decomposition of the weak
vertices. The amplitudes corresponding to $\Lambda\to B+\pi$
transitions, can be parametrized for $\Delta I=1/2$ transition as
$$
\aligned
\Lambda\to &n+\pi^0;\qquad f(\Lambda_0^0)\\
&f(\Lambda^0_0)=\frac{1}{\sqrt{3}}\alpha\hskip4cm|\Delta I|=1/2\\
\Lambda\to &p+\pi^-;\qquad f(\Lambda_0^-)\\
&f(\Lambda^0_-)=-\sqrt{\frac23}\alpha
\endaligned
\tag\apf-1
$$
Here we have displayed only particle content and isospin properties. The
same relations (\apf-1) hold for both $s-$wave and $p-$wave
contributions.

The $|\Delta I|=3/2$ transitions are parametrized as
$$
\aligned
g(\Lambda^0_0)&=\sqrt{\frac23}\beta\\
&\hphantom{xxxxxxxxxxxxxxxxxxxxxxxxxxxxxxxxxxxxxxxxx}
|\Delta I|=3/2\\
g(\Lambda^0_-)&=\sqrt{\frac13}\beta.
\endaligned
\tag\apf-2
$$
The weak amplitudes corrresponding to $N\to N +K$ transitions can be
parametrized as follows
$$
\aligned
n\to &n+K^0;\qquad f(n_0^0)\\
&f(n^0_0)=\delta\\
p\to &p+K^0;\qquad f(p_0^+)\\
&f(p^+_0)=\frac12(\xi+\delta)\\
&\hphantom{xxxxxxxxxxxxxxxxxxxxxxxxxxxxxxxxxxxxxxxxx}
|\Delta I|=3/2\\
p\to &n+ K^+;\qquad f(p_+^+)\\
&f(p^+_+)=\frac12(-\xi+\delta)
\endaligned
\tag\apf-3
$$
and
$$
\aligned
g(n_0^0)&=\frac12\epsilon\\
g(p_0^+)&=-\frac12\epsilon\\
g(p_+^+)&=-\frac12\epsilon
\endaligned
\tag\apf-4
$$
The only weak vertex from which $\eta$ is emitted corresponds to the
proces $\Lambda\to n+\eta$. Here only $|\Delta I|=1/2$
piece of $\Cal H_W$ can contribute. Thus the weak potential due
to the $\eta$ exchange satisfies the same selection rule.

The potential
due to a pion exchange, which corresponds toa diagram analogous to
Fig.\uvo.1, has the following particle (i.e. baryonic) contents:
$$
\aligned
(\overline n \Lambda)_{1\,w}\,&\kkitrk{\pi^0\,}{\pi^0}\,(\overline n n
)_{2\,S}\cdot\frac1{\sqrt{3}}\alpha \cdot (-g_{NN\pi})\\
(\overline n \Lambda)_{1\,w}\,&\kkitrk{\pi^0\,}{\pi^0}\,(\overline p p
)_{2\,S}\cdot\frac1{\sqrt{3}}\alpha \cdot
g_{NN\pi}\qquad\qquad\qquad\qquad
|\Delta I|=1/2\\
(\overline p \Lambda)_{1\,w}\,&\kkitrk{\pi^-}{\pi^-}\,(\overline n p
)_{2\,S}\cdot(-1)\cdot\sqrt{\frac23}\alpha \cdot (\sqrt{2}g_{NN\pi})
\endaligned
\tag\apf-5
$$
Here $1,2$ denotes the spin and coordinate dependence while $w$ and $S$
mean the weak and the strong vertex respectively. $g_{NN\pi}$ is the
strong coupling constant. The combination $\kkitrk{M}{M}$
symbolizes the meson propagator which appears as a Yukawa function in
the weak potential.

The $|\Delta I|=3/2$ piece is
$$
\aligned
(\overline n \Lambda)_{1\,w}\,&\kkitrk{\pi^0\,}{\pi^0}\,(\overline n n
)_{2\,S}\cdot\sqrt{\frac23}\beta \cdot (-g_{NN\pi})\\
(\overline n \Lambda)_{1\,w}\,&\kkitrk{\pi^0\,}{\pi^0}\,(\overline p p
)_{2\,S}\cdot\sqrt{\frac23}\beta \cdot g_{NN\pi}\qquad\qquad\qquad\qquad
|\Delta I|=3/2\\
(\overline p \Lambda)_{1\,w}\,&\kkitrk{\pi^-}{\pi^-}\,(\overline n p
)_{2\,S}\cdot\frac1{\sqrt{3}}\beta \cdot (\sqrt{2}g_{NN\pi})
\endaligned
\tag\apf-6
$$
The kaon exchange results in the combinations
$$
\aligned
(\overline n \Lambda)_{1\,S}\,&\kkitrk{K^0\,}{K^0}\,(\overline n n
)_{2\,w}\cdot\delta \cdot g_{NNK})\\
(\overline n \Lambda)_{1\,S}\,&\kkitrk{K^0\,}{K^0}\,(\overline p p
)_{2\,w}\cdot{\frac12}(\xi+\delta) \cdot g_{NNK}\qquad\qquad\qquad\qquad
|\Delta I|=1/2\\
(\overline p \Lambda)_{1\,S}\,&\kkitrk{K^-}{K^-}\,(\overline n p
)_{2\,w}\cdot\frac12(-\xi+\delta) \cdot g_{NNK}
\endaligned
\tag\apf-7a
$$
The comparison with Ref. [3] is easier if one introduces the
notation
$$
\aligned
\frac12(\xi+\delta)g_{NNK}&=d\\
\frac12(-\xi+\delta)g_{NNK}&=d\\
\delta\cdot g_{NNK}&=c+d.
\endaligned
\tag\apf-7b
$$
In (\apf-5,6,7) and in the following, the $\Lambda$ baryon is always in
the vertex $1$, irrespectively whether it is the strong or the weak
vertex. ( The complete potential, as discussed in Section \efp, is
symmetric in $1\leftrightarrow 2$.)

The $|\Delta I|=3/2$ kaon exchange pieces are
$$
\aligned
(\overline n \Lambda)_{1\,S}\,&\kkitrk{K^0\,}{K^0}\,(\overline n n
)_{2\,w}\cdot e,\\
(\overline n \Lambda)_{1\,S}\,&\kkitrk{K^0\,}{K^0}\,(\overline p p
)_{2\,w}\cdot{\frac12}(\xi+\delta) \cdot (-e) \qquad\qquad\qquad\qquad
|\Delta I|=3/2\\
(\overline p \Lambda)_{1\,S}\,&\kkitrk{K^-}{K^-}\,(\overline n p
)_{2\,w}\cdot(-e)
\endaligned
\tag\apf-7a
$$
Here $e\equiv \gamma g_{NNK}/2$.

The expressions (\apf-5-8) can be connected with (\efe-17) and
(\efe-20) by using the following isospin dependent quantities
$$
\aligned
\beta_1&=\left(\overline N {\bold 1}\pmatrix 0\\1\endpmatrix
\Lambda\right)_1(\overline N {\bold 1} N)_2\\
&{\hphantom{xxxxxxxxxxxxxxxxxxxxxxxxxxxxxxxxxxxxx}}\Delta I=1/2\\
\beta_{\tau}&=\left(\overline N \vec\tau\pmatrix 0\\1\endpmatrix
\Lambda\right)_1(\overline N {\vec\tau} N)_2\\
{}\\
\beta_{T}&=\left(\overline N (\vec T \chi)
\Lambda\right)_1(\overline N {\vec\tau} N)_2\qquad\qquad\qquad\qquad
\qquad\qquad\Delta I=3/2
\endaligned
\tag\apf-9a
$$
Here
$$
\gathered
\overline N=(\overline p,\overline n)\\
\overline N(\vec T\chi)\vec\tau =\sum_{m,r}
\kleb{1}{r}{1}{m}{3/2}{-1/2}\vec
t^r_1\cdot \vec\tau.
\endgathered
\tag\apf-10a
$$
The summation over $m$ goes over two isospin states $m=\pm1/2$ contained
in $N$. Thus
$$
\overline N(\vec T\chi)\vec\tau=\sqrt{\frac23}(\tau^-+\tau^3)
\tag\apf-10b
$$
The symbol $\Lambda$ carries strangeness and has no isospin dependence.
With (\apf-10) one obtains
$$
\aligned
\beta_1&=(\overline n\Lambda)_1[(\overline p p)_2+(\overline n n)_2]\\
\beta_{\tau}&=(\overline n\Lambda)_1[(\overline n n)_2-(\overline p p)_2]
+2(\overline p \Lambda)_1(\overline n p)_2\\
\beta_{T}&=\sqrt{\frac23}[(\overline p\Lambda)_1(\overline n p)_2
+(\overline n \Lambda)_1[(\overline p p)_2-(\overline n n)_2]
\endaligned
\tag\apf-9b
$$
The isospin dependence of the pion exchange contribution can be written
as
$$
V_{\pi}=\tilde A \vec\tau_1\cdot\vec\tau_2+\tilde B\vec T_1
\cdot\vec\tau_2.
\tag\apf-11a
$$
(This is not symmetrized). The factors $\tilde A$ and $\tilde B$ contain
all spatial and spin operators. In theoretical nuclear calculations the
operator (\apf-10a) is to be sandwiched between multiparticle baryon
states containing $N$ and $\Lambda$.

The correspondence with (\apf-5) and (\apf-6) is established if one
"reads" (\apf-11a) as
$$
\aligned
V_{\pi}&\longrightarrow \tilde A \beta_{\tau}+\tilde B\beta_{T}\\
&=(\overline n\Lambda)_1(\overline n n)_2\left[\tilde A-\sqrt{\frac23}
\tilde B\right]\\
&+(\overline n\Lambda)_1(\overline p p)_2\left[-\tilde A+\sqrt{\frac23}
\tilde B\right]\\
&+(\overline p\Lambda)_1(\overline n p)_2\left[2\tilde A+\sqrt{\frac23}
\tilde B\right]
\endaligned
\tag\apf-11b
$$
The kaon exchange contribution is
$$
\aligned
V_K&=\frac12\tilde C(1+\vec\tau_1\cdot\vec\tau_2)+\tilde D+
\tilde E\vec T_1\cdot\vec\tau_2\\
&\longrightarrow \left(\frac12 \tilde C+\tilde D\right)\eta_0+
\frac12\tilde C\eta_{\tau}+\tilde E\eta_T\\
&=(\overline n\Lambda)_1(\overline n n)_2(\tilde C+\tilde D-\tilde E)\\
&+(\overline n\Lambda)_1(\overline p p)_2(\tilde D+\tilde E)\\
&+(\overline p\Lambda)_1(\overline n p)_2(\tilde C+\tilde E)
\endaligned
\tag\apf-12
$$
Expressions (\apf-9b) can be inverted. By introducing the notation
$$
\aligned
a&=(\overline n\, \Lambda)(\overline p\, p)\\
b&=(\overline n\, \Lambda)(\overline n\, n)\\
c&=(\overline p\, \Lambda)(\overline n\, p)
\endaligned
\tag\apf-13
$$
one finds
$$
\alignedat 4
a&= &\frac12 \beta_1 &-\frac16 \beta_{\tau} &+\frac{1}{\sqrt{6}}\beta_T\\
b&= &\frac12 \beta_1 &+\frac16 \beta_{\tau} &-\frac{1}{\sqrt{6}}\beta_T\\
c&= & {}             &+\frac13 \beta_{\tau} &+\frac{1}{\sqrt{6}}\beta_T.
\endalignedat
\tag\apf-14
$$

\vfill
\mpikk
\eject
%
%
%

\noindent{\bf Appendix \apg: Shifted Yukawa Function}
\def\naspog{Appendix \apg: Shifted Yukawa Function}
\bigskip

It is usually stated [44,45] that
when a force is transmitted by a particle then the range of that force
depends on the mass of the intermediate particle. Ever since
Yukawa's seminal work [46]
the Yukawa potential
$$
V_Y(r)=\frac{e^{\mu r}}{r}
\tag \apg-1
$$
was a textbook feature [45], describing nucleon-nucleon interaction,
produced by one meson exchange.
It
turns out that the original statement (1) is somewhat modified in
hypernuclei [5] where one has nucleons and strange
particles.\footnote{$^{\sharp}$}{\devetrm For instance the
${}^{12}_{\hphantom{1}\Lambda}$C nucleus consists of 5 neutrons, 6
protons and the $\Lambda$-hyperon.}  In
that case the range of the force depends both on the
intermediate meson mass and the baryon masses.

Moreover, in order to connect the more general procedure with the
result (\apg-1) valid for the baryons which all
have equal masses,  one can alternatively  use an
interesting generalized definition of the delta-function [47].

A transitional amplitude $A_{f\,\Lambda}$ corresponding to the one pion
(or any meson) exchange has a generic form [45]
$$
A_{f\,\Lambda}=\frac N2 \int d^4x\,d^4y\int d^4k\frac{e^{-ik(x-y)}}
{k^2-\mu^2+i\epsilon}\cdot
\brik{f}{T(S(x)W(y))}{\Lambda}.
\tag \apg-2a
$$
Here
$$
\aligned
k^2&=k_0^2-\vec k\,{}^2\\
kx&=k_0x_0-\vec k\cdot\vec x,
\endaligned
\tag \apg-2b
$$
Here $S(x)$ and $W(x)$ are some baryon densities (scalar or
pseudoscalar) which are sources of the meson (pion) field, which mass is
$\mu$. Detailed form of those quantities need not concern us here. If
all baryons are nucleons $N$, then the calculation can be found, for
example, in ref. [45], p.213. In a more general
case the density $S(x)$ can contain a strange
baryon, for example $\Lambda$ hyperon [5].

The time ordered product in (2a) can be written as
$$
\aligned
\brik{f}{T(S(x)W(y))}{\Lambda}&=\theta(x-y)\sum_n\brik{f}{S(x)}{n}
\brik{n}{W(y)}{\Lambda}\\
&+\theta(y-x)\sum_s\brik{f}{W(y)}{s}
\brik{s}{S(x)}{\Lambda}.
\endaligned
\tag \apg-3
$$
Here the intermediate states
$$
\ket{n};\qquad S=0
\tag \apg-4
$$
have no strangeness while the states
$$
\ket{s};\qquad S=-1
\tag \apg-5
$$
must contain a strange baryon.

In (\apg-2a) one can integrate over times $(x_0,y_0)$. A useful identity is
$$
\brik{f}{S(x)}{n}=e^{i(E_f-E_n)x_0}\brik{f}{S(\vec x)}{n}.
\tag \apg-6
$$
An analogous expression hold for other matrix elements. Also
$$
\aligned
dx^0\,dy^0&=2d\xi\,d\eta;\\
x^0&=\xi+\eta\\
y^0&=-\xi+\eta
\endaligned
\tag \apg-7
$$
The integration $d\eta$ can be immediately carried out which results in
$$
\aligned
A_{f\,\Lambda}&=N\int d^3x\,d^3y\,d\xi\int d^4k\frac{e^{-2i\xi k_0+
i\vec k\cdot (\vec x-\vec y)}}{k^2-\mu^2+i\epsilon}\\
&\cdot 2\pi\delta(E_f-E_{\Lambda})\big[\theta(\xi)\sum_n e^{i\Delta_n\xi}
\alpha_n(\vec x,\vec y)+
\theta(-\xi)\sum_s e^{-i\Delta_s\xi}
\beta_s(\vec y,\vec x)\big].
\endaligned
\tag \apg-8a
$$
Here
$$
\aligned
\Delta_i&=E_f+E_{\Lambda}-2E_i\\
\alpha_n&=\brik{f}{S(\vec x)}{n}\brik{n}{W(\vec y)}{\Lambda}\\
\beta_s&=\brik{f}{W(\vec y)}{s}\brik{s}{S(\vec x)}{\Lambda},
\endaligned
\tag \apg-8b
$$
and $E_{k}$ are relativistic energies.

First one integrates over $d\xi$ obtaining
$$
\aligned
A_{f\,\Lambda}&=(2\pi i)N\delta(E_f-E_{\Lambda})\int d^3x\,d^3y
\int d^4 k\frac{
e^{i\vec k\cdot (\vec x-\vec y)}}{k^2-\mu^2+i\epsilon}\\
&\cdot\left[\sum_n\alpha_n\frac1{\Delta_n-2k_0+i\epsilon}+
\sum_s\beta_s\frac1{\Delta_s+2k_0+i\epsilon}\right].
\endaligned
\tag \apg-9
$$
Integration in the complex $k_0$ plane over the contours shown in
Fig.\apg.1 leads to
$$
\aligned
A_{f\Lambda}&=2\pi^2 \delta(E_f-E_{\Lambda})N\int d^3x\, d^3y\int d^3k\cdot
e^{i\vec k(\vec x-\vec y)}\\
&\cdot\left[\sum_n\frac{\alpha_n(\vec x,\vec y)}{\omega(\Delta_n-2\omega)}+
\sum_s\frac{\beta_s(\vec y,\vec x)}{\omega(\Delta_s-2\omega)}\right]\\
&\hphantom{=====}\omega=\sqrt{{\vec k\,}^2+\mu^2}.
\endaligned
\tag \apg-10
$$

\midinsert
\vbox{\medskip
\centerline{\epsfxsize=9cm\epsfbox{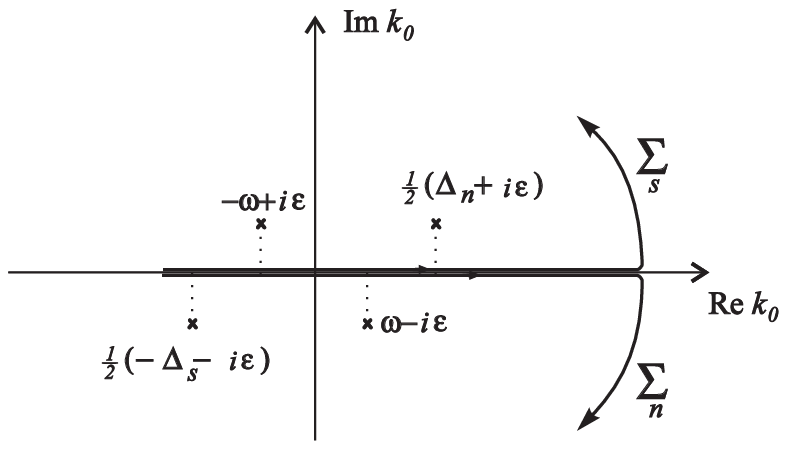}}
\medskip
\centerline{{\it Fig.\apg.1 - The contours in the $k_0$ plane. Here}}
\centerline{{\it $\omega^2={\vec k}^2+\mu^2$.}}
\medskip}
\endinsert

In the nonrelativistic limit the energy differences $\Delta_i$ can be
approximated by the corresponding baryon mass differences. Schematically
one can use the following baryonic contents:
$$
\alignedat 3
&\ket{f} &\qquad &\text{$K$ nucleons}\\
&\ket{\Lambda} &\qquad &\text{1 $\Lambda+(K-1)$ nucleons}\\
&\ket{s} &\qquad &\text{1 $\Lambda+(K-1)$ nucleons}\\
&\ket{n} &\qquad &\text{$K$ nucleons}.
\endalignedat
\tag \apg-11a
$$
Thus
$$
\aligned
E_f\ &\longrightarrow\ K\cdot m_N\\
E_{\Lambda}\ &\longrightarrow\ (K-1)\cdot m_N+m_{\Lambda}\\
E_{s}\ &\longrightarrow\ (K-1)\cdot m_N+m_{\Lambda}\\
E_n\ &\longrightarrow\ K\cdot m_N\\
\Delta_s&=E_f+E_{\Lambda}-2E_s\to -2\delta \\
\Delta_n&=E_f+E_{\Lambda}-2E_n\to 2\delta \\
 \delta &=(m_{\Lambda}-m_N)/2
\endaligned
\tag \apg-11b
$$
In such an approximation the factors depending on $\omega$ and
$\Delta_i$ can be taken out of the summations in (\apg-10). Furthermore
$$
\aligned
\sum_n\alpha_n(\vec x,\vec y)&=S(\vec x)W(\vec y)\\
\sum_s\beta_n(\vec x,\vec y)&=W(\vec y)S(\vec x)
\endaligned
\tag \apg-12
$$
As in the nonrelativistic approach $W$ and $S$ become operators in the
configuration space acting on (Schr\"odinger) wave functions [45,48],
their order is immaterial. Thus one obtains
$$
\aligned
A_{f\Lambda}&=(2\pi^2) N\delta(E_f-E_{\Lambda})\int d^3x\,d^3y\int
d^3k\cdot e^{i\vec k(\vec x-\vec y)}\\
&\cdot\frac12\left[(-1)\frac{1}{\omega(\delta+\omega)}+
\frac{1}{\omega(\delta-\omega)}\right] \brik{f}{O(\vec x,\vec
y)}{\Lambda}\\
&O(\vec x,\vec y)=S(\vec x)W(\vec y)
\endaligned
\tag \apg-13
$$
The integration over $\vec k$  gives the {\it shifted \/} Yukawa
function
$$
\aligned
&\int d^3k\cdot e^{i\vec k(\vec x-\vec y)}\frac12
\left[(-1)\frac{1}{\omega(\delta+\omega)}+
\frac{1}{\omega(\delta-\omega)}\right]=\\
&\int d^3k\frac{e^{i\vec k(\vec x-\vec y)}}{\omega^2-\delta ^2}=
(-2\pi^2)\frac{e^{-\epsilon r}}{r}\\
&\hphantom{xx}r=|\vec x-\vec y|\\
&\hphantom{xx}\epsilon=\sqrt{\mu^2-\delta^2}=\sqrt{\mu^2-\frac14(m_{\Lambda}-m_N)^2}.
\endaligned
\tag \apg-14
$$
Eventually one obtains a generic form
$$
A_{f\Lambda}=(-4\pi^4)N\int d^3x\,d^3y \frac{e^{-\epsilon r}}{r}
\brik{f}{O(\vec x,\vec y)}{\Lambda}
\tag \apg-15
$$
in which the Yukawa potential (function) depends on the effective mass
$$
\epsilon<\mu.
\tag \apg-16
$$
If one deals with nucleons only, then
$$
\aligned
\delta&=0\\
\epsilon &\to \mu,
\endaligned
\tag \apg-17
$$
and the standard textbook [45] form (\apg-1) is recovered.

This last result can be obtained also by starting from
the expression (\apg-8). In the
strong nucleon-nucleon interaction
the variable $k$  corresponds to
the invariant nucleon momentum transfer [45]
$$
\aligned
k^2&=(p-p'\,)^2\\
p&=(E_p,\vec p);\qquad E_p=\sqrt{{\vec p}\,{}^2+M_N^2}.
\endaligned
\tag \apg-18
$$
In the nonrelativistic limit, when $|\vec p|\ll M_N$ one can approximate
$$
\aligned
E_p&\simeq M_N\\
\delta \to E_p-E_p'&\simeq M_N-M_N=0\\
k^2&\simeq -{\vec k}\,{}^2\\
\int d^4k\frac{e^{-ik(x-y)}}{k^2-\mu^2+i\epsilon}&\longrightarrow
(-1)\int dk_0\cdot d^3 k\frac{e^{-2ik_0\xi}e^{i\vec k\cdot \vec r}}
{{\vec k\,}^2+\mu^2}\\
&=(-1)\pi \delta(\xi)\cdot \int d^3k\frac{e^{i\vec k\cdot \vec r}}
{{\vec k\,}^2+\mu^2}\\
&=(-1)\delta(\xi) 2\pi^3\frac{e^{-\mu r}}{r}.\\
&\hphantom{xxxxxx}r=|\vec x-\vec y|.
\endaligned
\tag \apg-19
$$
When this is introduced in (\apg-8) with $\ket{\Lambda}=\ket{i}$ (nucleon
state), one finds
$$
\aligned
A_{fi}&=N\int d^3x\,d^3y\, d\xi (-1)\delta(\xi)\frac{e^{\DS -\mu
r}}{r}\\
&\cdot(2\pi) \delta(E_f-E_i)\left[\theta(\xi)\sum_ne^{\DS i\Delta_n\xi}
\alpha_n(\vec x,\vec y)+
\theta(-\xi)\sum_s e^{\DS -i\Delta_s\xi}\beta_s(\vec x,\vec y)\right]
\endaligned
\tag \apg-20a
$$
with
$$
\Delta_i=E_f-E_i\to 0.
\tag \apg-20b
$$
In order to integrate over $\xi$ one uses the
identity [5]
$$
\int d\xi\,\theta(\xi)\,\delta(\xi)\cdot e^{\DS i\alpha\xi}=\frac12,
\tag \apg-21
$$
and eventually obtains
$$
A_{fi}=A_{f\,\Lambda=i}(\epsilon\to \mu),
\tag \apg-22
$$
as already mentioned above (\apg-17).
In a sense, that can be considered as an example for
the novel delta-function identity derived by ref. [47].

Although interesting as a matter of principle the shifted
Yukawa potential does not lead to some startling numerical differences.
The range of the potential is somewhat increased, as can be seen by
plotting the ratio
$$
\frac{e^{\DS -\epsilon r}}{e^{\DS -\mu r}}=e^{\DS(\mu-\epsilon)r}=X(r)
\tag \apg-23
$$
which is shown in Fig.\apg.2 (a). The shifted potential has an increased range.
With $\mu=m_{\pi}=0.70\,$fm${}^{-1}$ and
$\delta=(m_{\Lambda}-m_N)/2=0.45\,$fm${}^{-1}$ one finds: $r=0.5\,$fm\ $X=1.08$ and
$r=1\,$fm\ $X=1.17$ for example.

\midinsert
\vbox{\medskip
\centerline{\epsfxsize=14cm\epsfbox{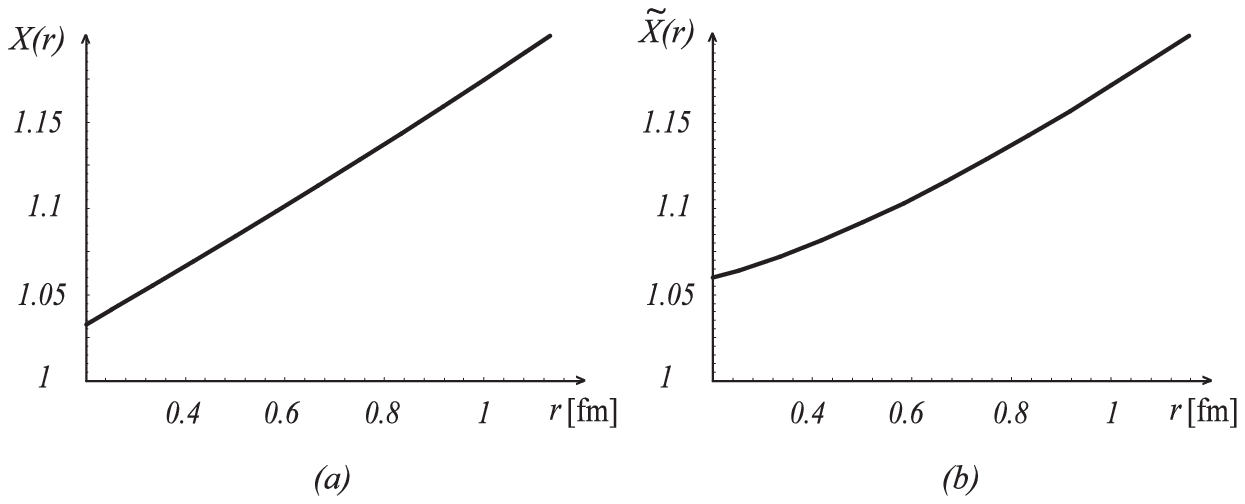}}
\medskip
\centerline{{\it Fig.\apg.2 - The ratio  (23) (a) and (26) (b) is plotted
as a function of $r$.}}
\medskip}
\endinsert

In the actual calculation with hypernuclei one introduces a monopole
form-factor at the each vertex [5]. The denominators in (\apg-14) and
(\apg-19)
are thus replaced by
$$
\frac 1{\vec k^2+\phi^2} \longrightarrow
\frac{(\Lambda ^2-\mu^2)}{(\vec k^2+\phi^2)(\vec k^2+\Lambda ^2)^2}
= W(\phi)\qquad (\phi=\epsilon,\  \mu).
\tag \apg-24
$$
The Fourier transform of that is
$$
\aligned
\Cal F\{W(\phi)\}=&2\pi^2\Big[\frac{\Lambda^2-\mu^2}{\Lambda^2-\phi^2}
\left(\frac{e^{-\DS \phi r}}{r}-\frac{e^{-\DS \Lambda r}}{r}\right)\\
&-\frac{\Lambda^2-\mu^2}{2\Lambda}\cdot e^{-\DS \Lambda r}\Big].
\endaligned
\tag \apg-25
$$
The ratio
$$
\tilde X(r)=\frac{\Cal F \{W(\epsilon)\}}{\Cal F \{W(\mu)\}}
\tag \apg-26
$$
is also plotted in Fig.\apg.2 (b) for $\Lambda=1.3\,$GeV [5].
One finds $\tilde X =1.09$ at 0.5\,fm for example.

\vfill
\eject

%
%

\noindent{\bf Appendix \aph: Effective $\widetilde a_{BB}$ Amplitude and Baryon
Pole Terms}
\def\naspog{Appendix \aph: Effective $\widetilde Q_{BB}$  Amplitude and Baryon
Pole Terms}

\vskip 1cm

Parity violating $A$ amplitudes obtain here the following contributions
$$
A=A_{CA}+A_{Sep}
\tag\aph-1
$$
The first corrensponds to formulae (\cur-20) to (\cur-26) while the
second one is calculated as described in (\sep-20). The contributions
$A$ (\aph-1) should be compared with the experimental result
$A_{\text{exp}}$, i.e.
$$
A_{\text{exp}}\sim A=A_{CA}+A_{Sep}.
\tag\aph-2
$$
The weak vertices in the baryon pole terms (see Fig.\pol.1) are
determined by the $A_{{CA}}$ which has a generic form
$$
A_{CA}\sim \frac{1}{f_{\pi}}\brik{B'}{H_W^{PC}}{B}=\frac1{f_{\pi}}
\widetilde a_{B'B}.
\tag\aph-3
$$
One actually uses
$$
\lim_{q\to 0} A=A_{CA}.
\tag\aph-4
$$
Thus in order to find $\widetilde a_{B'B}$ one must use an approximate
expression whose a generic form is
$$
\frac{1}{f_{\pi}}\tilde a_{B'B}\simeq \left[ A_{B'B/exp}-
A_{B'B/SEP}\right].
\tag\aph-5
$$

\vfill
\mpikk
\eject

%
%
%
\parskip 10pt
\baselineskip 15pt
\parindent 20pt
\centerline{\bf Acknowledgement}
\def\naspog{Acknowledgement}
\vskip 1cm

The authors acknowledge the support of Croatian ministry of science and
technology, grant 119222 and of ANPCyT (Argentina) under grant BID 1201/OC-AR
(PICT 03-04296) and of Fundaci\'{o}n Antorchas (Argentina) under grant Nr.
13740/01-111. F.K. and C.B. are fellows of the CONICET from Argentina.  

\vfill
\mpikk
\eject
%
%
%

\centerline{\bf References}
\def\naspog{References}
\vskip 1cm

\item{[1]} C. Barbero, D. Horvat, F. Krmpoti\' c, Z. Naran\v ci\' c, and
D. Tadi\' c, in preparation.
\medskip
\item{[2]} C. Barbero, D. Horvat, F. Krmpoti\' c, Z. Naran\v ci\' c, and
D. Tadi\' c, {\it Non-Mesonic Weak Decay of ${}^{12}_{\Lambda}\text{C}$,
submitted for publication.
\medskip
\item{[3]} J. Cohen, Prog. Part. Nucl. Phys. {\bf 25}, 139 (1990).
\medskip
\item {[4]} J. F. Dubach, G. B. Feldman, B. R. Holstein and L. de la Torre,
Ann. Phys. {\bf 249}, 146 (1996).
\medskip
\item {[5]} A. Parre\~no, A. Ramos, C. Bennhold, Phys. Rev.
C {\bf 56}, 339 (1997).\medskip
\item{[6]} H. Bando, T. Motoba and J. Zofka, Int. Jour. Mod. Phys.
{\bf A5}, 4021 (1990).
\medskip
\item{[7]} C. Bennhold, A. Parre\~no and A. Ramos, Few-Body
Systems Suppl. 9, 475 (1995).
\medskip
\item{[8]} E. D. Commins and
P. H. Bucksbaum, {\it  Weak Interactions of Leptons and Quarks\/}
(Cambridge University Press, Cambridge, 1983).
\medskip
\item {[9]} R. E. Marshak, Riazuddin, C. P. Ryan, {\it  Theory
of Weak Interactions in Particle Phyiscs\/} (Wiley Interscience,
New York, 1969);
\medskip
\item{[10]} V. S. Mathur and L. K. Pandit,in {\it Advances in
Particle Physics\/} ed. by R. Cool and R. E. Marshak ((Wiley Interscience,
New York, 1968), p. 383.
\medskip
\item{[11]} L. B. Okun, {\it Leptons and Quarks\/} (North Holland
Publ. Comp., Amsterdam, 1982)
\medskip
\item{[12]} F. Palmonari, Riv. N. Cim. 7 (9), 1 (1984).
\medskip
\item {[13]} J. F. Donoghue, E. Golowich and B. Holstein,
Phy. Rep. {\bf 131}, 319 (1986).
\medskip
\item{[14]} D. Tadi\' c and J. Trampeti\' c, Phys. Rev. D {\bf 23},
144 (1981).
\medskip
\item{[15]} S. P. Rosen and S. Pakvasa, in {\it Advances in
Particle Physics\/} ed. by R. Cool and R. E. Marshak ((Wiley Interscience,
New York, 1968), p. 473.
\medskip
\item{[16]} M. D. Scadron and D. Tadi\' c, Jour. Phys. G, to be
published.
\medskip
\item {[17]}  D. Horvat  and D. Tadi\' c, Z. Phys. C
{\bf 31}, 311 (1986); ibid. {\bf 35}, 231 (1987);
\medskip
\item{[18]} D. Horvat, Z. Naran\v ci\' c and D. Tadi\' c,
Z. Phys. C {\bf 38}, 431 (1988).
\medskip
\item{[19]} E. Fischbach and D. Tadi\' c, Phys. Rep. C {\bf 6}, 125
(1973).
\medskip
\item{[20]} B. Desplanques, J. F. Donoghue and B. R. Holstein, Ann.
Phys. 124, 449 (1980).
\medskip
\item{[21]} J. F. Donoghue, E. Golowich, W. A. Ponce and B. R.
Holstein, Phys. Rev. {\bf D} 21, 186 (1980).
\medskip
\item{[22]} M. J. Savage and R. P. Springer, Phys. Rev. C {\bf 53},
441 (1996), ibid. C 54(E), 2786 (1996).
\medskip
\item{[23]} M. K. Gaillard and B. W. Lee, Phys. Rev. Lett. {\bf 33}, 108
(1974); Phys. Rev. {\bf D} 10, 897 (1974);
G. Altarelli and L. Maiani, Phys. Lett. {\bf 52B}, 351 (1974);
A. I. Vainshtein, V. I. Zakharov and M. A. Shifman, Nucl. Phys. {\bf
B120}, 316 (1977).
\medskip
\item {[24]} J. F. Donoghue, E. Golowich and B. Holstein,
{\it Dynamics of The Standard Model\/}
(Cambridge University Press, New York, 1992).
\medskip
\item {[25]} O. Dumbrajs et al., Nucl. Phys. B 216, 277 (1983).
\medskip
\item{[26]} P. A. Carruthers, {\it Introduction to Unitary Symmetry\/}
(Interscience Publ., New York, 1966).
\medskip
\item{[27]} B. T. Feld, {\it Models of Elementary Particles\/}
(Blaisdell Publ. Comp., Waltham, 1969).
\medskip
\item{[28]} M. Gourdin, {\it Unitary Symmetries and Their Application to
High Energy Physics\/} (North Holland Publ. Comp., Amsterdam, 1967).
\medskip
\item{[29]} J. J. Sakurai, {\it Currents and Mesons\/} (Univ. of Chicago
Press, Chicago, 1969).
\medskip
\item{[30]} J. Gasser and H. Leutwyler, Phys. Rep. {\bf 87}, 77
(1982).
\medskip
\item{[31]} J. Gasser, H. Leutwyler and. M. E. Sainio, Phys. Lett.
{\bf 253}B, 252 (1991).
\medskip
\item{[32]} M. E. Sainio, in {\it Chiral Dynamics: Theory and
Experiment \/}, Proceedings of the Workshop at MIT, Cambridge, July
1994,
ed. by A. M. Bernstein and B. R. Holstein (Springer, New York, 1995).
\medskip
\item {[33]} J.D. Bjorken and S.D. Drell, {\it Relativistic Quantum
Fields\/} (McGraw-Hill, New York, 1964).
\medskip
\item{[34]} F. Gross, {\it  Relativistic Quantum Mechanics and Field
Theory \/} (John Wiley \& Sons, Inc., New York, 1993).
\medskip
\item{[35]} J. J. De Swart, Rev. Mod. Phys. {\bf 35}, 916 (1963).
\medskip
\item{[36]} P. Mc Namee and F. Chilton, Rev. Mod. Phys, {\bf 36}, 1005
(1964).
\medskip
\item{[37]} M. D. Scadron and M. Visinescu, Phys. Rev. D {\bf 28}, 1117
(1983); R. D. Springer, Phys. Lett. B 461, 167 (1999).
\medskip
\item{[38]} A. Le Yaouanc, L. L. Oliver, O. Pene and J.-C. Raynal, {\it
Hadron Transitions in The Quark Model\/} (Gordon and Breach, New York,
1988).
\medskip
\item{[39]} D. Palle and D. Tadi\' c, Z. Phys. {\bf C 23}, 301 (1984).
\medskip
\item{[40]} M. Milo\v sevi\' c, D. Tadi\'c and J. Trampeti\' c, Nucl.
Phys. {\bf B207}, 461 (1982).
\medskip
\item{[41]} D. Horvat, Z. Naran\v ci\' c, D. Tadi\' c, Phys. Rev. D {\bf
51}, 6277  (1995).
\medskip
\item{[42]} G. E. Brown and B. Weise, Phys. Rep. {\bf 22C}, 279 (1975).
\medskip
\item{[43]} H. Sugawara and F. von Hippel, Phys. Rev. {\bf 172},
1764 (1968).
\medskip
\item{[44]} J. J. Sakurai, {\it Modern Quantum Mechanics\/}
(Addison-Wesley  Publ. Comp., Redwood City, 1985);
\item{[45]}J. D. Bjorken and S. D. Drell, {\it Relativistic Quantum
Mechanics\/} (McGraw-Hill, New York, 1964); see p. 211.
\item{[46]} H. Yukawa, Proc. Phys.-Math. Soc., Japan, {\bf 17} (1935) 48.
\item{[47]} D. Zhang, Y. Ding and T. Ma, Am. J. Phys. {\bf 57}
(1989) 281.
\item{[48]} S. S. Schweber, {\it An Introduction to Relativistic Quantum
Field Theory\/} (Row, Peterson and Co., Evanston, Ill. (1961)).

\vfill
\mpikk
\bye
\enddocument